\documentclass[final,3p,times,twocolumn]{elsarticle}

\newif\iflong
\longtrue
\usepackage[greek,english]{babel} 
\usepackage{makeidx}  

\usepackage{ulem}
\usepackage{longtable}
\usepackage{amssymb}
\usepackage{multirow}
\usepackage{enumitem}
\usepackage{lscape}
\usepackage{subfigure}
\usepackage{caption}
\usepackage{graphicx}
\usepackage{float}
\usepackage{todonotes}
\restylefloat{table}
\usepackage{textcomp}
\usepackage{color}
\usepackage{soul}
\usepackage{verbatim}
\usepackage[font=small,labelfont=bf]{caption}
\usepackage{afterpage}
\usepackage{comment}
\usepackage{tablefootnote}
\usepackage[multiple]{footmisc}

\iflong
	\usepackage{url}
	\usepackage[T1]{fontenc}
	\usepackage{enumerate}
\fi

\definecolor{darkgreen}{rgb}{0,0.392,0}
\definecolor{darkred}{rgb}{0.545,0,0}

\newcommand{\sppbench}{\(\mathrm{SP^{2}Bench}\)}

\newcommand{\colbreak}{\hspace{20pt}}	
\usepackage{siunitx}
\definecolor{my-green}{rgb}{0,0.39,0}
\definecolor{my-red}{rgb}{0.57,0.05,0.05}

\iflong
	\journal{Web Semantics}
\fi

\begin{document}

\iflong
	\begin{frontmatter}
	
	\title{Evaluating Geospatial RDF stores Using the Benchmark Geographica 2\tnoteref{ref-thanks}
	\tnotetext[ref-thanks]{This work was supported in part by the European Commission project TELEIOS (257662), LEO (611141), MELODIES (603525) and ExtremeEarth (825258).}}
	
	\author{Theofilos Ioannidis}
	\ead{tioannid@di.uoa.gr}	
	\author{George Garbis}
	\ead{ggarbis@di.uoa.gr}
	\author{Kostis Kyzirakos}
	\ead{kkyzir@di.uoa.gr}
	\author{Konstantina Bereta}
	\ead{konstantina.bereta@di.uoa.gr}
	\author{Manolis Koubarakis}
	\ead{koubarak@di.uoa.gr}
	\address{Department of Informatics and Telecommunications, 
		National and Kapodistrian University of Athens, 
		University Campus, Ilissia, Athens 15784, Greece}

\else

	\title{Geographica: Benchmarking Scalable Geospatial RDF Stores and Frameworks
	\thanks{This work was supported in part by the European Commission project
	TELEIOS (257662) and LEO (611141).}}
	\titlerunning{Geographica: Benchmarking Scalable Geospatial RDF Stores}  
	\author{George Garbis \and Kostis Kyzirakos \and Manolis Koubarakis}
	\institute{National and Kapodistrian University of Athens, Greece\\
	\{ggarbis,koubarak\}@di.uoa.gr \\
	Centrum Wiskunde \& Informatica, Amsterdam, The Netherlands\\
	Kostis.Kyzirakos@cwi.nl}
	\authorrunning{G. Garbis, K. Kyzirakos and M.Koubarakis} 
	\tocauthor{George Garbis, Kostis Kyzirakos and Manolis Koubarakis}
	\maketitle              
\fi

\begin{abstract}
Since 2007, geospatial extensions of SPARQL, like GeoSPARQL and stSPARQL,
have been defined and corresponding geospatial RDF stores have been
implemented. In addition, some work on developing benchmarks for
evaluating geospatial RDF stores has been carried out. In this paper, we
revisit the Geographica benchmark defined by our group in 2013 which uses
both real world and synthetic data to test the performance and
functionality of geospatial RDF stores.
We present Geographica 2, a new version of the benchmark which extends
Geographica by adding one more workload, extending our existing workloads
and evaluating 5 more RDF stores. Using three different real workloads,
Geographica 2 tests the efficiency of primitive spatial functions in RDF
stores and the performance of the RDF stores in real use case scenarios, a
more detailed evaluation is performed using a synthetic workload and the
scalability of the RDF stores is stressed with the scalability workload.
In total eight systems are evaluated out of which six adequately support
GeoSPARQL and two offer limited spatial support.

\iflong
	\end{abstract}
	\begin{keyword}
		benchmarking \sep geospatial \sep RDF store \sep Linked Open Data \sep 
		GeoSPARQL \sep stSPARQL \sep scalability
	\end{keyword}
	\end{frontmatter}
\else
	\end{abstract}
	\keywords{benchmarking, geospatial, RDF store, Linked Open Data, GeoSPARQL,
	stSPARQL \sep scalability}
\fi

\section{Introduction}
\label{sec:introduction}
Many geospatial datasets have recently been added to the Web of data 
and geospatial extensions to SPARQL, such as GeoSPARQL and stSPARQL, have been defined.

GeoSPARQL \cite{ogc-geosparql} is a standard of the Open Geospatial Consortium (OGC)
for a SPARQL-based query language for geospatial data expressed in RDF.
GeoSPARQL defines a vocabulary (classes, datatypes and properties)
that can be used in RDF graphs to represent geographic features with vector geometries.

The query language stSPARQL \cite{iswc2012-strabon,eswc-temporal} is an extension of SPARQL 1.1
developed by our group for representing and querying geospatial data that changes
over time. Similarly to GeoSPARQL, the geospatial part of stSPARQL defines datatypes 
that can be used for representing in RDF the serializations of vector geometries 
encoded according to the widely adopted OGC standards Well Known Text (WKT)~\cite{herring2011opengis} and Geography Markup Language (GML)~\cite{portele2007opengis}. stSPARQL and GeoSPARQL also define extension functions from the OGC standard ``OpenGIS Simple Feature Access'' (OGC-SFA)~\cite{herring2011opengis} that can be used for querying vector geometries.

The query languages stSPARQL\footnote{\url{http://www.strabon.di.uoa.gr/files/stSPARQL_tutorial.pdf}} and GeoSPARQL were developed independently at around the same time, and have produced very similar representational
and querying constructs. A detailed comparison of stSPARQL and GeoSPARQL is given in~\cite{rr-survey}.

In parallel with the appearance of GeoSPARQL and stSPARQL, researchers have
been implementing geospatial RDF stores that support these SPARQL extensions (e.g., 
 Strabon~\cite{iswc2012-strabon}, Parliament~\cite{Kolas-Parliament}, and uSeekM). The earlier 
approach for the implementation was by extending existing RDF frameworks (e.g., Sesame) with limited geospatial
functionality and relying on state of the art spatially-enabled RDBMSs (e.g., PostGIS) for the storage and querying of geometries (e.g., Strabon and uSeekM with PostGIS). One reason that this hybrid approach had been successful is that the
relational realization of the OGC-SFA standard has been widely adopted by many
RDBMS for storing and manipulating vector geometries. The state of the art in
this area is summarized in the early survey paper~\cite{rr-survey}. 

However new highly competitive geospatial RDF stores appeared lately that belong to the NoSQL graph databases technology family e.g., GraphDB. In addition, as of mid-2018, some RDF Frameworks (e.g., RDF4J, formerly known as Sesame) advanced substantially in terms of GeoSPARQL support, the availability of indexing and search technologies and may make an attractive starting point for building geospatial RDF stores rich in terms of features and more efficient in terms of performance.

The above advances to the state of the art in query languages and implemented
systems has also been matched with work on evaluation and
benchmarking of geospatial RDF stores. Although there are
various benchmarks for spatially-enabled Relational Database Management Systems
(RDBMS)~\cite{sequoia,Patel97buildinga,alacarte,vespa,jackpine,dynamark}, there
are some publications \cite{kolas,iswc2012-strabon,geoknow-del2.1,geographica2013} that study the
performance of geospatial RDF stores but no widely-accepted benchmark exists.

The work described in \cite{kolas} has preceded the GeoSPARQL and stSPARQL proposals, therefore
it does not cover many of the features available in these languages. Only 
point and rectangle geometries and only few topological and non-topological 
functions are included in its workload.
Similarly, only the geospatial RDF store SPAUK~\cite{spauk}, which is a precursor to
Parliament, has been evaluated using this benchmark. \cite{kolas} uses
a synthetic workload only and does not consider real linked geospatial datasets
such as the ones that are available in the LOD cloud today.

In \cite{iswc2012-strabon} authors present the geospatial RDF store
Strabon\footnote{\url{http://www.strabon.di.uoa.gr/}} and they include a section that is an
evaluation targeted mostly to Strabon than a general evaluation benchmark. 
Both a real world workload and a synthetic one is used in \cite{iswc2012-strabon}.
The synthetic workload uses only point geometries and spatial selection 
queries, but it allows the study of performance in a controlled environment.

\cite{geoknow-del2.1} presents a benchmark based on \cite{kolas} and 
adapted to the technological advances at the time. \cite{geoknow-del2.1}
evaluates several geospatial RDF stores taking into account
the expressive power of GeoSPARQL and using real data from OpenStreetMap (OSM)\footnote{\url{https://www.openstreetmap.org/}} of
 various geometry types (points,
lines, polygons). Its workload covers the primary query types covered 
 in \cite{kolas} (spatial location queries, spatial range queries, spatial 
join queries, and nearest neighbor queries)
and additional query types, such as queries using non-topological
spatial functions, and negation and aggregation queries that use spatial filters.

Finally, the previous version of our benchmark \cite{geographica2013}, named Geographica\footnote{Geographica
	(Greek: \greektext Gewgrafik'a\latintext)
	is a 17-volume encyclopedia of geographical
	knowledge written by the Greek geographer, philosopher and historian Strabon
	(Greek: \greektext Str'abwn\latintext)
	in 7 BC. (\url{http://en.wikipedia.org/wiki/Geographica})} was a comprehensive proposal at the time
and has been used to evaluate RDF stores supporting GeoSPARQL and stSPARQL.
It comprises two workloads with their associated datasets and queries: a
\textit{real world} workload and a
\textit{synthetic} workload. The real world workload uses publicly available
linked geospatial data, covering a wide range of geometry types (e.g., points, lines,
polygons). This workload, follows the approach of the benchmark
Jackpine \cite{jackpine} and defines a micro benchmark and a macro benchmark.
The micro benchmark tests primitive spatial functions. The spatial
component of a system is tested with queries that use non-topological functions, spatial
selections, spatial joins and spatial aggregate functions. The macro
benchmark tests performance of selected RDF stores in typical
application scenarios like reverse geocoding, map search and browsing and a
real world use case from the Earth Observation (EO) domain.
For the synthetic workload of Geographica, a generator was developed that produces
synthetic datasets of various sizes and generates queries of varying thematic
and spatial selectivity. In this way, performance of 
geospatial RDF stores can be studied in a closely controlled environment. This workload follows the
rationale of earlier papers \cite{vespa,iswc2012-strabon,brodtRDF3Xgermanoi}.
For reasons of reproducibility, both workloads are publicly
available on the web site\footnote{\url{http://geographica2.di.uoa.gr/}} of the benchmark.

\iflong
The present article revisits~\cite{geographica2013} and offers
the following contributions:
\begin{itemize}
\item We present a new version of Geographica, called Geographica 2, which contains the following extensions. We extended the macro
part of the real world workload of Geographica~\cite{geographica2013} by adding two more application 
scenarios: the \texttt{geocoding scenario} and a scenario that involves the \texttt{computation of  statistics} for geospatial datasets.
The second, an important addition to the original benchmark is the \texttt{scalability workload}, which unveils the behavior of three stores (Strabon, GraphDB and RDF4J) in three key areas, storage space, bulk loading and query response time, all with respect to the number of triples in the dataset. All three aspects measured are critical for developing systems that work with large data sizes and discovering deficiencies can help choose the appropriate system for each case, and guide future research and system improvements in the areas of data storage, indexing strategies and query processing.
\item We include in our evaluation a qualitative comparison of geospatial RDF stores in order to stress the differences between them in terms of supported geospatial features and functionality. 
\item We also include in our evaluation RDF stores such as OpenLink Virtuoso which does not have substantial support of GeoSPARQL\footnote{\url{http://vos.openlinksw.com/owiki/wiki/VOS/VirtGeoSPARQLEnhancementDocs}} or that implement only limited geospatial functionality e.g., only support for points.
We did not include these systems in experiments presented in our previous work \cite{geographica2013},
because we concentrated on systems that are able to execute the complete 
benchmark. 
However, a performance comparison between generic RDF stores 
with limited geospatial capabilities (e.g., handling only point geometries) and 
geospatial RDF stores is quite interesting
as it gives an insight about the trade-off between targeted implementations that cover
a subset of geospatial features and implementations that cover most of the GeoSPARQL
standard. That is why in this article we added such a comparison and the performance of
the aforementioned fully geospatial RDF stores are compared with
Virtuoso and another proprietary RDF store, called here System Y, 
with limited geospatial functionality.
For this purpose, a subset of Geographica was used which covers only
point geometries.
\item In \cite{geographica2013}, 
we chose to test three well known open source
RDF stores that provide GeoSPARQL functionality, namely Strabon, Parliament and 
uSeekM. In this article, we additionally test a proprietary geospatial RDF store,  
called here System X\footnote{Two proprietary RDF stores are used in this manuscript. We refer to them as System X and System Y since their licenses do not allow us to reveal their names. However they are still included in our experiments, because it is interesting to have a comparison between open source
	systems, that are usually developed in academic environments and focus on extending
	the state of the art and proprietary systems, that mainly focus on serving real applications well.}, the free edition of the GraphDB v8.6.1 NoSQL graph database and the RDF4J v2.4.3 Semantic Framework. To the best
of our knowledge, these systems are the only ones that currently provide
support for a rich subset of GeoSPARQL and stSPARQL, so we did not include any other
system in the main part of Geographica.

\else
We chose to test the systems Strabon, Parliament and uSeekM and a proprietary 
geospatial RDF store (which we will call System X for reasons of anonymity). 
To the best
of our knowledge, these systems are the only ones that currently provide
support for a rich subset of GeoSPARQL and stSPARQL. Other RDF stores like
OpenLink Virtuoso, OWLIM and AllegroGraph, allow only the representation of
point geometries and provide support for a few geospatial 
functions~\cite{rr-survey}. 
The limited functionality provided by these systems did not allow us 
to include them in the experiments presented in this paper.
A comparison between generic RDF stores with limited geospatial capabilities and
geospatial RDF stores are given in the long version of this paper\footnote{\url{http://geographica.di.uoa.gr/files/Geographica-ISWC-2013-long-version.pdf}}.
\fi
\end{itemize}

The rest of the paper is organized as follows. Section~\ref{sec:background}
presents the main data models and query languages for linked geospatial data.
Section~\ref{sec:relatedwork}
presents previous related work. Section~\ref{sec:geographica-functional} presents
well-known geospatial RDF stores and compares them in terms of geospatial 
functionality that they offer.
The benchmark is described in Section~\ref{sec:benchmark} and its results are discussed in
Section~\ref{sec:benchmarkresults}.  Section~\ref{sec:resultPoints} discusses the performance of generic RDF stores with limited geospatial capabilities in
comparison to geospatial RDF stores providing full geospatial capabilities. Finally,
the contributions of the paper are summarized and future work is discussed in
Section~\ref{sec:conclusions}.

\section{Background}
\label{sec:background}
In this section, we introduce GeoSPARQL and stRDF/stSPARQL. 
GeoSPARQL allows the representation of
geographic data in RDF and querying it using an extension
of SPARQL. stRDF is an extension of RDF that allows 
the representation of geospatial linked data that evolves over time.
stSPARQL is an extension of SPARQL that permits querying
stRDF data taking into account its spatial and temporal
dimension.

\subsection{GeoSPARQL}
GeoSPARQL is a standard, developed by the OGC, that defines 
a core RDF/OWL vocabulary
and a set of SPARQL extension functions for representing and
querying linked geospatial data. 
GeoSPARQL follows a modular architecture, shown in 
Figure~\ref{fig:geosparql-arch}, that defines six conformance
classes. Each implementation may support one or more conformance
classes.

The \textit{Core} conformance class defines a basic RDFS/OWL vocabulary for
representing geospatial data. This vocabulary includes the class \texttt{SpatialObject}
and its subclasses \texttt{Feature} and \texttt{Geometry}. Features can have geometries
and geometries can be encoded by the OGC standards WKT and GML.
The \textit{Topology Vocabulary Extension} defines a vocabulary for
asserting topological relations between spatial objects. This conformance
class is parameterized so that an implementation can use any of the well-known
topological relation families: RCC8~\cite{DBLP:conf/kr/RandellCC92}, Egenhofer~\cite{DBLP:conf/fodo/Egenhofer89}, and OGC SFA.
The \textit{Geometry Extension} conformance class defines a vocabulary
for asserting information about geometry data and query functions operating on
geometry data. This class defines the appropriate RDFS datatypes for asserting
geometry data as literal values. A geometry literal can be encoded in WKT or in GML;
 this is defined by a parameter of the conformance class.
The \textit{Geometry Extension} conformance class also defines non-topological 
functions that operate on geometry data and return geometry or numeric data (e.g., the distance between two geometries).
The \textit{Geometry Topology Extension} conformance class defines topological query functions
that operate between two geometry literals and return if a topological relation holds between 
their corresponding geometries. According to parameters of the \textit{Geometry Topology Extension}, GeoSPARQL implementations
can support any of the geometry serializations (WKT, GML) and any of the aforementioned
topology relation families (RCC8, Egenhofer, OGC SFA). 
The \textit{RDFS Entailment Extension} conformance class defines a mechanism for
matching implicitly derived RDF triples in GeoSPARQL queries. 
Finally, the \textit{Query Rewrite Extension} conformance class defines rules
to support implication of direct topological predicates between features based 
on the geometries of these features. This is achieved by a set of RIF rules that expand 
direct topological predicates (from \textit{Topology} vocabulary) into a series of triple 
patterns and an invocation of the corresponding extension function (from \textit{Geometry 
Topology} vocabulary). For example a RIF rule asserts that if the function 
\texttt{geof:sfIntersects} holds between two geometry literals then the topological 
relation \texttt{geo:sfIntersects} holds among the corresponding features. Using these 
rules, queries that contain a topological relation between two variables standing for features
(e.g., \texttt{?x geo:sfIntersects ?y}) can be re-written into queries that contain topological functions 
standing for two literals (e.g., \texttt{geof:sfIntersects(?f1, ?f2)}).

\begin{figure}[t]
	\centering
	\includegraphics[width=0.3\textwidth]{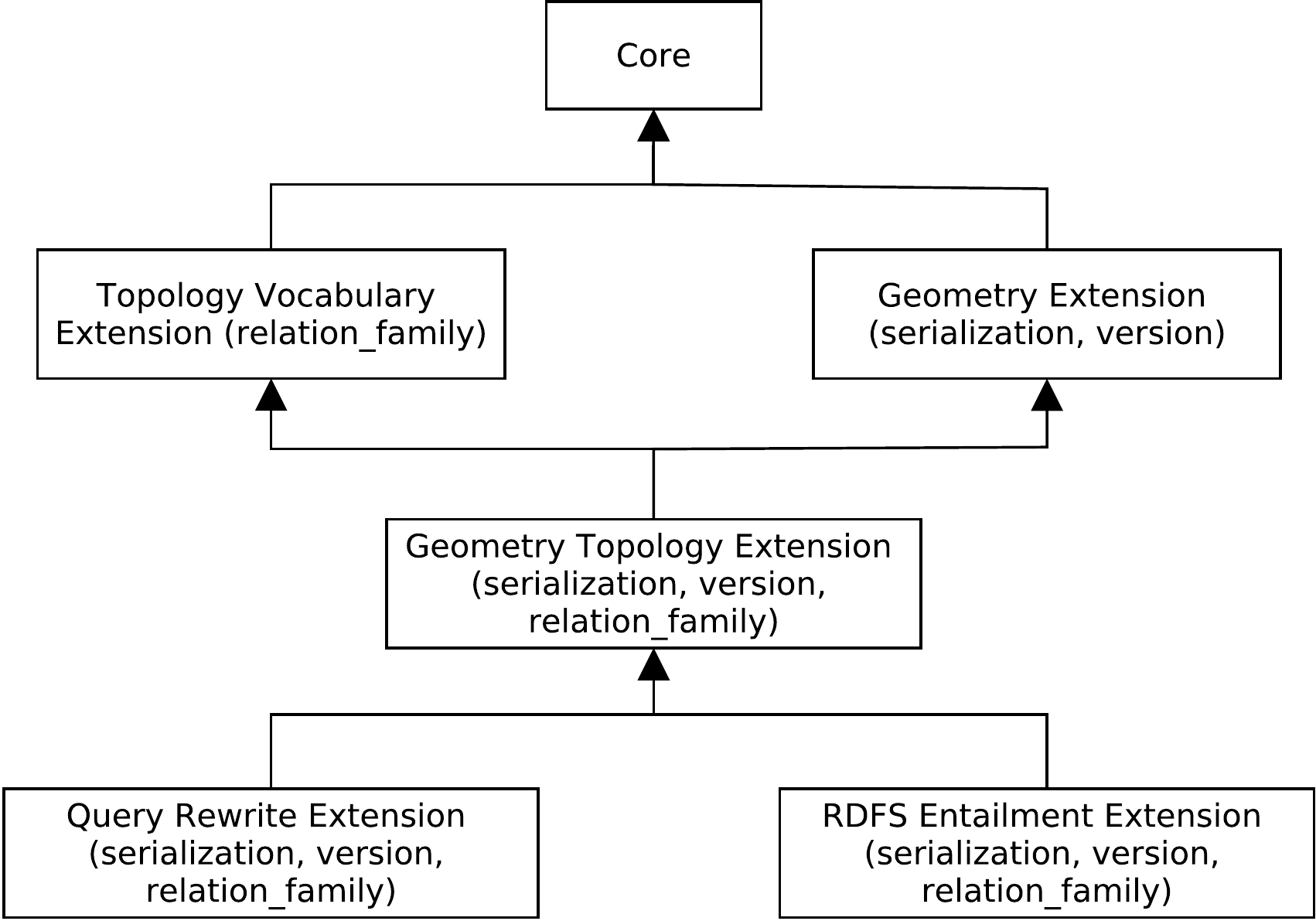}
	\caption{Conformance class dependency graph of GeoSPARQL}
	\label{fig:geosparql-arch}
\end{figure}

\subsection{stRDF and stSPARQL}
stRDF and stSPARQL are extensions of RDF and SPARQL that allow the representation
and querying of linked spatiotemporal data. stSPARQL has been developed by our group
at the same time as GeoSPARQL, and has resulted in a similar representation model. 
stSPARQL, like GeoSPARQL, defines two datatypes (\texttt{strdf:WKT}, 
\texttt{strdf:GML}) for encoding geometry literals and a 
set of functions that correspond to the functions of the 
\textit{Geometry} extension and the \textit{Geometry Topology} extension of
GeoSPARQL. In addition to these functions, stSPARQL defines directional relation functions 
that are based on the minimum bounding boxes of two geometries (e.g., if a geometry is 
strictly on the left of another geometry) and spatial aggregate functions that operate on 
sets of geometries and compute new geometry objects (e.g., the union of a set of geometries). Note that both GeoSPARQL and stSPARQL include functions that computes the union of two given geometry literals, stSPARQL additionally includes a function computes the union of a given set of geometry literals.

In addition to its geospatial features, stRDF has a temporal component which
can represent the \textit{valid time} of a triple and
stSPARQL defines a set of temporal functions for querying the valid time
of triples. The temporal component of stRDF and stSPARQL are described in \cite{eswc-temporal}
and it will not be considered in the rest of the paper.

\section{Related Work}
\label{sec:relatedwork}
\iflong
	This section discusses the most important benchmarks that are relevant
to Geographica. First benchmarks for SPARQL query 
processing are presented, followed by those from the geospatial relational databases area and, finally,
we concluded with benchmarks for querying linked geospatial data.

\subsection{Benchmarks for SPARQL Query Processing}
\label{sec:relatedwork:rdfbenchmarks}

A well-known benchmark for Semantic Web knowledge base systems is
the Lehigh University Benchmark (LUBM)~\cite{lubm}.
It tests scalability, efficiency and reasoning 
capabilities
of memory-based systems and systems with persistent storage.
Concerning reasoning capabilities three degrees are tested: (i) RDFS reasoning, 
(ii) partial OWL reasoning, and (iii) complete or almost complete OWL Lite reasoning.
The authors propose a benchmark
with fourteen queries over a large dataset that commits to an 
ontology describing the university domain. 
This is one of the first benchmarks for SPARQL query processing and its design 
is based on techniques applied to older database benchmarks. 
For example, its data are synthetically generated so that 
the data size can be arbitrarily large and
the selectivity and output size of each query can be predefined.
Finally, LUBM uses a set of predefined performance metrics, namely load time, 
repository size, and query response time and it also suggests two new 
metrics about completeness and soundness of the query evaluations.

The SPARQL performance benchmark (\sppbench{})~\cite{sppbench} is an RDF
benchmark directed towards a comprehensive performance evaluation of RDF
stores. The authors of this benchmark cover a wide spectrum of 
SPARQL features. They define queries with various SPARQL
operators (e.g., UNION, OPTIONAL, FILTER) and solution modifiers
(e.g., DISTINCT, ORDER BY, LIMIT) and they also test negation as failure queries.
Queries are grouped in two categories: (i) long path chains and (ii) bushy
patterns and they are designed so they are amenable to SPARQL optimization 
techniques (e.g., triple reordering, FILTER pushing). \sppbench{} 
 defines a data generator that produces datasets resembling the DBLP dataset.
Another SPARQL query processing benchmark is the Berlin SPARQL Benchmark 
(BSBM) \cite{bsbm}. This benchmark 
compares the performance of native RDF stores with the performance of
SPARQL-to-SQL rewriters. BSBM uses synthetic data that describes
an e-commerce use case. Different vendors offer 
products and reviews have been posted about these products by consumers. 
Unlike the 
systematic approach of \sppbench{}, the Berlin SPARQL Benchmark uses an 
application-based query mix that emulates the search and navigation pattern of a 
consumer looking for a product. Thus, the query mix
covers an adequate range of SPARQL features but not all of them. 
Since BSBM is application-oriented, it uses metrics 
defined for application scenarios and not for single queries, such
as query mixes per hour (QMpH) and queries per second (QpS).

The DBpedia SPARQL benchmark (DBPSB) \cite{dbpsb} follows a 
different approach and proposes a generic 
SPARQL benchmark creation procedure which is based on real application
data and query logs. DBPSB proposes a technique to create data of arbitrary size
that resembles real data. This technique enables increasing or decreasing
the size of a real RDF dataset so that generated data retains the
basic network characteristics (in and out degree) and other characteristics,
such as the number of classes and properties of the original data. 
Also, DBPSB proposes a query analysis technique to extract representative
queries of a set of real queries. The techniques of DBPSB
were applied in the use case of DBpedia\footnote{\url{https://wiki.dbpedia.org/}} but they can be applied
to any dataset and query log to produce a use case specific benchmark.

Finally, the Waterloo SPARQL Diversity Test Suite (WatDiv)~\cite{DBLP:conf/semweb/AlucHOD14} provides stress testing tools for RDF systems that face diverse queries and varied
workloads. It defines two classes of query features based on which it discusses the variability of the datasets and workloads in a SPARQL benchmark: (i) \texttt{structural} features such as triple pattern count, join vertex count, degree and type and (ii) \texttt{data-driven} features such as result cardinality, filtered triple pattern (f-TP) selectivity, basic graph patterns (BGPs) restricted f-TP selectivity and join-restricted f-TP selectivity.  The second part of ~\cite{DBLP:conf/semweb/AlucHOD14} includes an experimental evaluation of other SPARQL benchmarks with emphasis on identifying test cases that are not handled by these benchmarks.
The last part of ~\cite{DBLP:conf/semweb/AlucHOD14} is an experimental evaluation of five RDF stores of different architectures using WatDiv, demonstrating that none of the systems is sufficiently robust across a diverse set of queries.

\subsection{Benchmarks for Geospatial Relational Databases}
\label{spatialdbbenchmarks}

One of the first benchmarks for spatial databases is the SEQUOIA
benchmark~\cite{sequoia} which focuses on Earth Science use cases 
and which has been used for testing many
Geographic Information Systems (GIS). SEQUOIA uses real data (satellite raster data,
point locations of geographic features, land use/land cover polygons and data
about drainage networks covering the area of USA) and real queries.
It also considers different scales of datasets and use cases (e.g., local or 
national scale). The SEQUOIA benchmark is formed by 11 queries trying to 
cover the most usual tasks in Earth Science, like (i) data loading and 
building of respective indexes, (ii) raster data management,
(iii) selections based on spatial and non-spatial filters, (iv) spatial joins,
(v) and a recursive spatial query. 

SEQUOIA was later extended by
\cite{Patel97buildinga} to evaluate the geospatial DBMS Paradise.
In \cite{Patel97buildinga} DeWitt et al. study traditional database techniques and how these
techniques can be used (or extended to be used) in geospatial query processing.
SEQUOIA takes into account only points and polygons, while \cite{Patel97buildinga} 
also tests polylines and circles and broadens the tested functionality (e.g., it 
tests spatial aggregate functions).
Finally, a methodology, called resolution scaleup, is applied to scale up geospatial 
data. This technique simulates the zoom-in operation of map applications. Existing
spatial features are represented in more detail by adding more points to their
boundaries, and at the same time new smaller spatial features appear around the 
existing ones.

Rather than focusing on evaluating performance of systems, the 
\'{A} La Carte~\cite{alacarte} benchmark compares performance of spatial 
join techniques. In particular, \cite{alacarte} tests the following algorithms: 
nested loops, scan and index, and synchronized tree traversal.
A data generator is presented that generates rectangles with edges 
parallel to the axes. The \'{A} La Carte generator enables data 
of arbitrary size that can follow
various statistical distributions (uniform, normal and exponential). This
allows for the generation of realistic data in terms of spatial distribution. However,
the fact that generated rectangles have edges parallel to the axes does
not allow testing the full process of spatial evaluation in a DBMS.
The usual workflow of a spatial evaluation is composed of two steps. The first
step utilizes a spatial index, which is built according to the minimum
bounding boxes of geometries, to find candidate results. This step is
called \textit{filtering} step. The second step, which is called \textit{refinement} 
step, tests the actual geometries and discards false positives generated by
the filtering step. Using rectangles with edges parallel to the axes means that
their actual geometries are identical to their minimum bounding boxes. So,
the exact answer is already found by the filtering step that does not
generate any false positive.

In order to generate data and conduct experiments, 
\'{A} La Carte defines some 
statistical models for the generator that resembles typical cartographic applications.
These models are the following: (i) ``Biotopes'' simulates a biotope map where there
are few large rectangles uniformly distributed that may overlap but not to a large
degree, (ii) ``Cities'' simulates the distribution of cities and it
is composed of many small rectangles (modelled as squares) uniformly distributed
around the map, finally, (iii) a hybrid model is defined that resembles a word map. This 
model comprises two nested submodels. First, a ``Biotopes'' model creates the 
continents of the world and inside each continent there are rectangles modelled
by the ``Cities'' model.

A more complex data generator is used in VESPA~\cite{vespa} to compare 
PostgresSQL with the Rock \& Roll deductive object-oriented database system.
The data generator of VESPA produces spatial features that resemble
real maps. The spatial features that are produced by the VESPA generator
represent land ownership, states, land use, roads, streams, gas lines and 
points of interest. They are uniformly distributed, in contrast to spatial
features generated by the \'{A} La Carte generator, but  they are more complex 
than simple rectangles. The dataset consists not only of polygons but also of lines 
and points. 
The produced polygons are hexagons and triangles, so their edges are not parallel
to the axes and both filtering and refinement step of spatial joins can be tested.
Apart from
spatial selection and spatial analysis queries VESPA also tests updates and 
spatial aggregate queries which are not tested by previous benchmarks.

Finally, a more generic benchmark is Jackpine~\cite{jackpine}.
Jackpine defines two kinds of benchmarking, micro and macro. Micro benchmarking 
tests spatial functions in isolation, in order to evaluate the performance of 
systems in evaluating spatial selection, spatial join, and spatial analysis queries.
Macro benchmarking defines real application scenarios as series 
of queries and tests the performance of systems for evaluating the entire series 
of queries for each scenario. Tested scenarios range from simple ones, like geocoding 
and reverse geocoding, or more complex scenarios like flood risk and toxic spill analysis.

\subsection{Benchmarks for Geospatial RDF Stores}
The first published benchmark for querying geospatial data encoded in RDF has 
been proposed in~\cite{kolas}. \cite{kolas} extends LUBM to include spatial 
entities and test the performance of spatially  enabled RDF stores. 
The data generator of LUBM is extended so that each university, department or 
student gets a spatial extent (rectangle or point).
LUBM queries are extended to cover four primary types of spatial queries, 
namely spatial location queries, spatial range queries, spatial join queries, and
nearest neighbor queries. Range queries aim to test cases of various selectivity, 
while spatial joins aim to test whether the query planner selects a good plan
by taking into account the selectivity of the spatial and ontological part 
of each query.

Another systematic evaluation of geospatial RDF stores has been done
in \cite{iswc2012-strabon}. 
In the context of presenting the geospatial RDF store Strabon, experiments 
studying its performance were conducted. Strabon is compared with
Parliament~\cite{Kolas-Parliament} and an implementation on top of RDF-3X~\cite{rdf3x} that supports spatial 
queries. In this evaluation more emphasis is given to 
study Strabon itself rather than creating a benchmark for
various RDF stores. This is why different variations of Strabon are 
studied in order to demonstrate advantages and disadvantages of different
implementation choices.
Two workloads are used: one based on real world linked
data and a synthetic one. The first workload consists of eight real world 
queries that are either frequently used in Semantic Web applications
(e.g., DBpedia and LinkedGeoData endpoints) or they demonstrate
the spatial extensions of Strabon. This workload contains thematic queries
as well as spatial selection and spatial join queries and queries using
non-topological spatial functions.
The second workload  uses a modified
version of the data generator of \cite{brodtRDF3Xgermanoi} to generate spatial datasets with arbitrary size and predefined characteristics.
The data generator produces spatial data with point geometries and only spatial
query selections are studied.

Recently in \cite{geoknow-del2.1}, Patroumpas et al. have reviewed 
the state of the art
in managing geospatial data in the Semantic Web. \cite{geoknow-del2.1} starts by
presenting basic concepts and standards (e.g., GeoSPARQL) about geospatial data in 
the Semantic Web, then it presents the current state of the art 
geospatial RDF stores and a qualitative comparison between them. Finally, 
\cite{geoknow-del2.1} presents and performs an evaluation of the
performance of the geospatial RDF stores. For this evaluation, 
\cite{geoknow-del2.1} uses data from OpenStreetMap and it follows the guidelines 
of \cite{kolas} to define a query workload. This workload consists of basic 
queries that cover the four primary types of spatial queries that has been suggested 
in \cite{kolas} and geospatial analysis queries  that cover query types not studied 
in \cite{kolas}. These are queries that use 
non-topological spatial functions, combine thematic and spatial criteria, and 
aggregate and negation queries that use spatial filters.

Geographica goes beyond the above benchmarks  \cite{kolas,geographica2013,geoknow-del2.1} as it is the first benchmark that combines all of the following features.
It contains a real world workload that uses publicly available linked geospatial data, covering a wide range of geometry types (e.g., points, lines, polygons).
The real world workload follows the approach of the benchmark
Jackpine \cite{jackpine} and defines a micro benchmark and a macro benchmark.
The micro benchmark tests primitive spatial functions. The spatial
component of a system is tested with queries that use non-topological functions, spatial selections, spatial joins and spatial aggregate functions. The macro
benchmark tests performance of selected RDF stores in typical
application scenarios like reverse geocoding, map search and browsing and a
real world use case from the EO domain.
It also contains of a synthetic workload, using a generator that produces
synthetic datasets of various sizes and generates queries of varying thematic
and spatial selectivity. In this way, performance of 
geospatial RDF stores can be studied in a closely controlled environment. This workload follows the
rationale of earlier papers \cite{vespa,iswc2012-strabon,brodtRDF3Xgermanoi}.

\fi
\section{A Functional Comparison of Geospatial RDF Stores}                     
\label{sec:geographica-functional}                                              
This section presents all of the RDF stores known to the authors that implement some geospatial
functionality, and compares them in terms of 
the geospatial functionality that they offer.

Although GeoSPARQL is an OGC standard since 2012, 
it is not fully supported by any geospatial RDF store. Usually systems do not implement
the Query-Rewrite Extention. Also there are some RDF stores that provide geospatial 
capabilities which are limited to point geometries. 

A common problem area is CRS support. A coordinate reference system (CRS) also referred 
to as spatial reference system (SRS) is a coordinate system that is related to an object 
(e.g., the Earth) through a datum which specifies its origin, scale, and orientation.
Authorities that maintain partial or non fully compatible lists of CRSs include
OGC which maintains a set of CRS URIs\footnote{\url{http://www.opengis.net/def/crs/}} and 
the International Association of Oil and Gas Producers (IOGP)\footnote{\url{https://www.iogp.org/}}
which after the absorption of the European Petroleum Survey Group (EPSG) maintains the EPSG online registry of geodetic parameters\footnote{\url{https://www.epsg-registry.org/}}.

We organize our comparison according to the GeoSPARQL standard. We indicate which extensions of GeoSPARQL are supported by each RDF store, which spatial relation classes and geometry serialization formats are implemented, and whether multiple CRSs are supported. 
We have also included a selection of available geospatial extensions which are not part of GeoSPARQL, such as the use of geometry literals as objects in triple patterns, the spatial aggregate functions defined by stSPARQL~\cite{iswc2012-strabon} and three main spatial query classes that are used for querying points. A tabular view of this comparison can be found in Table~\ref{tab:functional-comparison}. The rest of the section essentially explains the contents of Table 1 by discussing in detail the functionality of each system.

\begin{table*}[!t]
	\tiny
	\centering
	\resizebox{\textwidth}{!}{
		\begin{tabular}{|c|c|c|c|c|c|c|c|c|}
			\hline
			System & Strabon & uSeekM & Parliament & System X & Virtuoso & System Y & GraphDB & RDF4J \\
			\hline \hline
			
			\multicolumn{9}{|c|}{GeoSPARQL} \\
			\hline
			Core & \color{darkgreen}yes & \color{darkgreen}yes & \color{darkgreen}yes & \color{darkgreen}yes & \color{darkred}no & \color{darkred}no & \color{darkgreen}yes & \color{darkgreen}yes \\
			\hline
			Topology Vocabulary & \color{darkred}no & \color{darkgreen}yes & \color{darkgreen}yes & \color{darkgreen}yes & \color{darkred}no & \color{darkred}no & \color{darkgreen}yes & \color{darkred}no \\
			\hline
			Geometry & \color{darkgreen}yes & \color{darkgreen}partial & \color{darkgreen}yes & \color{darkgreen}yes & \color{darkred}no & \color{darkred}no & \color{darkgreen}partial & \color{darkgreen}partial \\
			\hline
			Geometry Topology & \color{darkgreen}yes & \color{darkgreen}yes & \color{darkgreen}yes & \color{darkgreen}yes & \color{darkred}no & \color{darkred}no & \color{darkgreen}yes & \color{darkgreen}yes \\
			\hline
			Query Rewrite & \color{darkred}no & \color{darkred}no & \color{darkred}no & \color{darkred}no & \color{darkred}no & \color{darkred}no & \color{darkred}no & \color{darkred}no \\
			\hline
			RDFS Entailment & \color{darkred}no & \color{darkgreen}yes & \color{darkgreen}yes & \color{darkgreen}yes & \color{darkred}no & \color{darkred}no & \color{darkgreen}yes & \color{darkgreen}yes \\
			\hline \hline
			
			\multicolumn{9}{|c|}{Relation classes} \\
			\hline
			OGC-SFA & \color{darkgreen}yes & \color{darkgreen}yes & \color{darkgreen}yes & \color{darkgreen}yes & \color{darkred}no & \color{darkred}no & \color{darkgreen}yes & \color{darkgreen}yes \\
			\hline
			Egenhofer & \color{darkgreen}yes & \color{darkgreen}yes & \color{darkgreen}yes & \color{darkred}no & \color{darkred}no & \color{darkred}no & \color{darkgreen}yes & \color{darkgreen}yes \\
			\hline
			RCC8 & \color{darkgreen}yes & \color{darkgreen}yes & \color{darkgreen}yes & \color{darkred}no & \color{darkred}no & \color{darkred}no & \color{darkgreen}yes & \color{darkgreen}yes \\
			\hline \hline 
			
			\multicolumn{9}{|c|}{Geometry Serializations} \\
			\hline
			WKT & \color{darkgreen}yes & \color{darkgreen}yes & \color{darkgreen}yes & \color{darkgreen}yes & partial & \color{darkred}no & \color{darkgreen}yes & \color{darkgreen}yes \\
			\hline
			GML & \color{darkgreen}yes [GML-SF] & \color{darkred}no & \color{darkgreen}yes [GML-SF] & \color{darkred}no & \color{darkred}no & \color{darkred}no & \color{darkred}no & \color{darkred}no \\
			\hline \hline 
			
			CRS support & \color{darkgreen}yes & \color{darkred}no & \color{darkgreen}yes & \color{darkgreen}yes & partial & partial & \color{darkred}no & \color{darkred}no \\
			\hline \hline 
			
			\multicolumn{9}{|c|}{Functionality not described by GeoSPARQL} \\
			\hline
			Aggregation & \color{darkgreen}yes & \color{darkred}no & \color{darkred}no & \color{darkred}no & \color{darkred}no & \color{darkred}no & \color{darkred}no & \color{darkred}no \\
			\hline
			Geometry literal as Object & \color{darkred}no & \color{darkred}no & \color{darkred}no & \color{darkred}no & \color{darkred}no & \color{darkred}no & \color{darkgreen}yes & \color{darkred}no \\
			\hline \hline 
			
			\multicolumn{9}{|c|}{Specific functionality for RDF stores with limited geospatial capabilities} \\
			\hline
			Buffer & \color{darkgreen}yes & \color{darkgreen}yes & \color{darkgreen}yes & \color{darkgreen}yes & \color{darkgreen}yes & \color{darkgreen}yes & \color{darkgreen}yes & \color{darkgreen}yes \\
			\hline
			Distance & \color{darkgreen}yes & \color{darkgreen}yes & \color{darkgreen}yes & \color{darkgreen}yes & \color{darkgreen}yes & \color{darkgreen}yes & \color{darkgreen}yes & \color{darkgreen}yes \\
			\hline
			Bounding Box & \color{darkgreen}yes & \color{darkgreen}yes & \color{darkgreen}yes & \color{darkgreen}yes & \color{darkgreen}yes & \color{darkred}no & \color{darkgreen}yes & \color{darkgreen}yes \\
			\hline 
			
	\end{tabular}}
	\caption{Functionality of geospatial RDF stores}
	\label{tab:functional-comparison}
\end{table*}

\subsection{Geospatial RDF Stores that Conform to the GeoSPARQL Standard}
\label{sec:geographica-functional:full}

The system Strabon, which has been developed by our group, is a storage and query evaluation module for stRDF/stSPARQL~\cite{iswc2012-strabon}. 
Strabon extends the well-known RDF store Sesame, allowing it to manage both thematic and spatial data expressed in stRDF and stored in the PostGIS spatially enabled DBMS. 
The version 3.2.9 of Strabon fully implements 
stSPARQL, which provides 
the machinery of the OGC SFA standard
as well as spatial aggregation functions, other useful
spatial functions (e.g., directional relations) and temporal extension functions.
Given the close relationship between stSPARQL and GeoSPARQL~\cite{rr-survey},
it was straightforward to implement the relevant subset of GeoSPARQL in Strabon.
Strabon implements fully the Core, Geometry Extension and Geometry 
Topology Extension components of GeoSPARQL. It supports all three topological 
relation classes
defined by GeoSPARQL (OGC-SFA, Egenhofer, RCC8), both geometry
serializations (WKT, GML) and multiple CRS.

OpenSahara uSeekM\footnote{\url{https://opensahara.com/projects/useekm/}} 
also builds upon the RDF store Sesame. uSeekM is based
on the native store of Sesame to store and query thematic information and
it utilizes PostGIS for storing and querying spatial data.
uSeekM supports the majority of GeoSPARQL, namely the GeoSPARQL Core,
Topology Vocabulary Extension, Geometry
Topology Extension, RDFS Entailment Extension components and partially the Geometry Extension. uSeekM implements all 
three relation family classes (OGC SFA, Egenhofer, RCC8). Since the implementation does not use URIs for coordinate reference systems, it does not conform to requirement 20\footnote{\url{https://opensahara.com/issues/675}}, because \texttt{geof:getSRID()} returns an integer instead of a URI, and fails requirement 12\footnote{\url{https://opensahara.com/issues/675}} of the GeoSPARQL standard, because axis order interpretation in \texttt{geo:wktLiterals} is fixed lon-lat/x-y ordering and not derived by the spatial reference system of the literal. It supports only the WKT serialization for geometries in WGS84 CRS. Therefore, the Geometry Extension support is marked as partial. Finally, uSeekM implements some extension functions (not defined
by GeoSPARQL), which compute, for example,
the area of a geometry and the shortest line between two geometries accordingly. 
For spatial indexing uSeekM utilizes a PostGIS database to create an R-Tree-over-GiST~\cite{ramsey2004postgis}.
The optimizer will check whether the query evaluation will benefit from the extra index
and if so, the spatial part of the query is executed by PostGIS using the R-Tree and 
the rest of the query is executed by the native store of Sesame.

The RDF store Parliament\footnote{\url{http://parliament.semwebcentral.org/}}~\cite{Kolas-Parliament} implements most of the functionality of GeoSPARQL except the Query Rewrite Extension. Both WKT and GML serializations are supported as
well as multiple CRS and all three topological relation family classes.
Unlike Strabon and uSeekM, which detect spatial objects from their datatype, Parliament
works as follows. The RDF
graph is scanned for triples that contain \texttt{geo:asWKT} or \texttt{geo:asGML} predicates 
and for each matching triple Parliament creates a record for the geometry that is represented 
in the object of the triple and inserts it into its spatial index (a standard R-tree 
implementation). The query optimizer tries
to split SPARQL queries into multiple parts and produce an optimized query plan
between the spatial and thematic components of the query. The current version of 
Parliament (v2.7.4) concentrates on optimizing query patterns (using the 
Topology Vocabulary extension of GeoSPARQL) while it omits optimization for 
functions in the filter clause of a query. However, when the topological comparisons
are included in the triple patterns as predicates instead of the corresponding 
filter functions, the optimizer of Parliament takes both thematic and spatial dimensions
into consideration in order to produce a better plan.

The proprietary RDF store System X also supports the GeoSPARQL standard for representing and 
querying geospatial data in RDF. 
System X supports the Core, Topology Vocabulary Extension, Geometry Extension, Geometry 
Topology Extension, and RDFS Entailment Extension components of GeoSPARQL. 
Multiple CRS are supported but only the WKT geometry serialization and the OGC SFA relation family class are supported. 
Additionally, System X offers some specific functions that are not defined by
GeoSPARQL, such as computing the area or
the minimum bounding rectangle of a geometry, computing the union of two geometries, etc.
System X uses an R-Tree to index spatial data. 
When creating a spatial index, a user should define the CRS which will be used to 
create the spatial index, the minimum and maximum value for each dimension of data 
and a positive number indicating how close together two points must be to be 
considered the same point.

RDF4J\footnote{\url{http://rdf4j.org/}} is a Java-based RDF framework which provides for creating, parsing, reasoning and querying over linked data. It offers two core store NoSQL implementations. The Memory Store allows the creation of a very fast transactional RDF repository in main memory with optional persistence to disk. A more scalable alternative which allows the creation of transactional RDF repository of up to a hundred millions of triples is the Native Store, which uses direct disk I/O for persistence. Many of the systems examined in this paper have extended the precursor of RDF4J, Sesame, which additionally offered out of the box a third repository implementation, the RDBMS Store supporting persistence in PostgreSQL and MySQL database systems. Its architecture allows for constructing repositories in a layered approach using \texttt{Sails(Storage And Inference Layer)} adding storage and inference options. Starting from October 2018 and version 2.4.3, RDF4J offers adequate geospatial functionality and GeoSPARQL support through the use of LocationTech's  Spatial4J\footnote{\url{https://projects.eclipse.org/projects/locationtech.spatial4j}} and JTS\footnote{\url{https://projects.eclipse.org/projects/locationtech.jts}} libraries. RDF4J supports the Core, Geometry Topology Extension, RDFS Entailment Extension and partially the Geometry Extension components of GeoSPARQL. It supports all three topological relation classes (OGC-SFA, Egenhofer, RCC8), non-topological and common query functions and the WKT geometry serialization. Repositories can use the RDFS Inference Sail. Although GeoSPARQL is supported natively on all types of store implementations, geospatial querying on large datasets is advertized to benefit when enabling the Lucene Sail which spatially indexes a customizable list of fields that contain the \texttt{<http://www.opengis.net/ont/geosparql\#asWKT>} field by default.

GraphDB\footnote{\url{http://graphdb.ontotext.com/documentation/free/}} v8.6.1 (formerly known as OWLIM) is a NoSQL semantic graph database enhanced with geospatial capabilities and it is the flagship product of company Ontotext. GraphDB is implemented as a SAIL (Storage and Inference Layer) of the RDF4J Framework v2.3.2 and can support billions of triples per server. The supported semantics, which are RDF rule-entailment by default, can be configured through ruleset definitions such as RDFS with OWL Lite and OWL2 profiles RL and QL. GraphDB supports the Core, Topology Vocabulary Extension, Geometry Topology Extension, and RDFS Entailment Extension and partially the Geometry Extension components of GeoSPARQL. Only the WKT geometry serialization and the default WGS84 CRS are supported. It supports all three topological relation classes (OGC SFA, Egenhofer, RCC8). A provided extension which is not part of the GeoSPARQL specification is the ability to use geometry literals in the object position of triple patterns. For its geospatial capabilities it relies on a uSeekM implementation. The spatial index mechanism is controlled through an optional GeoSPARQL-plugin. 
The plugin supports two approximate matching indexing algorithms, a quad prefix tree~\cite{DBLP:journals/acta/FinkelB74}, which is the default option, and a geohash prefix tree~\cite{DBLP:conf/dexa/VanT15}, each one with a different range of accuracy values. 
 GraphDB also provides support for the \texttt{WGS84 Geo Positioning RDF vocabulary}\footnote{\url{https://www.w3.org/2003/01/geo/wgs84_pos}} which allows for the representation of latitude, longitude and altitude of features with geo-spatial index and basic operations such as distance calculation between points, filtering points within a rectangle, polygon or circle.

\subsection{RDF Stores with Limited Geospatial Capabilities}
\label{sec:geographica-functional:limited}

OpenLink Virtuoso\footnote{\url{http://virtuoso.openlinksw.com/}} provides geospatial
support for representing and querying two-dimensional point geometries.
Virtuoso allows geometries to be expressed either in WGS84 or a flat 
two-dimensional plane. 
Virtuoso does not support GeoSPARQL but it models geometries by typed literals like stSPARQL 
and GeoSPARQL. For this purpose, it introduces its own datatype \texttt{virtrdf:Geometry}. The 
value of such a literal is the WKT serialization of a point.
Virtuoso offers vocabulary for a subset of the ISO 13249 SQL/MM standard~\cite{DBLP:conf/btw/Stolze03} to
perform geospatial queries using SPARQL. 
For example, a user can ask for points in a region utilizing SPARQL functions 
corresponding to the
\texttt{st\_intersects}, \texttt{st\_contains}, and \texttt{st\_within} SQL/MM functions.
Note that these functions are extended with a third argument (called precision)
which specifies a maximum distance between two points such that the points will still
be considered to overlap with each other. Thus, these functions can support buffer 
queries exploiting the spatial index of Virtuoso. Virtuoso utilizes an R-tree 
index implemented as a table in the relational database component of Virtuoso.
It is worth mentioning that   we also had the opportunity to examine the development branch v7.2.6-rc1 of Virtuoso OpenSource Server available since early October 2018 which introduced some  features of GeoSPARQL. However, there was still not enough support for several of the features that we evaluate in our benchmark, so we decided that it would only be fair to continue our experiments,  once a more stable release of Virtuoso supporting these features becomes available. 

The proprietary RDF store System Y, also provides limited support for geospatial data.
System Y can store and query only points. Support is provided
both for Cartesian coordinate systems and for spherical coordinate systems 
but not for standard geographic CRS, like those maintained by the IOGP.
System Y supports only range queries (points within a rectangle or a circle)
utilizing either property 
functions\footnote{\url{https://www.w3.org/wiki/SPARQL/Extensions/Computed_Properties}} or a
 non SPARQL compliant syntax.

\section{The Benchmark Geographica 2}
\label{sec:benchmark}

	This section presents in detail the benchmark Geographica 2 which extends our earlier benchmark Geographica~\cite{geographica2013}. Section
\ref{sec:realworkload} presents its first part (the real world workload) while
Section \ref{sec:syntheticworkload} presents the second part (the synthetic
workload).

	\subsection{Real World Workload}
	\label{sec:realworkload}
		This workload aims at evaluating the efficiency of basic spatial 
		functions that a geospatial RDF store should offer. In addition, 
		this workload includes five real use case application scenarios.
		\subsubsection{Datasets}
		\label{sec:dataModel}
		In this section we describe the datasets that we use for the real world workload.
We have datasets that are part of the 
Linked Open Data Cloud, such as the Greek version of DBpedia and part of the GeoNames 
dataset referring to Greece.

DBpedia is a crowd-sourced knowledge base that contains structured information from Wikipedia. DBpedia contains also geographic information for its articles. For example a point position for a country, a city or a building. 

GeoNames\footnote{\url{http://www.geonames.org/}} is a crowd-sourced geographical database containing more than eleven 
million unique features. These placenames are classified according to a two-level
schema. The first level uses very generic categories. For example, a point may be
characterized as a waterbody or some kind of facility. The second level narrows down to very specific categories. For example a waterbody may characterized as a river, a lake, etc., while a facility may characterized as a bank, a hospital, etc. Each point in Geonames is a pair of latitude and longitude in CRS WGS84. More complex geometries (e.g., lines or polygons) are not included.

Since the spatial information of GeoNames and DBpedia is limited to points, datasets with
richer spatial information are also used in Geographica 2. 
LinkedGeoData\footnote{\url{http://linkedgeodata.org/}} (LGD) is a project which in 2009 made 
OSM data available on the Web as linked data~\cite{SLHA11}. We have included a part of the  
OSM dataset about the motorways and rivers of Greece. 

We also chose to use a
dataset containing geometries of Greek municipalities defined by the
Greek Administrative 
Geography\footnote{\url{http://www.linkedopendata.gr/dataset/greek-administrative-geography/}} (GAG) 
and the CORINE Land Cover\footnote{\url{http://www.linkedopendata.gr/dataset/corine-land-cover-of-greece}} (CLC) 
dataset for Greece which have 
complex polygons. 
The CLC dataset is made available by the European Environmental Agency for the 
whole of Europe and contains data regarding the land cover of European countries.
Both of these datasets with information about Greece have been published 
as linked data by us
in the context of the European project TELEIOS\footnote{\url{http://www.earthobservatory.eu/}}.

\begin{table*}[!t]
\tiny
\centering
\resizebox{\textwidth}{!}{
\begin{tabular}{|c|c|c|c|c|c|}
\hline
 &  &  &  & \textbf{\# of Lines} & \textbf{\# of Polygons} \\
 \textbf{Datasets} & \textbf{Size} & \textbf{Triples} & \textbf{\# of Points} & (max/min/avg \# of & (max/min/avg \# of \\
 &  &  &  & points per line) & points per polygon) \\ \hline
 GAG & 33MB & 4K & - & - & 325 (14K/4/192) \\ \hline
 CLC & 401MB & 630K & - & - & 45K (15K/4/171) \\ \hline
 OSM (only ways) & 29MB & 150K & - & 12K (1.6K/2/21) & - \\ \hline
 GeoNames & 45MB & 400K & 22K & - & - \\ \hline
 DBpedia & 89MB & 430K & 8K & - & - \\ \hline
 Hotspots & 90MB & 450K & - & - & 37K (4/4/4) \\ \hline
 Census & 3.3GB & 23M & - & 894K (262/2/6) & - \\ \hline
 
\end{tabular}
}
\caption{Dataset characteristics}
\label{table:datasets}
\end{table*}

Finally, Geographica 2 includes a dataset containing polygons that represent wild-fire hotspots. 
This dataset has been produced by the National Observatory of Athens (NOA) in 
the context of project TELEIOS by processing
appropriate satellite images as described in~\cite{edbt2012}. 
Each dataset is loaded in a separate named graph so that each query access only the part
of the dataset that is needed. 

All the aforementioned datasets were loaded in a common repository for each
RDF store and they were used for all experiments of the real world workload of Geographica
apart from the "Geocoding" scenario, that is described in~Section~\ref{sec:macro-def}.
This scenario requires detailed information about street addresses (e.g., zip code and building number) which is not provided in any of the above datasets.
So, the "Geocoding" scenario uses data about the street network of New York that is
publicly available as part of TIGER (Topologically Integrated Geographic Encoding and 
Referencing) products\footnote{\url{http://www.census.gov/geo/maps-data/data/tiger.html}} 
produced by the US Census Bureau\footnote{\url{http://www.census.gov/}}. This dataset
contains geometries of streets of New York as linestrings and address information like 
street name, zip code, building numbers, etc. It has been stored in a separate
repository for each RDF store and it was used only for the "Geocoding" scenario.

Table~\ref{table:datasets} describes important characteristics of the datasets. In this table the size of each dataset is presented
in MB and the number of contained triples. The size in MB is calculated from 
uncompressed text files in N-Triples syntax. Also, Table~\ref{table:datasets}
presents the type and number of geometries that each dataset contains. In parenthesis
we give the maximum, minimum and average number of points per geometry to give
an insight of the geometry complexity for each dataset.
		
		\subsubsection{Micro Benchmark}
		\label{sec:micro-def}
		\iflong
	The micro benchmark aims at testing the efficiency of primitive spatial functions 
	in state of the art geospatial RDF stores. Thus, it uses simple SPARQL queries which
	consist of one or two triple patterns and a spatial function. 
	In this way, the spatial module is stressed instead of the basic triple pattern matching
	module of RDF stores.
	First, simple spatial selections are tested. Next, more complex 
	operations such as spatial joins are tested. Spatial joins are tested using the topological 
	relations defined in stSPARQL~\cite{iswc2012-strabon} and the Geometry Topology 
	Extension component of GeoSPARQL. 
	
	Table~\mbox{\ref{table:microSpatialCombinations}} 
	summarizes the combinations between topological relations and geometry types that are tested by Geographica 2. In 
	Table~\mbox{\ref{table:microSpatialCombinations}\subref{table:microSpatialCombinations:selections}} 
	columns indicate the geometry type of the constant used for the spatial selections 
	and rows  indicate the geometry type of retrieved spatial features. In 
	Table~\mbox{\ref{table:microSpatialCombinations}\subref{table:microSpatialCombinations:joins}} 
	both columns and rows indicate the geometry types that participate in each join 
	query. In parenthesis the datasets that participate in every query are reported.
	The possible combinations of geometry types and topological relations are
	too many and it would be pointless to 
	exhaustively test all of them. Thus, we selected an interesting part of this 
	combinations based on previous work (e.g. \mbox{\cite{jackpine}}) and our 
	experience in building geographical applications. All topological 
	functions defined by the OGC SFA relation family and every 
	geometry type combination are included at least once. Mainly, we focus on topological 
	relations including polygons since polygon is the most complex 2-D geometry type 
	which can form many topological relations and is the most demanding geometry 
	type to handle.

		\begin{table*}[!t]
	\centering
	\scriptsize
	\subtable[]
	{
		\begin{tabular}{|c|c|c|c|}
			\hline
												& Query Point 			& Query Line			& Query Polygon	\\
			\hline
			\multirow{2}{*}{Point (GeoNames)}	& Within Buffer			& \multirow{2}{*}{-}	& Within		\\
												& In Distance			& 						& Disjoint		\\
			\hline
			\multirow{3}{*}{Line (OSM)}			& \multirow{3}{*}{-}	& Equals				& Intersects	\\
												&						& Crosses				& Disjoint		\\
												&						& 						& 				\\
			\hline
			Polygon (CLC)						& \multirow{3}{*}{-}	& Intersects			& Overlaps		\\
			\hspace{3.5em} (GAG)				&						& 						& Equals		\\
												&						& 						& 				\\
			\hline
		\end{tabular}
		\label{table:microSpatialCombinations:selections}
	}
	\subtable[]
	{
		\begin{tabular}{|c|c|c|c|}
			\hline
											 	& Point (DBpedia) 	& Line (OSM)			& Polygon (GAG)		\\
			\hline
			\multirow{2}{*}{Point (GeoNames)}	& Equals			& Intersects			& Intersects		\\
											 	& 					& 						& Within			\\
			\hline
			\multirow{3}{*}{Line (OSM)}		 	& \multirow{3}{*}{-}& \multirow{3}{*}{-}	& Intersects		\\
											 	& 					& 						& Within			\\
											 	& 					& 						& Crosses			\\
			\hline
			Polygon(CLC)						& \multirow{3}{*}{-}& \multirow{3}{*}{-}	& Within			\\
			\hspace{3.5em} (CLC)			 	& 					& 						& Overlaps			\\
			\hspace{3.5em} (GAG)			 	& 					& 						& Touches			\\
			\hline
		\end{tabular}
		\label{table:microSpatialCombinations:joins}
	}
	\caption{Topological relations tested in (a) spatial selections and (b) spatial joins}
	\label{table:microSpatialCombinations}
	\end{table*}

\fi
 
Apart from topological relations, the micro benchmark tests non-topological functions (e.g., \texttt{geof:buffer}), defined 
by the Geometry Extension of GeoSPARQL, which construct new geometry objects from existing ones. 
Additionally, a metric function for evaluating the area of a polygon is tested.
This function is not defined by GeoSPARQL, but it is supported by almost all tested 
geospatial RDF stores (Strabon, uSeekM, System X, GraphDB).
The aggregate functions \texttt{strdf:extent}, and \texttt{strdf:union} of stSPARQL
are also included in the evaluation although the GeoSPARQL standard does not define them. 
We include aggregate functions in Geographica since they are present in all geospatial
RDBMS, and we found them very useful in EO applications in the context of the project TELEIOS~\cite{edbt2012}.
A short description of queries used in the micro benchmark can be found in 
Table~\ref{table:microbenchmar1} and the full SPARQL queries can be found online\footnote{\url{http://geographica.di.uoa.gr/}}. 

\begin{table*}[!t]
\centering
{	
	\scriptsize
	\begin{tabular}{|p{1cm}|p{2cm}|l|}
		\hline
		\textbf{Query} & \textbf{Operation} & \textbf{Description} \\ \hline
		\multicolumn{3}{|l|}{\textbf{Non-topological construct functions}} \\ \hline
		Q1 & Boundary &	Construct the boundary of all polygons of CLC \\ \hline
		Q2 & Envelope &	Construct the envelope of all polygons of CLC \\ \hline
		Q3 & Convex Hull & Construct the convex hull of all polygons of CLC \\ \hline
		Q4 & Buffer & Construct the buffer of all points of GeoNames \\ \hline
		Q5 & Buffer & Construct the buffer of all lines of OSM  \\ \hline
		Q6 & Area & Compute the area of all polygons of CLC \\ \hline
		\multicolumn{3}{|l|}{\textbf{Spatial selections}} \\ \hline
		Q7 & Equals & Find all lines of OSM that are spatially equal with a given line \\ \hline
		Q8 & Equals & Find all polygons of GAG that are spatially equal a given polygon \\ \hline
		Q9 & Intersects & Find all lines of OSM that spatially intersect with a given polygon \\ \hline
		Q10 & Intersects & Find all polygons of CLC that spatially intersect with a given line \\ \hline
		Q11 & Overlaps & Find all polygons of CLC that spatially overlap with a given polygon \\ \hline
		Q12 & Crosses & Find all lines of OSM that spatially cross a given line \\ \hline
		Q13 & Within polygon & Find all points of GeoNames that are contained in a given polygon \\ \hline
		Q14 & Within buffer of a point & Find all points of GeoNames that are contained in the buffer of a given point \\ \hline
		Q15 & Near a point & Find all points of GeoNames that are within specific distance from a given point \\ \hline
		Q16 & Disjoint & Find all points of GeoNames that are spatially disjoint of a given polygon \\ \hline
		Q17 & Disjoint & Find all lines of OSM that are spatially disjoint of a given polygon \\ \hline
		\multicolumn{3}{|l|}{\textbf{Spatial joins}} \\ \hline
		Q18 & Equals & Find all points of GeoNames that are spatially equal with a point of DBpedia \\ \hline
		Q19 & Intersects & Find all points of GeoNames that spatially intersect a line of OSM \\ \hline
		Q20 & Intersects & Find all points of GeoNames that spatially intersect a polygon of GAG \\ \hline
		Q21 & Intersects & Find all lines of OSM that spatially intersect a polygon of GAG \\ \hline
		Q22 & Within & Find all points of GeoNames that are within a polygon of GAG \\ \hline
		Q23 & Within & Find all lines of OSM that are within a polygon of GAG \\ \hline
		Q24 & Within & Find all polygons of CLC that are within a polygon of GAG \\ \hline
		Q25 & Crosses & Find all lines of OSM that spatially cross a polygon of GAG \\ \hline
		Q26 & Touches & Find all polygons of GAG that spatially touch other polygons of GAG \\ \hline
		Q27 & Overlaps & Find all polygons of CLC that spatially overlap polygons of GAG \\ \hline
		\multicolumn{3}{|l|}{\textbf{Aggregate functions}} \\ \hline
		Q28 & Extension & Construct the extension of all polygons of GAG \\ \hline
		Q29 & Union & Construct the union of all polygons of GAG \\ \hline
	\end{tabular}
}	
\caption{Queries of the micro benchmark}
\label{table:microbenchmar1}
\end{table*}
		
		\subsubsection{Macro Benchmark}
		\label{sec:macro-def}
		The macro benchmark tests the performance of the selected RDF 
stores in three typical application scenarios, namely "Geocoding",
"Reverse Geocoding", and "Map Search and Browsing" and two more
sophisticated scenarios from the EO domain, namely "Rapid Mapping for Fire 
Monitoring" and "Computing Statistics of Geospatial Datasets".
Descriptions of the queries associated with these scenarios can be found in 
Table~\ref{table:macrobenchmark} and SPARQL templates used to generate these
queries are provided on the site of the benchmark.

\iflong
\textit{Geocoding.} Geocoding is the process of finding the coordinates of a feature based on other 
geographic data, such as street address, house number, city, country, etc. 
The simplest method
of geocoding is called address geocoding and is applied to street network data 
that contain street segments and address ranges for each segment. The address range
of a street segment is the minimum and maximum house numbers that are attributed
to this street segment. Usually, two address ranges are assigned to a street segment,
one for its left side and one for its right one. A geocoding query retrieves a street 
segment based on thematic criteria and then interpolates its geometry within
the address range to estimate the position of the given house number. Imagine a user who is 
looking for the Metropolitan Museum of Art in New York (address is
1000 5th Avenue, New York, 10028). A geocoding query will retrieve a street segment 
with name "5th Avenue", ZIP code equal to 10028 and an address range that contains 
even numbers including 1000 (e.g., from 998 to 1002). Then taking into account the 
spatial extent and the minimum and maximum numbers of this segment an estimation
of the position of the museum in 5th Avenue is calculated and returned to the user.   

Because neither GeoSPARQL nor any geospatial RDF store offer any sophisticated 
function for geometry interpolation, used queries perform a simple linear 
interpolation between the start and end points of a street segment. This 
scenario tests two identical queries that search the left and right sides of streets
and they return a point estimation for the given address number and the actual geometry 
of the street segment that is matched to the given address.
This scenario, uses the Census dataset that is described in 
Section~\ref{sec:dataModel}.
Address ranges are published by Census Bureau as ESRI shapefiles. Each shapefile
contains a relational table and each tuple of the table represents a street segment.
The main contents of the shapefile can be modelled as the relation: 
\texttt{StreetSegment(geo GEOMETRY, fullname VARCHAR, lfromhn NUMBER, ltohn NUMBER, rfromhn 
NUMBER, rtohn NUMBER, parityl VARCHAR, parityr VARCHAR, zipl VARCHAR, zipr VARCHAR)}. In this
relation \texttt{geo} represents the geometry of a road segment and \texttt{fullname} its name.
The minimum and maximum house numbers of the left (right) side of a road are represented by 
\texttt{lfromhn} (\texttt{rfromhn}) and \texttt{ltohn} (\texttt{rtohn}). One side of a road
usually has only odd or even house numbers. This is indicated by \texttt{parityl} and 
\texttt{parityr} that take the values "O" for only odd numbers, "E" for only even numbers,
and "B" if a road side has both odd and even house numbers. The zip code of a road
side is represented by \texttt{zipl} and \texttt{zipr}. Finally, in order to simplify the
linear interpolation computation, the following extra attributes are added: \texttt{minx NUMBER,
maxx NUMBER, miny NUMBER, maxy NUMBER}. These attributes represent the coordinates of the 
extreme points of a road segment.
This data was transformed into RDF in a 
straightforward way. For each tuple of the table an instance of the class 
\texttt{StreetSegment} was generated and for every column of the table a data property
that associates the \texttt{StreetSegment} instance with the relevant value from the
table column was created. This transformation resulted in 23 million triples and 1 million linestrings.
Also a list of addresses composed by a street name, a zip code, and a 
building number is exported. For each iteration of this scenario an address is randomly selected
by this list, the queries are produced using the corresponding SPARQL query templates, 
populating them with the street 
name, the zip code, the building number and the parity of the number (even or odd)
of the selected address and the estimated coordinates of this address is retrieved.
This random sequence of addresses is generated using
a pseudo-random number generator, which is initialized with the same seed
for every experiment run. The same process is repeated for the initialization of  each iteration of the other scenarios thus, the experiments of the macro benchmark are repeatable.
\fi
 
 \begin{table*}[!hbt]
	\centering
	\scriptsize
	\begin{tabular}{|p{1cm}|l|}
		\hline
		\textbf{Query} & \textbf{Description} \\ \hline
		\multicolumn{2}{|l|}{\textbf{Geocoding}} \\ \hline
		G1 & Geocode left side of roads (from Census) \\ \hline
		G2 & Geocode right side of roads (from Census) \\ \hline
		\multicolumn{2}{|l|}{\textbf{Reverse Geocoding}} \\ \hline
		RG1 & Find the closest populated place (from GeoNames) \\ \hline
		RG2 & Find the closest motorway (from OSM) \\ \hline
		\multicolumn{2}{|l|}{\textbf{Map Search and Browsing}} \\ \hline
		MSB1 & Find the co-ordinates of a given POI based on thematic criteria (from GeoNames) \\ \hline
		MSB2 & Find other POI in a given bounding box around these co-ordinates (from GeoNames) \\ \hline
		MSB3 & Find roads in a given bounding box around these co-ordinates (from OSM) \\ \hline
		\multicolumn{2}{|l|}{\textbf{Rapid Mapping for Fire Monitoring}} \\ \hline
		RM1 & Find the land cover of areas inside a given bounding box (from CLC) \\ \hline
		RM2 & Find primary roads inside a given bounding box (from OSM) \\ \hline
		RM3 & Find municipality boundaries inside a given bounding box (from GAG) \\ \hline
		RM4 & Find detected hotspots inside a given bounding box (from Hotspots) \\ \hline
		RM5 & Find coniferous forests inside a given bounding box which are on fire (from CLC and \\
		    & Hotspots) \\ \hline
		RM6 & Find road segments inside a given bounding box which may be damaged by fire (from \\
		    & OSM and Hotspots) \\ \hline
		\multicolumn{2}{|l|}{\textbf{Computing Statistics of Geospatial Datasets}} \\ \hline
		CS1 & Compute how many instances of each CLC class exist in a municipality (from CLC) \\ \hline
		CS2 & Compute how many instances of each GeoNames class exist in a municipality \\
		    & (from GeoNames) \\ \hline
		CS3 & Compute how many instances of each GeoNames class exist in areas characterized as \\
		    & ''Continuous Urban Fabric'' according to CLC (from GeoNames and CLC) \\ \hline
	\end{tabular}
	\caption{Queries of the macro benchmark}
	\label{table:macrobenchmark}	
\end{table*}
 
\textit{Reverse Geocoding.}
Reverse geocoding is the process of attributing a readable address or place name
to a given point. This scenario tests two nearest neighbor
queries which retrieve the nearest point (from GeoNames) and the nearest motorway 
(from OSM) of the given point. To achieve this nearest neighbor 
functionality the queries of this scenario sort retrieved geometries by their 
distance to the given point and select the first one. Every iteration of this scenario 
is initialized with a given point. This point is picked at random from a list of 
point coordinates extracted from GeoNames. 

\textit{Map Search and Browsing.}
This scenario demonstrates the queries that are typically used in Web-based mapping 
applications. A user first searches for points of interest based on thematic
criteria. Then, he/she selects a specific point and information about the area 
around it is retrieved (e.g., POI and roads).
Similarly to the "Reverse Geocoding" scenario, this scenario is initialized by picking at 
random a toponym from a list of toponyms extracted by GeoNames. 
The coordinates of this toponym is retrieved by the first query of the scenario.
Then, these coordinates are used to create an area of interest which is used by
the remaining two queries. These queries retrieve points of interest (from GeoNames)
and roads (from OSM) that lie inside this area.

\textit{Rapid Mapping for Wild Fire Monitoring.}
This scenario tests queries which retrieve map layers for creating a 
map that can be used by decision makers tasked with the monitoring
of wild fires. This application has been studied in detail in project
TELEIOS \cite{edbt2012} and the scenario covers its core querying needs.
First, spatial 
selections are used to retrieve basic information of interest
(e.g., roads, administrative areas, etc.). Second, more complex information can 
be derived using spatial joins and non-topological functions. For example, a 
user may be interested in those road segments damaged by fire. We
point out that this scenario is representative of many rapid mapping
tasks encountered in EO applications.
Again, a list of areas, and the relevant time, where fire occured has been compiled by
data from the real fire monitoring application and used to randomly initialize each iteration
of this scenario.

\iflong
\textit{Computing Statistics of Geospatial Datasets.}
This scenario concentrates on generating a high level overview of geospatial 
datasets by calculating summary statistics (e.g., how many fields are identified 
as agricultural by a 
dataset) and discovering correlations between  different datasets describing 
the same geographical area (e.g., how many farms  in Crete, according to GeoNames, lie 
in areas that are identified as agricultural areas by CLC). 
Since geospatial datasets are produced in many ways (e.g., contributed by users,
produced by experts using surveys, satellite images, aerial photographs, etc.) such 
overviews and comparisons are meaningful and interesting for specialists.
An example of user-contributed  data is GeoNames. In this dataset, users provide 
information about points on a map and a two level schema, with various classes, is used 
to characterize these geographic features with broad terms (e.g., 
administrative division, waterbody, road) or more specific terms (e.g., village, 
lake, tunnel). On the other hand, specialists, such as geographers and cartographers, 
have compiled information originated from aerial photographs, topographic maps, satellite 
images, etc. to create the CLC dataset, that provides information about  
the land cover in European countries using a more targeted schema with broader terms 
(e.g., urban fabric, agricultural area, etc.). Despite the fact that these two datasets
contain different kinds of information, the comparison between them can help to 
evaluate the consistency between these datasets in order to validate the
provided information. For example a useful query would be to discover the  kind of geographic features, according to GeoNames, that are contained in areas of CLC with specific land use. It is 
expected that urban areas should contain more features identified as roads, buildings, 
bus/metro stations etc., while agricultural areas should contain more geographic features 
identified as farms, irrigated fields, plantations, etc.
In project TELEIOS we investigated such a scenario in collaboration with 
the German Aerospace Center (DLR).

DLR used the knowledge discovery and data mining framework for satellite images
presented in \cite{albanipilot} to identify semantic classes of geographic features 
(e.g., parking area, port, etc.) in radar images from the archive of the TerraSAR-X satellite.
Since the knowledge discovery and data mining framework relies on semi-supervised 
machine learning techniques, the comparison of the DLR classification with 
classifications of other datasets can prove important for the
training phase of these techniques but also for evaluating the effectiveness of 
the framework.

This scenario is composed of three queries that represent two main query categories
that were useful in the DLR use case of the project TELEIOS. The first category (first level statistics)
computes statistics about one dataset (e.g., compute the distribution of CLC classes in a 
city). The second category (second level statistics) helps to investigate possible 
correlations between datasets by computing statistics
that involve two datasets (e.g., compute how many instances of each GeoNames class lie in 
areas characterized as "continuous urban fabric" by CLC). 
Such queries are usually applied for a specific area (e.g., 
a city, a country). A list of the minimum bounding rectangles
of all Greek municipalities  have been created  and for each iteration of this scenario the queries
are applied to a randomly selected bounding rectangle.
The main characteristic of 
these queries is that they compute aggregations over spatial selections of a dataset 
and spatial joins between 
two datasets. The results of the queries can later be visualized (e.g., on a chart) 
to provide insights on the correlations of these datasets. Since the 
classification that was produced using the techniques developed by DLR is not freely available,
we decided to use the publicly available dataset GeoNames to keep our experiments
easily reproducible. This scenario, also, compares GeoNames with information from CLC which is
also publicly available.
\fi

	\subsection{Synthetic Workload}
	  \label{sec:syntheticworkload}
		The synthetic workload of Geographica 2 relies on a generator that produces synthetic datasets of various sizes and instantiates query templates that can produce queries with varying thematic and spatial selectivity. In this way, the
		evaluation of geospatial RDF stores can be performed in a controlled environment in
		order to measure their performance with great precision. The synthetic generator is a
		component of Geographica 2 and is distributed freely as open-source software.
		\subsubsection{Datasets}
		\label{sec:syntheticDataset}
		The workload generator produces synthetic datasets
of arbitrary size that resemble features on a map. As in VESPA~\cite{vespa},
the produced datasets model the following geographic features: states in a country,
land ownership, roads and points of interest.
For each dataset, we
developed a minimal
ontology
 that follows a general version
of the schema of OSM and uses GeoSPARQL ontologies and vocabularies. 
In Figure~\ref{fig:sytheticDatasetOntology} the developed
ontology for representing points of interest is presented. 
As in \cite{brodtRDF3Xgermanoi,iswc2012-strabon}, every feature
(i.e., point of interest) is assigned a number of thematic tags each of which consists of
a key-value pair of strings. Each feature is tagged with key $1$, every
other feature with key $2$, every fourth feature with key $4$,
etc. up to key $2^k, k \in \mathbb{N}$.
This tagging makes it possible to select different parts of the entire
dataset in a uniform way, and perform queries of various thematic
selectivities. For example, if we selected all points of interest tagged with
key $1$, we would retrieve all available points of interest, if we
selected all points of interest tagged with key  $2$, we would retrieve half of
them, etc.

Every feature has a spatial extent that is modelled using the GeoSPARQL
vocabulary. The spatial extent of the land ownership dataset constitutes a
uniform grid of $n \times n$ hexagons. The land ownership dataset forms the
basis for the spatial extent of all generated datasets since the size of each
dataset is given relatively to the number $n$. By modifying the number of
hexagons along an axis, datasets of arbitrary size can be produced. As we will see
in the following section, this enabled us to adjust the selectivity of the
spatial predicates appearing in queries in a uniform way too.

\begin{figure*}[t]
	\centering
	\subfigure[][\shortstack{\\ Ontology for Points of Interest}]{
		\includegraphics[scale=0.3078]{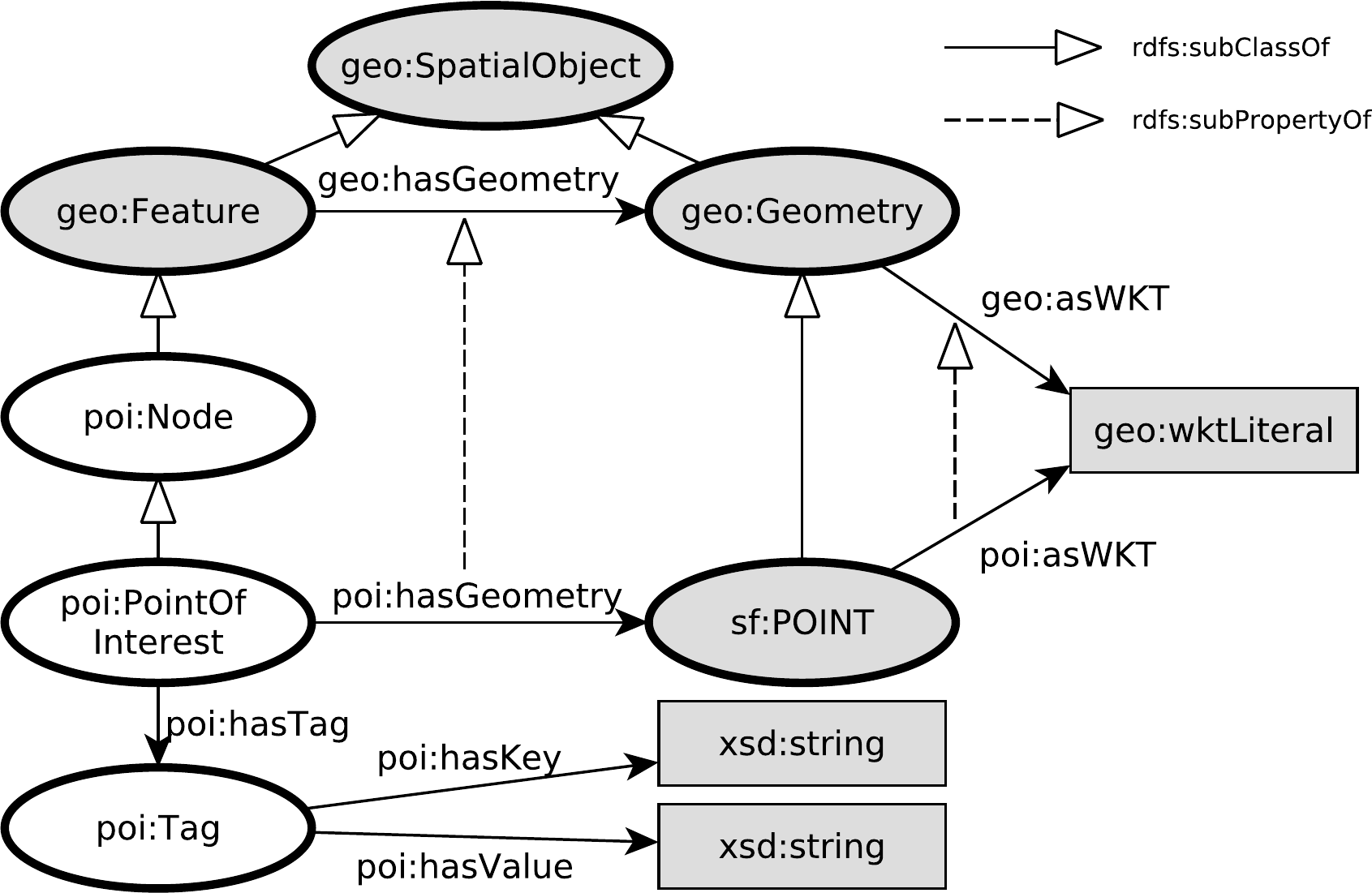}
		\label{fig:sytheticDatasetOntology}}
	\colbreak
	\subfigure[][Visualization of the geometric part of the synthetic dataset]{
		\includegraphics[scale=0.1458]{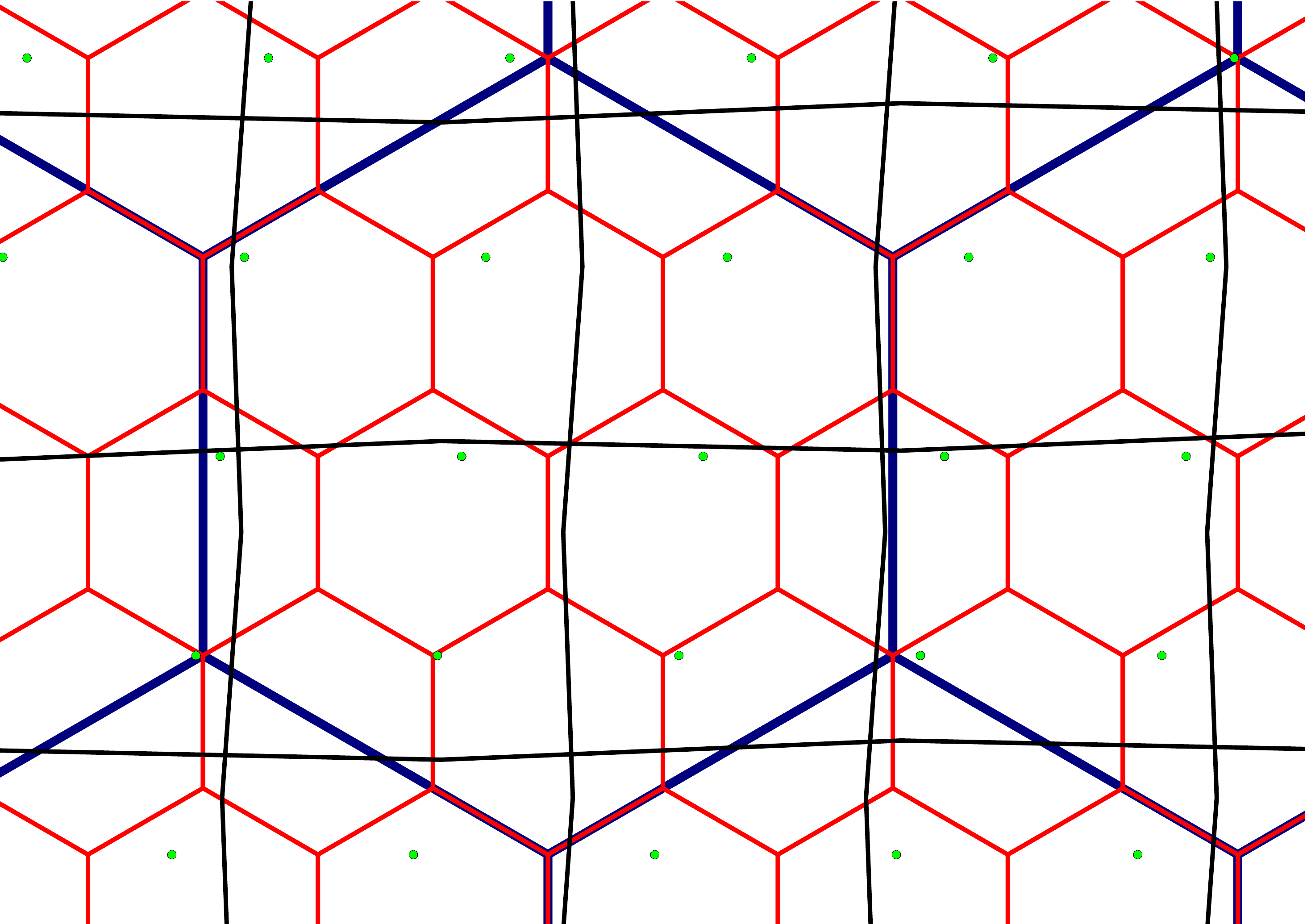}
		\label{fig:sytheticDatasetVisualization}}
	\caption{Synthetic Dataset}
	\label{fig:sytheticDataset}
\end{figure*}

As in \cite{vespa}, the generated land ownership dataset consists of $n^2$
features with hexagonal spatial extent, where each hexagon is placed  uniformly on a $n
\times n$ grid. The cardinality of the land ownerships is $n^2$.
The generated state dataset consists of $(\frac{n}{3})^2$ features with
hexagonal spatial extent, where each hexagon is placed  uniformly on a
$\frac{n}{3} \times \frac{n}{3}$ grid. The cardinality of the state
geometries is $(\frac{n}{3})^2$.
The generated road dataset consists of $n$ features with sloping line geometries.
Half of the line geometries are roughly horizontal and the other half are
roughly vertical. Each line consists of $\frac{n}{2}+1$ line segments.
The cardinality of the road geometries is $n$.
The generated point of interest dataset consists of $n^2$ features with point
geometries which are uniformly placed on $n$ sloping, evenly spaced, parallel
lines. The cardinality of the point of interest geometries is $n^2$. In
Figure~\ref{fig:sytheticDatasetVisualization}, a sample of the
generated geometries is presented.
		
		\subsubsection{Queries}
		\label{sec:synthetic-def}
		The synthetic workload generator produces SPARQL queries corresponding
to spatial selection and spatial joins using the two query templates
presented in Table~\ref{tbl:syntheticQueryTemplates}.

The query template, presented in Table~\ref{tbl:syntheticQueryTemplatesSelect},  used for producing SPARQL queries corresponding to spatial
selections is identical to the query template used in
\cite{brodtRDF3Xgermanoi,iswc2012-strabon}. In this query template, parameter
\texttt{THEMA} is one of the values used when assigning tags to a feature and
parameter \texttt{GEOM} is the WKT serialization of a rectangle.
As in \cite{iswc2012-strabon}, we define the \textit{thematic selectivity} of
a query as the fraction of the total features of a
dataset that are tagged with a key equal to \texttt{THEMA}. For example, by
altering the value of \texttt{THEMA} from 1 to 2, the thematic
selectivity of the query is reduced by selecting half the nodes it previously did.  
We define the \textit{spatial selectivity} of a query as the fraction of the total features for which the topological
relations defined by parameter \texttt{FUNCTION} holds between each of them and the rectangle
defined by parameter \texttt{GEOM}. By modifying the value of the parameter namespace \texttt{ns} we
specify the dataset and the corresponding type of geometric information that
is examined by an instance of the query template.

The query template, presented in Table~\ref{tbl:syntheticQueryTemplatesJoin}, used for producing SPARQL queries corresponding to spatial
joins involves two datasets identified by the values of the parameter namespaces
\texttt{ns1} and \texttt{ns2}. In this query template, parameters \texttt{THEMA1}
and \texttt{THEMA2} control the thematic selectivity of the query. The value of parameter
\texttt{FUNCTION} defines the topological relation that must hold between
instances of the two datasets that are involved in an instance of the query
template. 
Parameter \texttt{FUNCTION} can be instantiated with every function defined in the Geometry
Topology Extension component of GeoSPARQL. In our experiments, as described in 
Section~\ref{sec:resultsSyntheticQueries}, \texttt{geof:sfIntersects}, 
\texttt{geof:sfTouches}, \texttt{geof:sfWithin} were used.
For example, by instantiating the query template (b) with the values
\texttt{poi} for \texttt{ns1}, \texttt{state} for \texttt{ns2}, \texttt{1} for
\texttt{THEMA1}, \texttt{2} for \texttt{THEMA2} and \texttt{geof:sfWithin} for
\texttt{FUNCTION}, we get a SPARQL query that asks for all generated points of
interest that are inside half of the generated states.

\begin{table*}[!t]
  \centering
  \caption{Query templates for generating SPARQL queries corresponding to (a)
	  spatial selections, and (b) spatial joins.}
	  \label{tbl:syntheticQueryTemplates}	
	  \hspace{-1em}	  
	\subtable[]
	  {
	  	\includegraphics[scale=1]{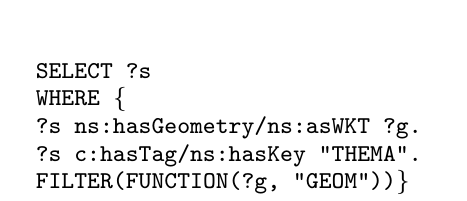}
			\label{tbl:syntheticQueryTemplatesSelect}
		}
	  \subtable[]
	  {	
		  \includegraphics[scale=1]{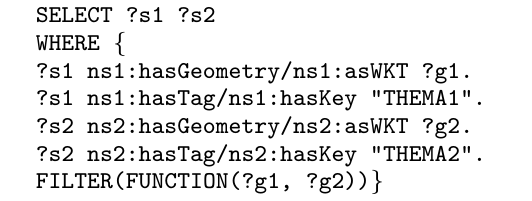}		
		\label{tbl:syntheticQueryTemplatesJoin}	
		}
\end{table*}

These query templates allow us to generate SPARQL queries with great diversity
regarding their spatial and thematic selectivity, thus stressing the optimizers
of the geospatial RDF stores that we test and evaluating their effectiveness in
identifying efficient query plans. 

	\subsection{Scalability Workload}
		\label{sec:scalabilityworkload}
		\textit{Scalability} has been an important metric regarding the evaluation of various kinds of systems, such as multiprocessor, network, database and distributed systems  ~\cite{hill1990scalability, DBLP:conf/wosp/Bondi00, DBLP:conf/serp/CrockettMRBR05,garcia2008experimental,appuswamy2013scale,cattell2011scalable,hu2014toward,hausenblas2007performance}.  
		
		In the context of this paper, the scalability experiment  aims at discovering the limits of the systems under test as the number of triples in the dataset increase. Each system is tested against six increasingly bigger, proper subsets of the reference dataset. For each system-dataset combination we measure \textit{(1) the repository size} on disk, \textit{(2) the bulk loading time} taking into consideration the limitations of loading methods of each system and \textit{(3) the response time} in three queries which represent a spatial selection, a heavy spatial join with high spatial selectivity and a lighter spatial join with lower spatial selectivity.

		\subsubsection{Datasets}
		\label{sec:scalabilityDataset}

%
%

\begin{table*}[ht]
\tiny
\centering
\resizebox{0.5\textwidth}{!}{
\begin{tabular}{|c|r|r|r|}
	\hline
	\textbf{Datasets}    & \multicolumn{1}{c|}{\textbf{Country}} & \multicolumn{1}{S|}{\textbf{Triples (M)}} & \multicolumn{1}{S|}{\textbf{Size (MB)}} \\ \hline
	\multirow{6}{*}{OSM} & Wales                                 & 6.56                                  & 1,206                               \\ \cline{2-4}
	& Scotland                              & 15.78                                  & 2,913                               \\ \cline{2-4}
	& Greece                                & 15.22                                  & 2,877                               \\ \cline{2-4}
	& N. Ireland                            & 15.27                                  & 3,240                               \\ \cline{2-4}
	& England                               & 104.21                                  & 18,965                               \\ \cline{2-4}
	& Germany                               & 326.48                                  & 59,002                               \\ \hline
	CLC-2012             & \textit{39 countries}                          & 16.60                                  & 11,283                               \\ \hline
	\multicolumn{2}{|r|}{\textbf{Totals}}           & 483.52                                 & 99,486                               \\ \hline
\end{tabular}
}
\caption{Scalability workload: reference dataset sources}
\label{table:refScalabilityDatasetDescription}
\end{table*}
\begin{table*}[]
\tiny
\centering
\resizebox{0.65\textwidth}{!}{
\begin{tabular}{|r|r|r|r|r|}
\hline
\textbf{Dataset} & \textbf{\# of Features} & \textbf{\# of Points} & \textbf{\# of Lines} & \textbf{\# of Polygons} \\ \hline
10K     & 1,135          & 587          & 0           & 900            \\ \hline
100K    & 12,166         & 6,623        & 4,239       & 2,531          \\ \hline
1M      & 118,161        & 46,781       & 45,238      & 29,200         \\ \hline
10M     & 1,038,739      & 317,865      & 328,630     & 427,842        \\ \hline
100M    & 10,259,959     & 904,677      & 2,058,386   & 7,553,440      \\ \hline
500M    & 48,623,878     & 5,520,767    & 15,771,932  & 23,390,220     \\ \hline
\end{tabular}
}
\caption{Scalability datasets basic characteristics}
\label{table:scalabilityDatasetsBasicDescription}
\end{table*}

\begin{table*}[]
\tiny
\centering
\resizebox{0.85\textwidth}{!}{
\begin{tabular}{|r|l|r|r|r|r|r|r|}
	\hline
	\multicolumn{1}{|c|}{\textbf{lgo:has\_code}}                                     & \multicolumn{1}{c|}{\textbf{lgo:has\_fclass}}                           & \multicolumn{1}{c|}{\textbf{10K}} & \multicolumn{1}{c|}{\textbf{100K}} & \multicolumn{1}{c|}{\textbf{1M}} & \multicolumn{1}{c|}{\textbf{10M}} & \multicolumn{1}{c|}{\textbf{100M}} & \multicolumn{1}{c|}{\textbf{500M}} \\ \hline
	\begin{tabular}[c]{@{}r@{}}1001\\ (used in SC2, SC3)\end{tabular}                & city                                                                    & 1                                 & 1                                  & 7                                & 14                                & 84                                 &  232                                  \\ \hline \hline
	5601                                                                             & railway\_station                                                        & 15                                & 284                                & 284                              & 669                               & 1,194                              &  8,449                                  \\ \hline
	5621                                                                             & bus\_stop                                                               & 4                                 & 4,416                              & 4,416                            & 22,337                            & 35,555                             &  503,455                                  \\ \hline
	5622                                                                             & bus\_station                                                            & 36                                & 46                                 & 46                               & 98                                & 425                                &  2,647                                  \\ \hline
	5641                                                                             & taxi                                                                    & 7                                 & 43                                 & 43                               & 217                               & 886                                &  5,798                                  \\ \hline
	5661                                                                             & ferry\_terminal                                                         & 4                                 & 18                                 & 18                               & 153                               & 583                                &  1,508                                  \\ \hline
	\begin{tabular}[c]{@{}r@{}}5601,5621,5622,5641,5661\\ (used in SC3)\end{tabular} &                                                                         & 66                                & 4,807                              & 4,807                            & 23,488                            & 38,643                             &  521,857                                  \\ \hline \hline
	\begin{tabular}[c]{@{}r@{}}(5001-5999) - \{5260\}\\ (used in SC2)\end{tabular}   & \begin{tabular}[c]{@{}l@{}}transportation\\ except parking\end{tabular} & 66                                & 4,875                              & 11,412                           & 264,199                           & 1,978,632                          &  16,151,652                                  \\ \hline
\end{tabular}
}
\caption{Value distribution of \textit{lgo:has\_code} in scalability datasets}
\label{table:scalabilityDatasetFilteringCodes}
\end{table*}

\textit{Reference Dataset Characteristics.} The reference dataset created has an approximate size of 500 million triples. The OSM data concern the following list of countries: \textit{Wales, Scotland, Greece, Northern Ireland, England } and \textit{Germany}. The feature classes selected are: \textit{buildings, landuse, natural, places, points of interest, railways, roads, traffic, transport, water} and \textit{waterways}. CLC-2012 is the 2012 version of the CLC dataset presented earlier. Its data covers the 33 European Environment Agency member countries and six cooperating countries.

The reference dataset has been assembled in the following order: OSM (Wales, Scotland, Greece, Northern Ireland, England, Germany), CLC-2012. Table~\ref{table:refScalabilityDatasetDescription} presents the order and size of each part of the data comprising the reference dataset.

\textit{Reference Dataset Requirements.} For the scalability experiment we needed to design a reference dataset from which we could create six datasets of increasing size. This reference dataset had to satisfy the following requirements: \textit{(1)} contain real life data, \textit{(2)} be realistically big for the given infrastructure, \textit{(3)} contain features of multiple types of geometries (points, lines, polygons), \textit{(4)} have an as homogeneous feature class distribution as possible among the six datasets, meaning \textit{(4a)} avoid having fewer feature classes in the smaller datasets so that thematic filtering and by extension spatial filtering would behave in a predictable and unbiased manner and \textit{(4b)} guarantee that as datasets become bigger, a similar assortment of feature classes will be present, but with more instances per class, \textit{(5)} if possible, include data with highly complex geometries to stress even more the store's geospatial capabilities, \textit{(6)} the different data sources must have an overlapping spatial extent in order for spatial comparisons to be meaningful.

\textit{Reference Dataset Design and Creation.} Requirements \textit{(1)}, \textit{(2)}, \textit{(3)} were met by a subset of the OSM dataset since it is big enough, with real life data and has features with all the main types of geometries. Requirement \textit{(5)} was met  by the CORINE Land Cover 2012 (CLC-2012) \footnote{\url{https://land.copernicus.eu/pan-european/corine-land-cover/clc-2012}}  dataset since it contains very detailed geometries such as burned forest areas. The spatial extent of the OSM countries selected fall well within the spatial extent of the CLC-2012 dataset, therefore requirement \textit{(6)} was satisfied. In order to meet requirement \textit{(4)}, we had to also take into account that the OSM dataset has the same feature classes per country, each data file describes one feature class per country and that these files differ greatly in size. Therefore, in order to satisfy \textit{(4b)} we had to use the OSM dataset first and the CLC-2012 second, and for OSM we needed to start from countries with smaller number of triples. In order to satisfy \textit{(4a)} we had to sort the files of each country based on their file size in ascending order before concatenating them. 

\textit{Scalability workload characteristics.} By selecting six subsets containing 10K, 100K, 1M, 10M, 100M and 500M triples of the reference dataset, we created the corresponding six scalability datasets which were used in the scalability benchmark. The basic characteristics 
of the datasets (e.g.,  features and geometries) are described in Table~\ref{table:scalabilityDatasetsBasicDescription}. The property \url{<http://data.linkedeodata.eu/ontology#has_code>} \textit{(lgo:has\_code)} was used for thematic filtering in scalability join queries and Table~\ref{table:scalabilityDatasetFilteringCodes} shows the distribution of this property's values and the value ranges used.

		\subsubsection{Queries}
		\label{sec:scalability-def}
		\begin{table*}[ht]
	\centering
	\tiny
	\resizebox{0.75\textwidth}{!}{
	\begin{tabular}{|p{1cm}|l|}
		\hline
		\textbf{Query} & \textbf{Description} \\ \hline
		SC1 & Find all geometries that intersect with the given polygon \\ \hline
		SC2 & Find all transportation-related features (except parkings) \\
		    & within a city \\ \hline
		SC3 & Find all bus stops, bus stations, railway stations, taxis and ferry terminals \\
		    & within a city \\ \hline
	\end{tabular}
}
	\caption{Queries of the scalability benchmark}
	\label{table:scalabilityQueriesList}	
\end{table*}
To find an appropriate set of queries we took into consideration multiple factors, such as: \textit{(a)} the performance of the systems under test with the smaller workloads was presented in previous sections, which showed that only few stores can perform well with spatial joins (see Section ~\ref{sec:benchmarkresults}), \textit{(b)} spatial joins are extremely heavy when it involves polygons, \textit{(c)} the datasets from 10M triples onward contain a very high number of polygon geometries, \textit{(d)} thematic selectivity should be used only if necessary and without breaking the expected load scaling as datasets got bigger, 
\textit{(e)} queries should provide some narrative and avoid useless Cartesian products between unrelated geometries.

Factors \textit{(a), (b), (c)} led to the decision of introducing thematic selectivity in the spatial join query to have some stores that could successfully run it. Factor \textit{(e)} led to the decision of having a fixed thematic selectivity of the first part of the join to the feature class \texttt{city} (\textit{lgd:has\_code 1001}). Factor \textit{(d)} obliged us to find an appropriate feature class range for the thematic selectivity of the second part of the join, which was the group of feature classes that represent \texttt{transportation} (\textit{lgd:has\_code 5001-5999}) with the exception of \textit{'parking'} which had a high number of instances. Even this thematic selectivity proved high enough to create difficulties to some RDF stores and we provided an additional lighter variant of the join query by limiting the thematic selectivity of the second part of the join to the following list of feature classes: \textit{railway\_station (5601)}, \textit{bus\_stop (5621)}, \textit{bus\_station (5622)}, \textit{taxi stands (5641)} and \textit{ferry\_terminals (5661)}. These two join queries are SC2 and SC3 respectively and Table~\ref{table:scalabilityDatasetFilteringCodes} presents a quantitative view of the expected number of spatial operations for each join query and each dataset.

SC1 query is a spatial selection that uses a polygon literal to filter geometries both from OSM and CLC-2012 datasets. Major cities of countries of the OSM dataset such as \textit{Athens, Thessaloniki, Munich, London, Edinburgh, Belfast} and \textit{Cardiff} were used as the polygon's vertices, thus asserting that it covers areas from all countries of the selected OSM dataset and areas from an augmented set of countries of the CLC-2012 dataset that include France, Italy, Austria, Belgium, etc. 
 
 The queries used for this test are listed in Table~\ref{table:scalabilityQueriesList}.
		\subsubsection{Systems}
		\label{sec:scalability-systems}
		For this test we choose the following three systems to participate: \textit{Strabon, GraphDB} and \textit{RDF4J}. Each of these systems has: (a) adequate support of GeoSPARQL, (b) is a good representative of a different design flavor of RDF store, and (c) is actively supported by the corresponding team. Strabon is a hybrid system using Sesame RDF framework and PostgreSQL RDBMS extended with PostGIS geospatial capabilities. RDF4J is an RDF Framework that supports GeoSPARQL. GraphDB is an RDF Store based on RDF4J which it extends with specialized libraries for its geospatial capabilities, among other things. As mentioned previously, our intention  was to include the latest  beta version of OpenLink Virtuoso which is a well established RDF Store  but as explained in ~Section~\ref{sec:geographica-functional:limited} our preliminary evaluation results showed that we should wait for a stable release that will cover bigger part of the features that are evaluated in our benchmark.	
			
\section{Benchmark Results}
\label{sec:benchmarkresults}
	\iflong
		This section presents the results of running Geographica against six
		geospatial RDF stores. As mentioned earlier, we test the 
		open source systems Strabon v3.2.9, uSeekM v1.2.1, Parliament v2.7.4, 
		a proprietary RDF store, called here System X, GraphDB v8.6.1 and RDF4J v2.4.3. 
		
	\fi

	\subsection{Experimental Setup}
	\label{sec:experimentalsetup}
	This section describes the setup of the experiments used to evaluate the
selected triple stores. The machine that was used to run the benchmark is
equipped with two Intel Xeon E5620 processors with 12MB L3 cache running at 2.4 GHz, 32 GB
of RAM and a RAID-5 disk array that consists of four disks. Each disk has 32 MB
of cache and its rotational speed is 7200 rpm.
The experiments of Geographica have been performed on an Ubuntu 12.04 
installation, however System X is not officially supported on Ubuntu systems. 
Alternatively, System X comes with its own Linux distribution that also provides
a dedicated volume manager and a file system. Therefore, this distribution was used 
for System X experiments. Also, System X supports
parallelism in query execution. Thus, System X was tested
in two different modes; a mode where queries are executed in a single
process (indicated as ``Ser.'' in tables and figures) and a mode (indicated as 
``Par.'' in tables and figures) where the parallel query feature of System X is 
used. This way, we can assess about how much parallelism can speed
up query evaluation.

Each query in the micro, synthetic and scalability benchmarks was run three
times on cold and warm caches. For warm caches, each query ran once before measuring
the response time, in order to warm up the caches.
We measured the response time of each query by measuring the elapsed
time from submitting the query until a complete iteration over the
results had been completed. The response time of each query was measured and the
median of each measurement is reported. 
The experiments of the macro benchmark have a slightly different setup, 
each scenario ran many times (with different initialization
each time, as described in Section~\ref{sec:macro-def}) for one hour without
cleaning the caches and the average time for a complete execution
of all queries of each scenario are reported.
The time limit for real world and synthetic benchmarks was set to one hour for all queries, 
while for the scalability benchmark queries it was set to twenty four hours.

Strabon and uSeekM utilize PostgreSQL enhanced with PostGIS as a spatially-enabled
relational back-end. For these systems, an instance of PostgreSQL 9.2 with
PostGIS 2.0 was used. Because the default settings of PostgreSQL are rather conservative, 
it was tuned to make better use of the system resources. First, the system
configuration file \texttt{sysctl.conf} was edited to increase the amount of available shared
memory (e.g., increasing the kernel parameter \texttt{kernel.shmmax}) and the maximum
number of files that can be opened (e.g., increasing the file system parameter 
fs.file-max).
Second, the PostgreSQL configuration file \texttt{postgresql.conf} was edited.
PostgreSQL was enabled to exploit the increased shared memory. Also, we would like to
avoid resource intensive operations, like Write-Ahead Logging checkpoints. Thus by
editing parameters like, \texttt{checkpoint\_segments}, and \texttt{wall\_level}
we force such operations to happen less frequently and consume less resources than usual.
Finally, some parameters were edited so that the query evaluation planner produces better
query evaluation plans by avoiding genetic query optimization techniques, merging 
sub-queries into upper queries, and reordering joins.
A detailed report of the configuration parameters used is given on the web site of the benchmark.

For every dataset of Geographica, a unique property is used to connect geometries with 
their serialization (e.g. CLC we use the property 
\texttt{clc:asWKT}), and this property is defined as a subproperty of the property 
\texttt{geo:asWKT} that is defined by GeoSPARQL.
Parliament is able to identify and index a triple that represents the serialization of a 
geometric object only when the property \texttt{geo:asWKT} is used. 
As a result, the RDFS reasoning capabilities of Parliament have to be enabled so that it 
performs forward chaining during data loading and indexes the geometry using the spatial 
index as well. Strabon, uSeekM, System X, GraphDB and RDF4J do not perform any reasoning on the input 
data. Specifically for RDF4J the Lucene index option had been enabled since it was explicitly and 
unreservedly suggested as a geospatial optimisation in the official documentation. However further tests
reveiled that the Lucene spatial index increased costs in the repository size and load time do not provide 
any substantial benefit in most scenarios but instead deteriorate the query response times. Support questions 
confirmed that this is probably a performance issue,\footnote{\url{https://github.com/eclipse/rdf4j/issues/1281}}\footnote{\url{https://github.com/eclipse/rdf4j/issues/1160}} threrefore RDF4J was tested in two different modes;
One with the Lucene Sail enabled (indicated as ''\textit{Lucene enabled}'' in tables) and one with no Lucene Sail. 
The RDF4J results for both experimentation modes have been included in all tables but only the non-Lucene indexed results were
included in figures and charts, since they were the best ones. 
We also encountered technical issues connecting with the GraphDB Free runtime to GraphDB repositories which had the GeoSPARQL plugin enabled, thus we had to disable the plugin for all tests. In this mode no geospatial data is indexed and no GeoSPARQL predicates are handled but only queries with GeoSPARQL functions which are always enabled.

	\subsection{Real World Workload}
		\subsubsection{Dataset Storage}
		\label{sec:resultsRealStore}
		This section discusses the time required by each system to store
and index the datasets of the real world workload, as shown in Table~\ref{table:StoringTime}.
Also, Table~\ref{table:RepositorySize} reports the size in MB of the repositories 
created by each RDF store. 
\iflong
	\begin{figure*}[!t]
	\small
	\begin{minipage}[b][][b]{\textwidth}
		\begin{minipage}[t][][t]{\textwidth}
			\centering
			\begin{tabular}{|c|r|r|r|r|r|r|r|}
	\hline

	& \textbf{Strabon} & \textbf{uSeekM} & \textbf{Parliament} & \textbf{System X} & \textbf{GraphDB}  & \multicolumn{2}{c|}{\textbf{RDF4J}}  \\ 
	\cline{7-8}
	& & & & & & & \textit{Lucene} \\
	\textbf{Workload} & & & & & & & \textit{enabled} \\
	\hline
	\textbf{Real world} & 220 & 214  & 250 & \color{darkred} 531 & 91 & \color{darkgreen}82 & 198 \\
	\textbf{Census} & 1,255 & 1,675 & 1,085 & 895 & \color{darkgreen} 358 & 785 &\color{darkred} 3,819 \\
	\textbf{Synthetic} & 221 & 406 & \color{darkred} 462 & 270 & \color{darkgreen}118 & 160 & 250 \\
	\hline
\end{tabular}
			\captionof{table}{Storing times (sec.)}
			\label{table:StoringTime}      
		\end{minipage}
		\newline
		\vspace*{1 cm}
		\newline
		\begin{minipage}[b][][b]{\textwidth}
			\centering    
			\begin{tabular}{|c|r|r|r|r|r|r|r|}
	\hline
 & \textbf{Strabon} & \textbf{uSeekM} & \textbf{Parliament} & \textbf{System X} & \textbf{GraphDB} & \multicolumn{2}{c|}{\textbf{RDF4J}}  \\ 
\cline{7-8}
& & & & & & & \textit{Lucene} \\
\textbf{Workload} & & & & & & & \textit{enabled} \\
	\hline
	\textbf{Real world} & 1,181 & 997 & 1,508 & {\color{darkred} 2,591} & 696 & \color{darkgreen} 625 & 1,220\\
	\textbf{Census} & 5,221 & 2,952 & 4,087 & {\color{darkred} 6,598} & \color{darkgreen} 1,765 & 2,199 & 6,235\\
	\textbf{Synthetic} & 1,271 & 1,211 & 1,763 & {\color{darkred}{17,088}} & {\color{darkgreen}{513}} & 632 & 763 \\
	\hline
\end{tabular}	
			\captionof{table}{Repository sizes (MB)}
			\label{table:RepositorySize}
		\end{minipage}
	\end{minipage}
	\end{figure*}
\fi 
 
Strabon uses a storing scheme which is called "per-predicate" scheme. This scheme 
creates a relational table, in the underlying
DBMS, for every unique predicate in the input data. These tables are called 
\textit{predicate tables} and store pairs of subject and object that are 
associated with a specific predicate.
This storing scheme may lead to the creation of many predicate tables and 
consequently high storing times, if the input data contains a lot of predicates. 
Apart from incremental loading methods, Strabon provides a bulk loader which 
produces CSV files that emulate this ``per-predicate'' scheme and copies them 
into PostgreSQL. The Strabon bulk loader merges in a single relational table, 
which is called \textit{triple table}, triples containing predicates that are 
rarely used in the input data. Thus, the number of created predicate tables is 
reduced together with the required storing time. The storing times of Strabon are, 
still, affected by the number 
of predicates used in a dataset. The real world dataset contains various different 
datasets that also contain a lot of predicates. So, Strabon needs more time than 
uSeekM to store the real world dataset.
 
uSeekM needs slightly less time than Strabon to store the real world dataset, because
it is based on the native repository of Sesame which is known to be the most 
efficient implementation of Sesame for average sized datasets. This happens because
uSeekM stores geometry literals in PostGIS which is more time consuming than storing
data in the native repository of Sesame. If the input data does not contain any 
geometry literals, then uSeekM is entirely based on the native repository of 
Sesame and achieves much better
storing times. Because the Census dataset contains much more geometry literals than the rest real world datasets and it uses much less predicates, uSeekM needs more time to store it than Strabon.

Parliament is slower than uSeekM and Strabon at storing the real world 
workload datasets, as it requires more time to 
perform forward chaining on the input dataset in order to index its geometry literals,
as described in Section~\ref{sec:experimentalsetup}. However this overhead becomes less important in the case of the Census dataset and Parliament needs less time to store it than Strabon and uSeekM.

System X provides two bulk loading methods. The 
first, which is based on SQL operations inside its underlying RDBMS,  is 
designed to provide fast loading but does not support large literals. The second 
uses a Java API and supports large literals but needs more time to store a dataset. 
For storing the real world workload the Java bulk loading method was used, because of 
the many large literals that some datasets contain. For example, the CLC dataset
contains the longest literal that has size 9.3 MBs. Thus, System X needs 
at least twice more time than the other RDF stores to store and index data 
from the real world workload. The Census dataset does not contain big literals, so the SQL bulk loader of System X was used to store it and easily outperformed Strabon, uSeekM and Parliament.

GraphDB has several load methods of which two can be considered as bulk loaders since their are designed 
for offline loading of data sets, directly serializing RDF data into the internal indexes. \textit{LoadRDF} is fast 
but as load data variety grows a small degradation occurs because of page splits and tree 
rebalancing. \textit{PreLoad} is ultra fast with no speed degradation, because of
its two-phase load design, which allows it to first process in memory all RDF data creating multiple
GraphDB repository images and later on sorting and merging these into the final repository image. Both 
tools have the option of enabling parallel multithreaded operation. From our preliminary tests it was
clear that PreLoad was the fastest tool of the two in all datasets but the smallest ones. Therefore the tool of choice
was PreLoad tool for all workloads with the parallel option enabled. In the real world dataset which is the smallest it recorded the second best time very close to RDF4J's and in all other datasets it outperformed most of the other systems by a factor greater 
than x2.

RDF4J has a single method for loading data and it records the best time for the real world dataset
and the second best for the census dataset. With the Lucene index enabled it performs better than uSeekM, Parliament, System X and Strabon's bulk loader. However for the bigger 
census dataset the Lucene indexing cost becomes very high and thus RDF4J needs more than double the time compared to uSeekM which is the second slowest system.

Regarding storage space, System X is the most demanding RDF store while GraphDB and RDF4J are again the most efficient ones. 
System X requires a lot of storage space mainly for semantic indexes and also big literals 
(e.g., for the CLC dataset), that are stored as BLOB (binary large object) in its internal RDBMS. 
RDF4J statement indexes are B-trees with 4-letter composite index keys in various combinations (S=statement, P=predicate, O=object, C=context). By default there are two main indices SPOC, POSC and we enabled the context COSP index only for datasets with multiple graphs, such as the real world dataset.
RDF4J with the Lucene spatial indexing has high storage requirements for the real world and census datasets which
have the most complex geometries. With the Lucene indexing disabled, 
RDF4J needs the least storage space for the real world dataset and is the second best for the Census dataset. 
uSeekM stores most of the data into the native store of Sesame which does not require a lot 
of storage space and only triples with spatial literal are stored into PostgreSQL. So,
it needs less space. 
For GraphDB there are two main statement indices POS and PSO, the context index CPSO which was enabled for datasets with multiple graphs. The GeoSPARQL plugin was not enabled, which helped allocating the least amount of space. 
Strabon and Parliament have average space requirements. Strabon stores all data in PostgreSQL while Parliament uses customized binary files 
to store triples and indexes and it uses a Berkley 
DB\footnote{\url{http://www.oracle.com/us/products/database/berkeley-db/overview/index.html}} file to implement the resource dictionary. 

		\subsubsection{Micro Benchmark}
		\label{sec:resultsRealMicro}
		The query response times of the micro benchmark with cold caches are shown in Table~\ref{table:RealResponseTimes_Cold}
and the corresponding results with warm caches are shown in Table~\ref{table:RealResponseTimes_Warm}.
The two tables are very similar in terms of how the systems are performing hence we
do not discuss these tables separately below.
\begin{table*}[!t]
\tiny
\centering
\resizebox{\textwidth}{!}{
\begin{tabular}{|m{40pt}|l|c|c|c|c|c|c|c|c|}
\hline
\multirow{3}{*}{\textbf{Type}} & \multirow{3}{*}{\textbf{Query}} & \multicolumn{8}{c|}{\textbf{Cold caches} (sec.)} \\
\cline{3-10}
&	& \multirow{2}{*}{\textbf{Strabon}}	& \multirow{2}{*}{\textbf{uSeekM}} & \multirow{2}{*}{\textbf{Parliament}} & \multicolumn{2}{c|}{\textbf{System X}}	& \multirow{2}{*}{\textbf{GraphDB}}	& \multicolumn{2}{c|}{\textbf{RDF4J}} \\
\cline{6-7} \cline{9-10}
&	& & & & Parallel & Serial & & & \textit{Lucene enabled} \\
\hline \hline                                   		                                		                                			                                                                                                    		                                		                                				                                	            		                                
\multirow{6}{40pt}{\textbf{Non topological construct functions}}                		                                			                                                                                                    		                                		                                				                                	            		                                
& \textbf{Q1} & 42.33 & 38.11 & 152.71 & 62.58 & \color{darkred}293.85 & \color{darkgreen}29.68	& 37.24 & 37.34 \\ \cline{2-10}
& \textbf{Q2} & 22.48 & 21.47 & 90.23 & 44.02 & \color{darkred}204.65 & \color{darkgreen}14.55 & 18.41 & 18.75 \\ \cline{2-10}
& \textbf{Q3} & 29.48 & 27.06 & 98.56 & 45.86 & \color{darkred}213.47 & \color{darkgreen}19.42 & 24.57 & 24.56 \\ \cline{2-10}
& \textbf{Q4} & 7.65 & 3.22 & 23.16 & 19.62 & \color{darkred}309.00 & \color{darkgreen}3.20 & 3.32 & 3.48 \\ \cline{2-10}
& \textbf{Q5} & 14.68 & \color{darkgreen}4.17 & 21.63 & 23.60 & \color{darkred}236.60 & 7.45 & 4.52 & 4.64 \\ \cline{2-10}
& \textbf{Q6} & 23.82 & 19.58 & \mbox{-} & 39.87 & \color{darkred}199.25 & \color{darkgreen}13.53 & \mbox{-} & \mbox{-} \\ \cline{2-10}
\hline \hline                                   		                                		                                			                                                                                                    		                                		                                				                                                	                                                   
\multirow{11}{40pt}{\textbf{Spatial selections}} 
& \textbf{Q7} & \color{darkgreen}0.36 & 1.22 & 2.42 & long string & long string & \color{darkred}5.33 & 3.64 & 3.97 \\ \cline{2-10}
& \textbf{Q8} & \color{darkgreen}0.42 & 0.57 & \color{darkred}7.69 & long string & long string & 2.04 & 1.76 & 1.78 \\ \cline{2-10}
& \textbf{Q9} & \color{darkgreen}0.83 & 1.27 & 35.03 & long string & long string & 28.28 & 40.46 & \color{darkred}40.60 \\ \cline{2-10}
& \textbf{Q10} & \color{darkgreen}0.73 & 1.51 & \color{darkred}76.85 & long string & long string & 22.24 & 25.40 & 25.24 \\ \cline{2-10}
& \textbf{Q11} & \color{darkgreen}2.66 & 2.96 & \color{darkred}195.87 & long string & long string & 114.29 & 164.48 & 164.84 \\ \cline{2-10}
& \textbf{Q12} & 0.79 & \color{darkgreen}0.55 & 2.39 & \color{darkred}8.87 & 6.07 & 1.02 & 0.65 & 0.67 \\ \cline{2-10}
& \textbf{Q13} & \color{darkgreen}0.82 & 0.89 & 63.14 & long string & long string & 49.67 & \color{darkred}72.89 & 72.20 \\ \cline{2-10}
& \textbf{Q14} & \color{darkgreen}0.50 & 2.29 & \color{darkred}24.34 & 13.33 & 11.35 & 4.13 & 1.85 & 1.90 \\ \cline{2-10}
& \textbf{Q15} & 0.50 & 0.99 & 3.44 & 10.24 & \color{darkred}10.27 & 0.93 & \color{darkgreen}0.44 & 0.48 \\ \cline{2-10}
& \textbf{Q16} & \color{darkgreen}2.79 & 5.52 & 63.20 & long string & long string & 50.61 & \color{darkred}72.86 & 72.34 \\ \cline{2-10}
& \textbf{Q17} & 3.06 & \color{darkgreen}1.60 & 35.89 & long string & long string & 28.31 & \color{darkred}40.41 & 40.06 \\ \cline{2-10}
\hline \hline                                   		                                		                                			                                                                                                    		                                		                                				                                	                    		                                               
\multirow{10}{40pt}{\textbf{Spatial joins}}     		                                		                                			                                                                                                    		                                		                                				                                	                    		                                               
& \textbf{Q18} & \color{darkgreen}4.52 & 2233.73 & 2880.20 & \color{darkred}\mbox{>1h} & 14.51 & 942.89 & 2894.56 & 2885.72 \\ \cline{2-10}
& \textbf{Q19} & \color{darkgreen}1272.54 & \color{darkred}\mbox{>1h} & \color{darkred}\mbox{>1h} & \color{darkred}\mbox{>1h} & \color{darkred}\mbox{>1h} & \color{darkred}\mbox{>1h} & \color{darkred}\mbox{>1h} & \color{darkred}\mbox{>1h} \\ \cline{2-10}
& \textbf{Q20} & \color{darkgreen}115.93 & \color{darkred}\mbox{>1h} & \color{darkred}\mbox{>1h} & \color{darkred}\mbox{>1h} & 396.29 & \color{darkred}\mbox{>1h} & \color{darkred}\mbox{>1h} & \color{darkred}\mbox{>1h} \\ \cline{2-10}
& \textbf{Q21} & \color{darkgreen}113.26 & \color{darkred}\mbox{>1h} & \color{darkred}\mbox{>1h} & \color{darkred}\mbox{>1h} & 409.54 & \color{darkred}\mbox{>1h} & \color{darkred}\mbox{>1h} & \color{darkred}\mbox{>1h} \\ \cline{2-10}
& \textbf{Q22} & \color{darkgreen}26.33 & \color{darkred}\mbox{>1h} & \color{darkred}\mbox{>1h} & internal error & internal error & \color{darkred}\mbox{>1h} & \color{darkred}\mbox{>1h} & \color{darkred}\mbox{>1h} \\ \cline{2-10}
& \textbf{Q23} & \color{darkgreen}26.29 & \color{darkred}\mbox{>1h} & \color{darkred}\mbox{>1h} & internal error & internal error & \color{darkred}\mbox{>1h} & \color{darkred}\mbox{>1h} & \color{darkred}\mbox{>1h} \\ \cline{2-10}
& \textbf{Q24} & \color{darkgreen}26.66 & \color{darkred}\mbox{>1h} & \color{darkred}\mbox{>1h} & internal error & internal error & \color{darkred}\mbox{>1h} & \color{darkred}\mbox{>1h} & \color{darkred}\mbox{>1h} \\ \cline{2-10}
& \textbf{Q25} & \color{darkgreen}342.87 & \color{darkred}\mbox{>1h} & \color{darkred}\mbox{>1h} & \color{darkred}\mbox{>1h} & 1,629.45 & \color{darkred}\mbox{>1h} & \color{darkred}\mbox{>1h} & \color{darkred}\mbox{>1h} \\ \cline{2-10}
& \textbf{Q26} & 343.30 & 534.61 & 2040.00 & 909.18 & \color{darkred}\mbox{>1h} & 466.86 & 326.22 & \color{darkgreen}324.79 \\ \cline{2-10}
& \textbf{Q27} & \color{darkgreen}343.72 & \color{darkred}\mbox{>1h} & \color{darkred}\mbox{>1h} & internal error & \color{darkred}\mbox{>1h} & \color{darkred}\mbox{>1h} & \color{darkred}\mbox{>1h} & \color{darkred}\mbox{>1h} \\ \cline{2-10}
\hline \hline                                   		                                		                                			                                                                                                    		                                		                                				                                              	                                            
\multirow{2}{40pt}{\textbf{Aggregate functions}}		                                		                                			                                                                                                    		                                		                                				                                              	                                                    
& \textbf{Q28} & \color{darkgreen}3.56 & \mbox{-} & \mbox{-} & \mbox{-} &  \mbox{-} & \mbox{-} & \mbox{-} & \mbox{-} \\ \cline{2-10}
& \textbf{Q29} & \color{darkgreen}258.35 & \mbox{-} & \mbox{-} & \mbox{-} & \mbox{-} & \mbox{-} & \mbox{-} & \mbox{-} \\ \cline{2-10}
\hline
\end{tabular}}
\caption{Response times (cold) - real world workload}
\label{table:RealResponseTimes_Cold}
\end{table*}
\begin{table*}[!t]
	\tiny
	\centering
	\resizebox{\textwidth}{!}{
		\begin{tabular}{|m{40pt}|l|c|c|c|c|c|c|c|c|}
			\hline
			\multirow{3}{*}{\textbf{Type}} & \multirow{3}{*}{\textbf{Query}} & \multicolumn{8}{c|}{\textbf{Warm caches} (sec.)} \\
			\cline{3-10}
			&	& \multirow{2}{*}{\textbf{Strabon}}	& \multirow{2}{*}{\textbf{uSeekM}} & \multirow{2}{*}{\textbf{Parliament}} & \multicolumn{2}{c|}{\textbf{System X}}	& \multirow{2}{*}{\textbf{GraphDB}}	& \multicolumn{2}{c|}{\textbf{RDF4J}} \\
			\cline{6-7} \cline{9-10}
			&	& & & & Parallel & Serial & & & \textit{Lucene enabled} \\
			\hline \hline                                   		                                		                                			                                                                                                    		                                		                                				                                	            		                                
			\multirow{6}{40pt}{\textbf{Non topological construct functions}}                		                                			                                                                                                    		                                		                                				                                	            		                                
& \textbf{Q1} & 41.36	& 36.25	& 132.67 & 57.45 & \color{darkred}296.96 & \color{darkgreen}27.36 & 36.93 & 37.20	\\ \cline{2-10}
& \textbf{Q2} & 21.06 & 19.35	& 70.62	& 35.35 & \color{darkred}187.71 & \color{darkgreen}13.26 & 17.80 & 17.96 \\ \cline{2-10}
& \textbf{Q3} & 27.73 & 24.13 & 79.40 & 38.25 & \color{darkred}206.70 & \color{darkgreen}18.17 & 23.94 & 24.11 \\ \cline{2-10}
& \textbf{Q4} & 7.00 & 3.08 & 19.67 & 19.76 & \color{darkred}334.93 & \color{darkgreen}2.15 & 3.11 & 3.12 \\ \cline{2-10}
& \textbf{Q5} & 13.78 & 5.00 & 19.58 & 17.41 & \color{darkred}233.53 & 6.66 & \color{darkgreen}4.45 & 4.47 \\ \cline{2-10}
& \textbf{Q6} & 21.06 & 18.35 & \mbox{-} & 31.58 & \color{darkred}160.40 & \color{darkgreen}12.67 & \mbox{-} & \mbox{-} \\ \cline{2-10}
			\hline \hline                                   		                                		                                			                                                                                                    		                                		                                				                                                	                                                   
			\multirow{11}{40pt}{\textbf{Spatial selections}} 
& \textbf{Q7} & \color{darkgreen}0.01 & 0.02 & 1.36 & long string & long string & \color{darkred}4.66 & 3.49 & 3.46 \\ \cline{2-10}
& \textbf{Q8} & 0.06 & \color{darkgreen}0.05 & \color{darkred}5.84 & long string & long string & 1.64 & 1.56 & 1.55 \\ \cline{2-10}
& \textbf{Q9} & 0.16 & \color{darkgreen}0.05 & 34.09 & long string & long string & 27.95 & \color{darkred}41.07 & 40.77 \\ \cline{2-10}
& \textbf{Q10} & 0.13 & \color{darkgreen}0.10 & \color{darkred}57.18 & long string & long string & 21.24 & 24.32 & 24.14 \\ \cline{2-10}
& \textbf{Q11} & 2.03 & \color{darkgreen}1.29 & \color{darkred}175.98 & long string & long string & 114.95 & 164.78 & 163.44 \\ \cline{2-10}
& \textbf{Q12} & 0.38 & \color{darkgreen}0.03 & 1.20 & \color{darkred}4.02 & 3.99 & 0.67 & 0.57 & 0.57 \\ \cline{2-10}
& \textbf{Q13} & 0.13 & \color{darkgreen}0.04 & 59.60 & long string & long string & 49.65 & \color{darkred}72.59 & 71.92 \\ \cline{2-10}
& \textbf{Q14} & \color{darkgreen}0.03 & 1.63 & \color{darkred}19.97 & 4.12 & 4.03 & 3.11 & 1.69 & 1.68 \\ \cline{2-10}
& \textbf{Q15} & \color{darkgreen}0.12 & 0.30 & 0.56 & 3.94 & \color{darkred}3.97 & 0.30 & 0.26 & 0.26 \\ \cline{2-10}
& \textbf{Q16} & 2.19 & \color{darkgreen}1.96 & 59.85 & long string & long string & 49.68 & \color{darkred}72.63 & 71.97 \\ \cline{2-10}
& \textbf{Q17} & 2.62 & \color{darkgreen}0.86 & 34.39 & long string & long string & 27.66 & \color{darkred}40.28 & 39.93 \\ \cline{2-10}
			\hline \hline                                   		                                		                                			                                                                                                    		                                		                                				                                	                    		                                               
			\multirow{10}{40pt}{\textbf{Spatial joins}}     		                                		                                			                                                                                                    		                                		                                				                                	                    		                                               
& \textbf{Q18} & \color{darkgreen}3.98 & 2504.24 & 2875.02 & \color{darkred}\mbox{>1h} & 11.53 & 903.33 & 2972.27 & 2969.07 \\ \cline{2-10}
& \textbf{Q19} & \color{darkgreen}1284.62 & \color{darkred}\mbox{>1h} & \color{darkred}\mbox{>1h} & \color{darkred}\mbox{>1h} & \color{darkred}\mbox{>1h} & \color{darkred}\mbox{>1h} & \color{darkred}\mbox{>1h} & \color{darkred}\mbox{>1h} \\ \cline{2-10}
& \textbf{Q20} & \color{darkgreen}105.39 & \color{darkred}\mbox{>1h} & \color{darkred}\mbox{>1h} & \color{darkred}\mbox{>1h} & 365.01 & \color{darkred}\mbox{>1h} & \color{darkred}\mbox{>1h} & \color{darkred}\mbox{>1h} \\ \cline{2-10}
& \textbf{Q21} & \color{darkgreen}107.76 & \color{darkred}\mbox{>1h} & \color{darkred}\mbox{>1h} & \color{darkred}\mbox{>1h} & 364.32 & \color{darkred}\mbox{>1h} & \color{darkred}\mbox{>1h} & \color{darkred}\mbox{>1h} \\ \cline{2-10}
& \textbf{Q22} & \color{darkgreen}25.20 & \color{darkred}\mbox{>1h} & \color{darkred}\mbox{>1h} & internal error & internal error & \color{darkred}\mbox{>1h} & \color{darkred}\mbox{>1h} & \color{darkred}\mbox{>1h} \\ \cline{2-10}
& \textbf{Q23} & \color{darkgreen}25.01 & \color{darkred}\mbox{>1h} & \color{darkred}\mbox{>1h} & internal error & internal error & \color{darkred}\mbox{>1h} & \color{darkred}\mbox{>1h} & \color{darkred}\mbox{>1h} \\ \cline{2-10}
& \textbf{Q24} & \color{darkgreen}25.37 & \color{darkred}\mbox{>1h} & \color{darkred}\mbox{>1h} & internal error & internal error & \color{darkred}\mbox{>1h} & \color{darkred}\mbox{>1h} & \color{darkred}\mbox{>1h} \\ \cline{2-10}
& \textbf{Q25} & \color{darkgreen}341.04 & \color{darkred}\mbox{>1h} & \color{darkred}\mbox{>1h} & \color{darkred}\mbox{>1h} & 1,577.89 & \color{darkred}\mbox{>1h} & \color{darkred}\mbox{>1h} & \color{darkred}\mbox{>1h} \\ \cline{2-10}
& \textbf{Q26} & 341.28 & 534.15 & 2030.42 & 955.21 & \color{darkred}\mbox{>1h} & 465.35 & 325.69 & \color{darkgreen}324.04 \\ \cline{2-10}
& \textbf{Q27} & \color{darkgreen}342.06 & \color{darkred}\mbox{>1h} & \color{darkred}\mbox{>1h} & internal error & \color{darkred}\mbox{>1h} & \color{darkred}\mbox{>1h} & \color{darkred}\mbox{>1h} & \color{darkred}\mbox{>1h} \\ \cline{2-10}

			\hline \hline                                   		                                		                                			                                                                                                    		                                		                                				                                              	                                            
			\multirow{2}{40pt}{\textbf{Aggregate functions}}		                                		                                			                                                                                                    		                                		                                				                                              	                                                    
& \textbf{Q28} & \color{darkgreen}2.92 & \mbox{-} & \mbox{-} & \mbox{-} & \mbox{-} & \mbox{-} & \mbox{-} & \mbox{-} \\ \cline{2-10}
& \textbf{Q29} & \color{darkgreen}258.00 & \mbox{-} & \mbox{-} & \mbox{-} & \mbox{-} & \mbox{-} & \mbox{-} & \mbox{-} \\ \cline{2-10}
			\hline
	\end{tabular}}
	\caption{Response times (warm) - real world workload}
	\label{table:RealResponseTimes_Warm}
\end{table*}

\textit{Non-topological construct functions.}
First, the results of evaluating the queries with non-topological functions are 
reported. Computing the area of polygons (Query Q6) was tested only in uSeekM, 
Strabon, System X and GraphDB since Parliament and RDF4J do not offer such functionality.
For this class of queries, GraphDB and RDF4J are the fastest systems followed closely
by uSeekM which does very well in \texttt{geof:buffer()} calculations. uSeekM does not utilize 
PostgreSQL for evaluating these queries, but it is faster because it uses the native store 
of Sesame which is known to be more efficient, for small datasets, than Sesame 
implementations on top of an RDBMS, like Strabon. Strabon performance is average
while Parliament and System X perform the worse.  Parliament needs three or four times more time to evaluate the 
non-topological functions. Finally, System X, when it runs in serial mode, needs 
considerably more time to evaluate non-topological queries. System X stores spatial 
literals in lexical form instead of a dedicated binary geometric type. This means 
that each time a spatial function is evaluated, the spatial literals must be 
transformed from lexical to geometry form. This causes a great overhead, especially 
for the real-world workload that use complex geometries. However, when System X
runs in parallel mode this overhead is distributed among all available
processors and System X evaluates these queries up to sixteen times faster.

We also observe that none of the RDF stores exploits the warm caches when 
evaluating non-topological functions. This is because the non-topological functions 
used in this set of queries are computationally intensive (especially when complex 
geometries are used) and the time spent in the CPU dominates I/O time.

\textit{Spatial selections.}
In the case of spatial selections, Strabon and uSeekM have similar response times 
while Strabon is the fastest system in most cases. Both systems choose to start 
the query evaluation process by evaluating the spatial part of a query in PostGIS 
using the spatial index that is available. uSeekM continues by evaluating the rest 
of the query using the native store of Sesame. This adds a small overhead compared 
to Strabon which evaluates the whole query in PostgreSQL and utilizes a unified 
dictionary encoding scheme for both thematic and spatial information. 
GraphDB has an average performance which is expected since it does not use
its spatial indexing capabilities while RDF4J and Parliament are at the low end
of the list. RDF4J performs very well in queries that
deal with points and lines, average with polygon equality test and very low for all other
operations with polygons. Parliament, depending on the 
query, it may need even three orders of magnitude more time that Strabon and uSeekM 
to evaluate a spatial selection. This happens because the query optimizer of Parliament 
does not take into consideration filters containing  GeoSPARQL functions, so it 
evaluates the spatial predicate exhaustively over the results of the thematic part 
of the query.
System X is optimized for relatively simple spatial literals, while 
the tested spatial selections receive as parameter quite complex polygons 
and linestrings. For example, the WKT serialization of some polygons are even
10 KB long. Thus, System X returned an error or raised an exception 
(written as "long string" in Tables~\ref{table:RealResponseTimes_Cold},~\ref{table:RealResponseTimes_Warm}) for most of the spatial selection queries and it responded to only Query Q12 which uses
a small linestring and queries Q14 and Q15 which filter using points.

Let us now consider queries Q14 and Q15, that are semantically equivalent, but they
are evaluated in different ways. Both ask for points that
have a given distance from a given point. However, Query Q14 creates the buffer
of a given point with radius $r$ and asks for points which are within this
buffer, while Query Q15 asks for points that lie within distance less than $r$ from the
given point. uSeekM and Parliament evaluate both queries by starting with the 
thematic part of the query and then they evaluate exhaustively the spatial 
operations without using the spatial index. GraphDB follows the same path since it does 
not have a spatial index. Query Q14 is evaluated slower, than Query 15, by these systems, because 
calculating the distance between two points is much cheaper than computing the 
buffer of a point and evaluating the corresponding point-in-polygon operation. 
Strabon follows a similar process for responding to Query Q15. However, for query Q14, 
Strabon calculates the buffer of the given polygon, and uses it to probe the spatial 
index for discovering points that lie inside the constructed polygon. This choice is 
a good one and the response time of query Q14 is the same as the one of query Q15.
Finally, System X evaluates query Q14 with a similar process to Strabon, and regarding query 
Q15 it is the only system that uses an internal distance function that is able to 
perform index search instead of evaluating the distance filter over all intermediate 
results. So, it achieves similar times for both queries Q14 and Q15. For these
selection queries, parallelism does not have significant impact on response times, that
are very low independently of whether parallelism is being used or not.

\textit{Spatial Joins.}
In the case of spatial joins, uSeekM, Parliament, GraphDB and RDF4J are able to 
evaluate only queries Q18 and Q26 given the time limit of one hour.
Parliament, GraphDB and RDF4J do not take into account GeoSPARQL extension functions during the optimization 
of a query, resulting in query process that evaluate separately the graph patterns corresponding to 
different graphs, compute the Cartesian product between them, and then apply the
spatial predicate to the result of the Cartesian product. This strategy is very
costly, thus they are not able to respond to most of the spatial joins within the time limit. 
uSeekM, also, does not utilize PostGIS for evaluating 
spatial joins. Similarly to Parliament, it applies the spatial predicate to the 
result of the Cartesian product of the graph patterns. Strabon avoids evaluating 
Cartesian products by identifying graph patterns that are related only through 
the spatial predicate and pushes 
the evaluation of the spatial join in PostGIS, thus resulting in very good response times.
Strabon has the best performance in all queries with the exception of Q26.
In Q26 RDF4J performs the best mainly because the GAG graph is small and the thematic selectivity of the query is high. In Q18 the thematic 
selectivity is high while the spatial selectivity very low resulting in a similar performance with Parliament's.
System X also uses its native RDBMS to evaluate spatial joins and, when running
in serial mode, it avoids Cartesian products. But, because of implementation limitations 
(e.g., the overhead of transforming complex geometries from strings into geometry types) 
it needs more time than Strabon and it also has some time outs.
System X did not responded to Queries Q22, Q23, Q24 and Q27 because of an internal exception.
It is interesting that when running in parallel mode, the optimizer of System X
prefers to ignore the spatial index and compute a Cartesian product for evaluating
most of the spatial joins. These query execution methods, even if they run in parallel mode, 
lead to higher response times, than the respective methods of serial mode, 
and time outs.

\textit{Aggregate functions.}
Finally, spatial aggregations are tested only in Strabon since it is the only system 
that supports such functions. Query Q28 which computes the minimum 
bounding box that contains all geometries of the GAG dataset is much faster than 
Query Q29 which computes the union of the same geometries since the former operation 
is much cheaper than the latter one which is computationally expensive.

A general comment about RDF4J is that in the micro benchmark the Lucene index did not
make any difference in the query response times.

		\subsubsection{Macro Benchmark}
		\label{sec:resultsRealMacro}
		The results of the macro benchmark are shown in Table~\ref{table:MacroResults}.
In this table the average time needed for a complete iteration of all
queries of each scenario is reported. 

The "Geocoding" scenario includes only thematic queries
that retrieve geographic information. Thus, uSeekM evaluates the whole queries
in the native Sesame store achieving very fast response times. RDF4J performs second best
and GraphDB which is also based on RDF4J follows close by. System X also has very
fast response times both in serial and parallel mode and it is the 
fourth fastest RDF store for this scenario. RDF4J with the Lucene index 
performs average because it is not used in these queries.
Strabon uses its underlying RDBMS and has slower response times, while 
Parliament is the slowest RDF store in this scenario. 

The "Reverse Geocoding" scenario has two queries which use the 
function \texttt{distance} to sort retrieved geometries and select the first result 
that is closest to a given point. GraphDB performs the best in this scenario and is 
followed closely by RDF4J, uSeekM and RDF4J with Lucene. Parliament also has a fast 
response in this scenario, but it is 3 to 4 times slower than the systems of the first group. 
On the other hand, System X and Strabon, which are based on an internal RDBMS, need at least an order of 
magnitude more time to respond to a whole iteration of this scenario.

In order to respond to these nearest-neighbor queries of this scenario, all RDF 
stores compute the distance of every retrieved geometry from the given point, 
then they sort these values in ascending order and select the first geometry that 
corresponds to the minimum distance. Strabon, especially, inserts every value 
computed by the function \texttt{distance} 
into the respective dictionary encoding table. As more nearest-neighbor queries 
are posed, this dictionary table is getting bigger and bigger 
and the performance of Strabon is deteriorating. So, its average iteration time 
is very high in this scenario. On the contrary, the other systems discard the 
intermediate distance values, so they achieve faster response times. 
 
The "Map Search and Browsing" scenario has one thematic query and two spatial 
selection queries. As described in Section~\ref{sec:resultsRealMicro}, 
uSeekM and Strabon are efficient in evaluating spatial selections and they have good performance 
in this scenario as well, followed closely by GraphDB. RDF4J performance in both modes is average while
System X (in serial mode) performs the worst in this scenario because of Query MSB3, 
which asks for complex geometries (linestrings) and gives a lot of results. System X needs, on average, 
120 seconds to respond to Query MSB3. In parallel mode the same query needs only 
around 14 seconds, so the average performance of System X improves.
Strabon and Parliament spend most of the time 
in evaluating this query, as well. On the contrary, uSeekM spends more time in 
evaluating query MSB1, because it generates the Cartesian product between two 
triples. But, it has very fast response times so it is still the faster system in 
this scenario.

The "Rapid Mapping for Fire Monitoring" scenario is the most demanding
one. It comprises three spatial selections queries, but also two complex queries 
which include spatial joins and construct new geometries (boundary and intersection). 
Only Strabon could serve this scenario since all other stores exceeded 
the time limit of one hour during evaluating queries of this scenario. Parliament, uSeekM, 
GraphDB and RDF4J timed out while evaluating Query RM6, and System X while evaluating queries 
RM4 and RM6. These two queries are also the most time consuming for Strabon as well 
because they produce many results.

Finally, the "Computing Statistics of Geospatial Datasets" scenario tests computing 
aggregations over simple spatial selections or spatial joins of geospatial datasets.
In this scenario uSeekM is the fastest system that needs, on average, less than a second 
to respond to all three queries of this scenario. The second fastest system is Strabon that
needs about 5 seconds to respond to a full iteration of this scenario. GraphDB, RDF4J and System X
have an average performance. Parliament is the slowest store which spends the most of 
the time in evaluating Query CS1, 
while for the rest of the systems Query CS2
which contains a spatial join is the most time consuming. Finally, System X has 
similar performance both in parallel and serial mode.

\iflong
		\begin{table*}[!ht]
	\tiny
	\centering
	\resizebox{\textwidth}{!}{
		\begin{tabular}{|m{40pt}|c|c|c|c|c|c|c|c|} 
			\hline
			\multirow{2}{*}{\textbf{Scenario}} & \multirow{2}{*}{\textbf{Strabon}} & \multirow{2}{*}{\textbf{uSeekM}} & \multirow{2}{*}{\textbf{Parliament}} & \multicolumn{2}{c|}{\textbf{SystemX}} & \multirow{2}{*}{\textbf{GraphDB}} & \multicolumn{2}{c|}{\textbf{RDF4J}} \\ \cline{5-6}  \cline{8-9}
			& & & &	Par. & Ser. & & & \textit{Lucene enabled} \\
			\hline
			Geocoding & 29.40 & \color{darkgreen}0.05 & \color{darkred}63.26 & 3.04 & 7.71 & 1.51 & 1.37 & 10.25  \\ 
			\hline
			Reverse	Geocoding & \color{darkred}65 & 0.77 & 2.6 & 15.07 & 15.16 & \color{darkgreen}0.60 & 0.73 & 0.85 \\ 
			\hline
			Map Search and Browsing & 0.9 & \color{darkgreen}0.6 & 22.2 & 22.14 & \color{darkred}124.49 & 1.03 & 4.09 & 4.80 \\ 
			\hline
			Rapid Mapping for Fire Monitoring & \color{darkgreen}207.4 & - & - & - & - & - & - & - \\ 
			\hline
			Computing Statistics of Geospatial Datasets & 4.55 & \color{darkgreen}0.4 & \color{darkred}63.49 & 22.68 & 19.40 & 17.04 & 20.43 & 21.99 \\ 
			\hline
		\end{tabular}}
		\captionof{table}{Average iteration times - macro scenarios (sec.)}		
		\label{table:MacroResults}
\end{table*}

\else
			input{tabular_tex/results_macro.tex}
	\end{figure}
\fi

	\subsection{Synthetic Workload}
	Let us now discuss experiments that were run using a synthetic
	workload that was produced by the generator presented in
	Section~\ref{sec:benchmark}.
	A dataset was generated by setting $n=512$ and $k=9$, where $n$ is the number
	used for defining the cardinalities of the generated geometries, and $k$ is
	the number used for defining the cardinalities of the generated tag values.
	This dataset produced using the synthetic generator contains $262,144$ land
	ownership instances, $28,900$ states, $512$ roads and $262,144$ points of
	interest. All features are tagged with key $1$, every other feature with key
	$2$, etc. up to key $512$. The resulting dataset consists of 3,880,224 triples
	and its size is 745 MB.
		\subsubsection{Dataset Storage}
		\label{sec:resultsSyntheticStore}
		Table~\ref{table:StoringTime} presents the time required by each system to store and index the synthetic dataset and Table~\ref{table:RepositorySize} presents the required storage space. 

The synthetic dataset has fewer predicates and more geometries than the real world one. 
GraphDB is the fastest system because of the parallel multi-threaded operation of the PreLoad tool and because the GeoSPARQL plugin was disabled.
RDF4J performs very well and is placed second. With the additional cost of Lucene spatial indexing, it still performs close to Strabon's time and that is because the synthetic dataset is relatively small.
uSeekM requires more time than Strabon for storing the dataset, since it stores it in a Sesame native store and then it stores triples with geometric information in PostGIS as well. This overhead is significant compared to the total time required for storing the dataset, but leads to better response times of uSeekM in case of evaluating a query with low spatial selectivity, as discussed in Section~\ref{sec:resultsSyntheticQueries} (see Figures~\ref{fig:SyntheticIntersects1Cold}-~\ref{fig:SyntheticWithin512Warm}).
As already explained in Section~\ref{sec:resultsRealStore}, Parliament needs more time to store the synthetic dataset as well as the real world dataset because it performs forward chaining on the input data. 
The synthetic dataset does not contain any huge literals so we were able to use the SQL bulk loader of System X to store this data. As in the case of the Census dataset, the SQL bulk loader achieves much better storare times than the Java API used to store the real world workload dataset. 

Regarding storage space, GraphDB requires the least space because it uses the two basic indexes POS and PSO and the GeoSPARQL plugin was disabled. RDF4J is a little more demanding because it uses by default the more complex SPOC, POSC
indexes. Also, the SQL bulk loader of System X and the bulk loader of Strabon, which also utilizes SQL operations to load data, achieve similar fast storage. Strabon, uSeekM and Parliament increase their storage demands to cater for the increase
in number of facts and geometries. System X has the highest storage demands because of the semantic indexes and big
literals.

		\subsubsection{Queries}
		\label{sec:resultsSyntheticQueries}
		We used the query template presented in
Table~\ref{tbl:syntheticQueryTemplatesSelect} in order to produce SPARQL
queries corresponding to spatial selections that ask for land ownerships which
intersect a given rectangle, and points of interest that are within a given
rectangle. The given rectangle is generated in such a way that the spatial
predicate of the query holds for 0.01\%, 10\%, 25\%, 50\%, 75\%
of all the features of the respective dataset. In addition,
the query template was instantiated using the extreme values \texttt{1} and \texttt{512} of the
parameter \texttt{THEMA} for selecting either all or approximately
0.02\%  \mbox{ } of the total features  of a dataset.
The response time of each system for responding to this query
template are presented in
Figures~\ref{fig:SyntheticIntersects1Cold}-\ref{fig:SyntheticWithin512Warm}.

We implemented the query template presented in
Table~\ref{tbl:syntheticQueryTemplatesJoin} in order to produce SPARQL queries
corresponding to spatial joins that ask for land ownerships that intersect a
state, touching states and points of interest that are located inside a state.
We also implemented this query template using all combinations of the extreme
values \texttt{1} and \texttt{512} for the parameters \texttt{THEMA1} and
\texttt{THEMA2}.
The response time of each system executing
template are presented in
Figures~\ref{fig:SyntheticJoinIntersects}-\ref{fig:SyntheticJoinWithin}.

\textit{Spatial selections.}
By examining Figures~\ref{fig:SyntheticIntersects1Cold}-~\ref{fig:SyntheticWithin512Warm}, 
we observe that Strabon has very good performance overall in spatial selections.
uSeekM has low response times when few features satisfy the spatial predicate but when
more features satisfy the spatial predicate the response time increases. Parliament has very high response times in most cases regardless of the spatial or thematic selectivity of the queries. 
In most cases, System X has average performance when running in parallel mode and
it is the second fastest RDF store after Strabon. When running in serial mode
System X has worse performance.
GraphDB has the second best performance after Strabon and scores the best times in the low thematic selectivity queries. It has low sensitivity with spatial selectivity of the queries. These two features are also shared by RDF4J which scores very high in the \texttt{"512 tag"} group of queries and is insensitive to query spatial selectivity changes, but in the \texttt{"1 tag"} group of queries scores low, being better than Parliament and serial mode System X.

\begin{figure*}[!t]
  \centering
	  \subfigure
	  {\includegraphics{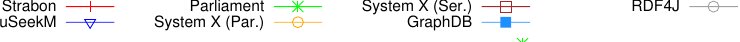}}
	  \\
	   \setcounter{subfigure}{0}
	  \subfigure[][\shortstack{Intersects\\ \normalsize{ tag 1, cold caches}}]
	  {\includegraphics[scale=1.2]{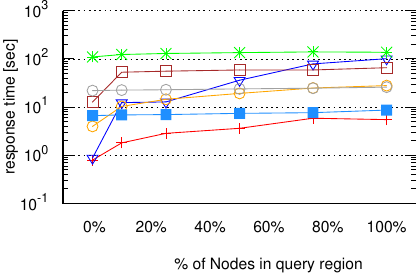}
	  \label{fig:SyntheticIntersects1Cold}}
	  \subfigure[][\shortstack{Intersects\\ \normalsize{ tag 512, cold caches}}]
	  {\includegraphics[scale=1.2]{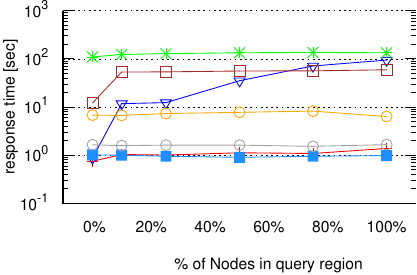}
	  \label{fig:SyntheticIntersects512Cold}}
	  \\
	  \subfigure[][\shortstack{Within\\ \normalsize{ tag 1, cold caches}}]
	  {\includegraphics[scale=1.2]{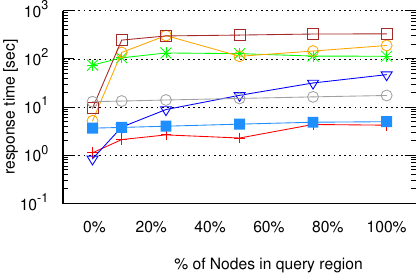}
	  \label{fig:SyntheticWithin1Cold}}
	  \subfigure[][\shortstack{Within\\ \normalsize{tag 512, cold caches}}]
	  {\includegraphics[scale=1.2]{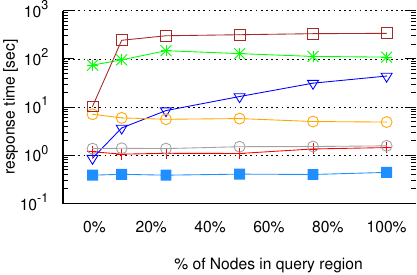}
	  \label{fig:SyntheticWithin512Cold}}
	  \\
	  \subfigure[][\shortstack{Intersects\\ \normalsize{ tag 1, warm caches}}]
	  {\includegraphics[scale=1.2]{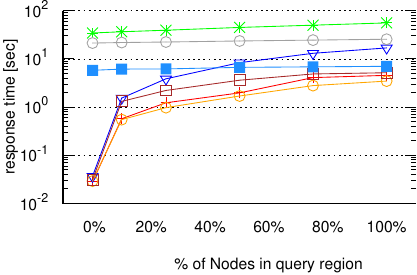}
	  \label{fig:SyntheticIntersects1Warm}}
	  \subfigure[][\shortstack{Intersects\\ \normalsize{ tag 512, warm caches}}]
	  {\includegraphics[scale=1.2]{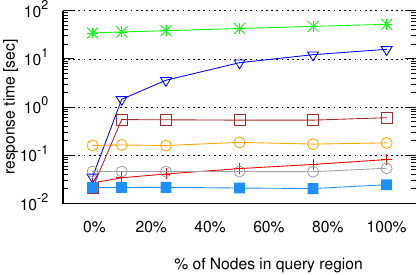}
	  \label{fig:SyntheticIntersects512Warm}}
	  \\
	  \subfigure[][\shortstack{Within\\ \normalsize{ tag 1, warm caches}}]
	  {\includegraphics[scale=1.2]{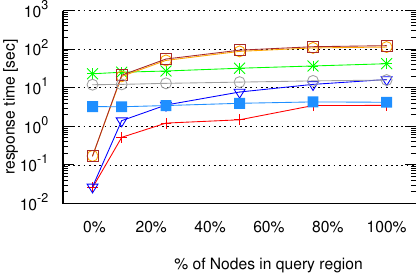}
	  \label{fig:SyntheticWithin1Warm}}
	  \subfigure[][\shortstack{Within\\ \normalsize{ tag 512, warm caches}}]
	  {\includegraphics[scale=1.2]{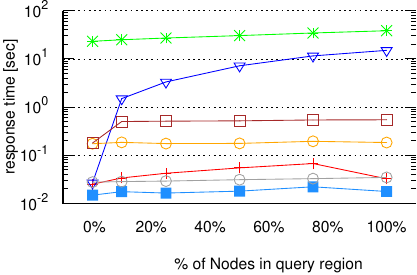}
	  \label{fig:SyntheticWithin512Warm}}
	\caption{Response times - synthetic workload (selections)}
	\label{fig:SyntheticResponseTimesSelections}
\end{figure*}

Strabon uses PostgreSQL (extended with PostGIS) to execute a SPARQL query. PostGIS
has been enhanced with spatial selectivity estimation capabilities, 
from versions 2.x onwards. 
As a result, when a query selects only few geometries, PostgreSQL always
starts with execution of the spatial predicate using the spatial index,
thus resulting in few intermediate results and good response times. While the
spatial selectivity increases and more geometries satisfy the spatial predicate,
the optimizer of PostgreSQL chooses different query execution methods. For example, when the
value of the parameter \texttt{THEMA} is $1$ (Figures
\ref{fig:SyntheticIntersects1Cold}, \ref{fig:SyntheticWithin1Cold},
\ref{fig:SyntheticIntersects1Warm}, \ref{fig:SyntheticWithin1Warm}) and the
value of the parameter \texttt{GEOM} is such that all geometries satisfy the spatial
predicate, PostgreSQL ignores the spatial index and performs a sequential scan on
the table storing the geometries for evaluating the spatial predicate.
Similarly, when the value of the parameter \texttt{THEMA} is $512$ (Figures
\ref{fig:SyntheticIntersects512Cold}, \ref{fig:SyntheticWithin512Cold},
\ref{fig:SyntheticIntersects512Warm}, \ref{fig:SyntheticWithin512Warm})
and the value of the parameter \texttt{GEOM} is such that all
geometries satisfy the spatial predicate, PostgreSQL starts with the execution of
the thematic selection that produces few intermediate results since only
0.02\% \mbox{ } of the features satisfy the thematic predicate,
resulting in good query response times.

Regarding uSeekM, its performance is not affected by the
thematic selectivity of the query. For spatial selections, uSeekM always starts
with the spatial predicate in PostGIS and then continues the query
execution in the native Sesame store. As a result, regardless of the thematic
selectivity, the response time of uSeekM is low when few features satisfy the spatial 
predicate and increases when the number
of features with geometries that satisfy the given spatial predicate increases.

Regarding Parliament, its performance is not affected by
the thematic or by the spatial selectivity of a query. Parliament always starts
by executing the non-spatial part of a query and then executes the thematic
filter and the spatial predicate exhaustively on the intermediate
results. Thus, the thematic and spatial selectivity of a query do not affect
its response time.

System X, like Strabon, is capable of estimating the selectivity of both the spatial and 
the thematic part of a query and to select correct query execution paths. 
Especially when it runs in parallel mode, System X has fast 
response times when it starts by executing thematic filters (Figures \ref{fig:SyntheticIntersects512Cold}, \ref{fig:SyntheticWithin512Cold}, 
\ref{fig:SyntheticIntersects512Warm}, \ref{fig:SyntheticWithin512Warm}) and outperforms
uSeekM and Parliament, which select wrong query execution paths. 
However, its response times 
get higher when it starts with executing spatial predicates (Figures 
\ref{fig:SyntheticIntersects1Cold}, \ref{fig:SyntheticWithin1Cold}, 
\ref{fig:SyntheticIntersects1Warm}, \ref{fig:SyntheticWithin1Warm}), even if this choice is 
correct, and results get mixed. This means that System X, which stores geometries in lexical 
form and uses an internal function to execute spatial predicates, is slower 
in executing spatial predicates that uSeekM and Strabon, which utilize PostGIS.
Also, the performance of System X in parallel mode compared to its performance in
serial mode improves more when it starts by executing the thematic part of a 
query than  when it starts with the spatial part. This indicates that filtering 
operations on actual lexical values are better parallelized by System X than 
filtering on spatial values.

In \cite{iswc2012-strabon}, similar experiments have been performed for evaluating
the performance of Strabon in spatial selection queries where spatial and thematic 
selectivity of queries can be controlled. In these experiments only point geometries
were used and an older version of Strabon was tested. This older version of Strabon 
utilized a PostGIS version prior to 2.x which lacks the capability to estimate the 
spatial selectivity of a query. Experimental results described in \cite{iswc2012-strabon} showed 
that the absence of dynamic estimation of spatial selectivity can lead to wrong 
query execution paths and increase the response time of a system. This happens
because the system is not able to correctly select the part (thematic or spatial) of 
the query  that produces less intermediate results.
The importance of dynamic estimation of spatial selectivity becomes more obvious 
in experiments of Geographica where Strabon and, in many cases, System X 
outperform uSeekM because they are able to select correct 
query execution paths. For example, in cases where the thematic selectivity 
of a query is low while the spatial selectivity increases 
(Figures~\ref{fig:SyntheticIntersects512Cold}, \ref{fig:SyntheticWithin512Cold}, 
\ref{fig:SyntheticIntersects512Warm}, \ref{fig:SyntheticWithin512Warm}) both
Strabon and System X begin the query execution with the thematic part of the
query and achieve lower response times than uSeekM that always executes first
the spatial part of a query.
 
 \begin{figure*}[!t]
 	\centering
 	\subfigure
 	{\includegraphics[scale=0.7]{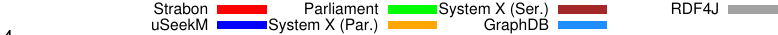}}
  	\setcounter{subfigure}{0}	
 	\\
 	\subfigure[][Intersects]
 	{\includegraphics[scale=0.7]{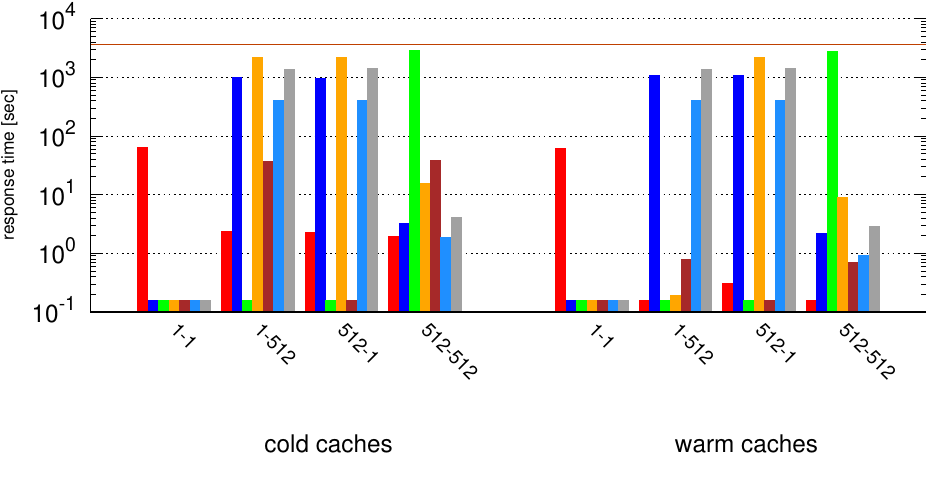}
 		\label{fig:SyntheticJoinIntersects}}	
 	\subfigure[][Touches]
 	{\includegraphics[scale=0.7]{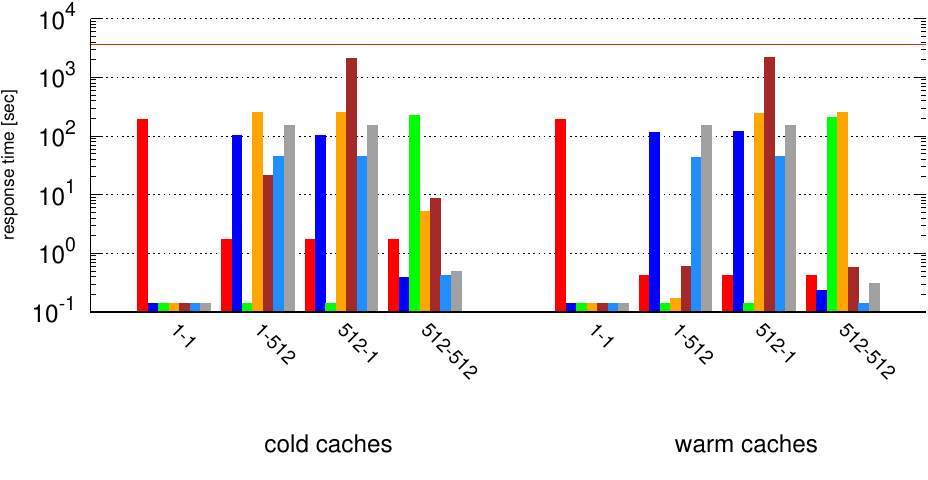}
 		\label{fig:SyntheticJoinTouches}}
 	\\
 	\subfigure[][Within]
 	{\includegraphics[scale=0.7]{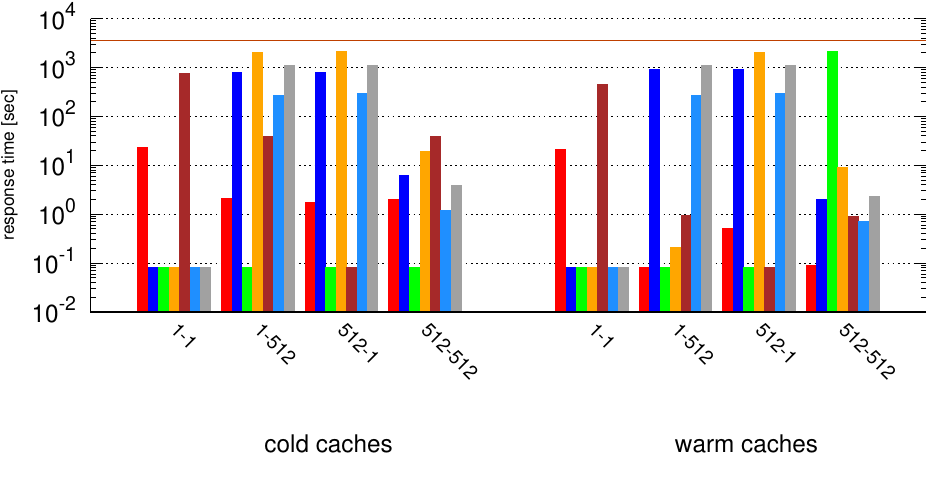}
 		\label{fig:SyntheticJoinWithin}}
 	\\
 	\caption{Response times  - synthetic workload (joins)}
 	\label{fig:SyntheticResponseTimesJoins}
 \end{figure*}
 
\textit{Spatial Joins.}
In the case of spatial joins 
(Figures~\ref{fig:SyntheticJoinIntersects}-~\ref{fig:SyntheticJoinWithin}), 
Strabon is the fastest system in most cases and the only one that responded to
every query within the time limit of one hour. uSeekM, System X, GraphDB and RDF4J executed most of the spatial joins, given the one hour time limit, but they needed
more time than Strabon. Finally, Parliament responded, within the time limit, 
the spatial joins only when parameters \texttt{THEMA1} and \texttt{THEMA2}
are equal to 512.

Strabon relies on the optimizer of PostgreSQL which takes into account the 
thematic selectivity of the
queries and selects good query paths, thus Strabon is the only system that is
able to respond to the spatial joins given the one hour timeout when the parameters 
\texttt{THEMA1} and \texttt{THEMA2} are equal to $1$.

System X, running in parallel mode, did not respond to any join, with parameters \texttt{THEMA1}
and \texttt{THEMA2} equal to 1, within the time limit of one hour. While, running 
in serial mode an internal exception occurred when evaluating functions 
\texttt{geof:sfIntersects} (Figure~\mbox{\ref{fig:SyntheticJoinIntersects}}) and 
\texttt{geof:sfWithin} (Figure~\mbox{\ref{fig:SyntheticJoinWithin}}). 
Also, System X (running in serial mode) needed more than one hour to evaluate the
joins with parameters \texttt{THEMA1} equal to 512 and \texttt{THEMA2} equal to 1 using
functions \texttt{geof:sfIntersects} and \texttt{geof:Within}.
Regarding spatial joins, System X relied on the estimated cardinality of the 
first variable
of the spatial predicate to decide whether to use the spatial index or not. System X
chooses to use the spatial index when the first spatial variable has high cardinality 
(\texttt{THEMA1=1}), regardless of the second spatial variable (consequently the parameter
\texttt{THEMA1}). Similarly, when the cardinality of the first spatial variable is low
(\texttt{THEMA1=512}) System X ignores the spatial index, it computes the Cartesian product
of the given triple patterns and evaluates the spatial filter over the intermediate 
results. This strategy is not always effective because it ignores the cardinality of the 
second spatial variable in planning the evaluation of a query. For example, in Figures \mbox{\ref{fig:SyntheticJoinIntersects}} and 
\mbox{\ref{fig:SyntheticJoinWithin}} 
evaluating the spatial joins with parameters \texttt{THEMA1}=1 and \texttt{THEMA2}=512
needs the same or less time than evaluating the joins with parameters \texttt{THEMA1}=512
and \texttt{THEMA2}=512 even though the latter are more selective.

Finally, uSeekM, GraphDB, RDF4J and Parliament produce the Cartesian
product between the graph patterns that are joined through the spatial
predicate, and evaluate the spatial predicate afterwards. This strategy is very
costly, thus Parliament is not able to respond to most spatial joins given the
one hour timeout and the other systems are more than two orders of magnitude slower than Strabon.
However, in Figure~\ref{fig:SyntheticJoinTouches} \texttt{512-512} it is shown that uSeekM, GraphDB and RDF4J outperform Strabon. Strabon stores all geometries in a single table, so the
evaluation of the spatial predicate $Touches$ on this table returns not only the
geometries of states that touch each other, but the touching geometries of land
ownerships as well. The touching geometries of land ownerships are discarded
later on, but this overhead proves to be more costly than producing a
Cartesian product and evaluating the spatial predicate afterwards.

	\subsection{Scalability Workload}	
		\subsubsection{Dataset Storage}
		\label{sec:resultsScalabilityStore}
		This section discusses the time required by each system to store
and index the datasets of the scalability workload, as shown in Table~\ref{table:ScalabilityStoringTime}.
Also, Table~\ref{table:ScalabilityRepositorySize} reports the size of the repositories 
created by each RDF store. We also reveal weak spots of each process and the workarounds employed to overcome them.
\begin{table*}[!t]
	\center
	\scriptsize
\begin{tabular}{|l|l|c|c|c|c|c|c|c|c|}
	\hline
	\multicolumn{2}{|c|}{\multirow{2}{*}{\textbf{System}}}  & \textbf{10K triples}  & \textbf{100K triples} & \textbf{1M triples}   & \textbf{10M triples}  & \multicolumn{2}{c|}{\textbf{100M triples}} & \multicolumn{2}{c|}{\textbf{500M triples}} \\ \cline{3-10} 
	\multicolumn{2}{|c|}{}                                      & \textit{Single chunk} & \textit{Single chunk} & \textit{Single chunk} & \textit{Single chunk} & \textit{Single chunk} & \textit{10M chunk} & \textit{Single chunk}  & \textit{10M chunk} \\ \hline
	\multirow{2}{*}{\textbf{Strabon}} & \textit{Default loader} & 7,00 sec & 19,00 sec & 292,00 sec & - & - & - & - & - \\ \cline{2-10} 
	& \textit{Bulk Loader} & - & - & 73,54 sec & 943,74 sec & 1.993,00 sec & - & 25.650,00 sec* & - \\ \hline
	\multicolumn{2}{|l|}{\textbf{GraphDB}}  & \color{darkred}47,10 sec & \color{darkgreen}4,15 sec & 64,59 sec & \color{darkgreen}157,95 sec & \color{darkgreen}1.515,64 sec & - & \color{darkgreen}7.173,00 sec & - \\ \hline
	\multirow{2}{*}{\textbf{RDF4J}} &  & \color{darkgreen}1.37 sec & 5.01 sec & \color{darkgreen}43.85 sec & 363.64 sec & X  & 4,252.30 sec & X & 169,063.759 sec \\  \cline{2-10} 
	& \textit{Lucene enabled} & 4,73 sec & \color{darkred}23,61 sec & \color{darkred}298,91 sec & \color{darkred}2.212,38 sec & X  & \color{darkred}19.201,34 sec & X & \color{darkred}357.970,89 sec \\
	\hline
	
\end{tabular}
\caption{Storing times (sec.)}
\label{table:ScalabilityStoringTime}   
\end{table*}
\begin{table*}[!t]
	\center
	\scriptsize
	\begin{tabular}{|l|l|c|c|c|c|c|c|c|c|}
		\hline
		\multicolumn{2}{|c|}{\textbf{System}} & \textbf{10K triples}  & \textbf{100K triples} & \textbf{1M triples}   & \textbf{10M triples}  & \multicolumn{2}{c|}{\textbf{100M triples}} & \multicolumn{2}{c|}{\textbf{500M triples}} \\ \hline
		\multicolumn{2}{|c|}{} & \textit{Single chunk} & \textit{Single chunk} & \textit{Single chunk} & \textit{Single chunk} & \textit{Single chunk} & \textit{10M chunk} & \textit{Single chunk}  & \textit{10M chunk} \\ \hline
		\multicolumn{2}{|l|}{\textbf{Strabon}}                      & \color{darkred}19 MB                 & \color{darkred}51 MB                 & 367 MB                & 3.1 GB                & \color{darkred}28 GB                 & -                  & 134 GB                 & -                  \\ \hline
		\multicolumn{2}{|l|}{\textbf{GraphDB}}                      & 16 MB                 & 25 MB                 & \color{darkgreen}132 MB                & \color{darkgreen}1.11 GB               & \color{darkgreen}10.9 GB               & -                  & \color{darkgreen}54 GB                  & -                  \\ \hline
		\multirow{2}{*}{\textbf{RDF4J}} & & \color{darkgreen}3.1 MB & \color{darkgreen}14 MB & 135 MB & 1.23 GB & X & 11.82 GB & X & 59.10 GB \\ \cline{2-10}
		& \textit{Lucene enabled} & 4.6 MB & 35 MB & \color{darkred}378 MB & \color{darkred}3.2 GB & X & 25 GB & X & \color{darkred}145 GB \\ \hline
		\end{tabular}
	\caption{Repository Sizes (MB, GB)}
	\label{table:ScalabilityRepositorySize}   
	\end{table*}

In order to load the data with Strabon we tried both the default loader and the bulk loader and used the most appropriate one. The default Sesame based loader is efficient for datasets of up to 100K triples but in the 1M triples the indexing costs of the PostgreSQL+PostGIS start becoming very high resulting in a time similar to the slowest response by RDF4J with Lucene indexing enabled and \texttt{x7} slower than the fastest response by RDF4J.
Strabon's bulk loader, which is a tailor-made tool, has two issues that we had to take into consideration. First, it has an initial overhead which is only worth when loading files with more than one million triples. That is why we choose not to use it for the two smaller datasets. The sweet spot for switching between the two loaders is somewhere around 1M triples. Second, it requires a memory size near the size of the dataset to load. The reason for this problem is that the RDF library used
needs to parse the entire input file in one step before storing it persistently. This situation did not allow the bulk
loader to import the 500M triple dataset and obliged us to use a server with 128 GB RAM in order to perform the first stage of the import which creates a set of CSV files. These files were afterwards transferred back to the reference server and we continued with the second step which was importing the CSV files into PostGIS database. The time \(25.650,00 sec = 14.000 sec + 11.250 sec\) of the two steps therefore is ``dirty'' and reported for completeness purposes only.

The default loader of RDF4J could not support files with more than approximately 15M-triples, therefore the 100M and 500M triples  datasets were split into 10M triples chunks and in these cases the total amount of time needed to diggest all chucks is reported. It scores acceptable times up to the 10M triples dataset and from there on its performance rapidly deteriorates. With Lucene spatial indexing the performance is extremely high for datasets over 1M triple. 

The two phase import design of GraphDB's bulk loader, the use of parallel thread execution and the very good choice of default values for its highly parametric configuration allowed it to load all datasets, but the smallest one, in a fraction of the time compared to all other systems. 

With respect to the resulting repository sizes, RDF4J is the most efficient system for the two smallest datasets and GraphDB is marginally better than RDF4J in the larger ones. The small improvement over RDF4J is that GraphDB uses POS and PSO indexes for statements and the literal index while RDF4J has SPOC, POSC. Since the GeoSPARQL plugin returned ambiguous results\footnote{\url{https://jira.ontotext.com/browse/GRAP-143}}, it was disabled and this contributed greatly to the efficient performance times achieved, as the repositories are smaller. Enabling the GeoSPARQL plugin increases the repository size by an average of 40\%. RDF4J with Lucene enabled has the same storage requirements as Strabon which more than double than GraphDB.

From Tables~\ref{table:ScalabilityStoringTime} and ~\ref{table:ScalabilityRepositorySize} it is clear that GraphDB is the most scalable system in terms of the initial loading time of datasets and the final repository size. 
		
		\subsubsection{Queries}
		\label{sec:resultsScalabilityQueries}
		The results of the scalability benchmark are shown in Figure~\ref{fig:ScalabilityResponseTimes}
where the response time of each query is reported for both cold and warm caches.
RDF4J with Lucene spatial indexing was slower and only RDF4J standard mode results are included.
Strabon is clearly the fastest system in all dataset, query and cache combinations but the smallest 
dataset with cold caches where RDF4J is the fastest. RDF4J is also faster than GraphDB in the most demanding 
SC2 spatial join query where high thematic selectivity results in full table scan to produce the result. The
same holds to a lesser degree for queries SC1 and SC3 but for the two smallest datasets. GraphDB on the 
other hand is much faster than RDF4J in the moderate SC3 spatial join query which has a smaller thematic 
selectivity and is clear that filters results on the thematic part before proceeding with the evaluation
of the spatial part. For the 500M triples dataset, GraphDB did not complete query SC2 in the 24 hour limit
while RDF4J failed to complete queries SC2 and SC3.

\begin{figure*}[!t]
	\centering
	\subfigure{\includegraphics{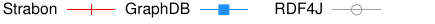}}
	\setcounter{subfigure}{0}
	\subfigure{\includegraphics{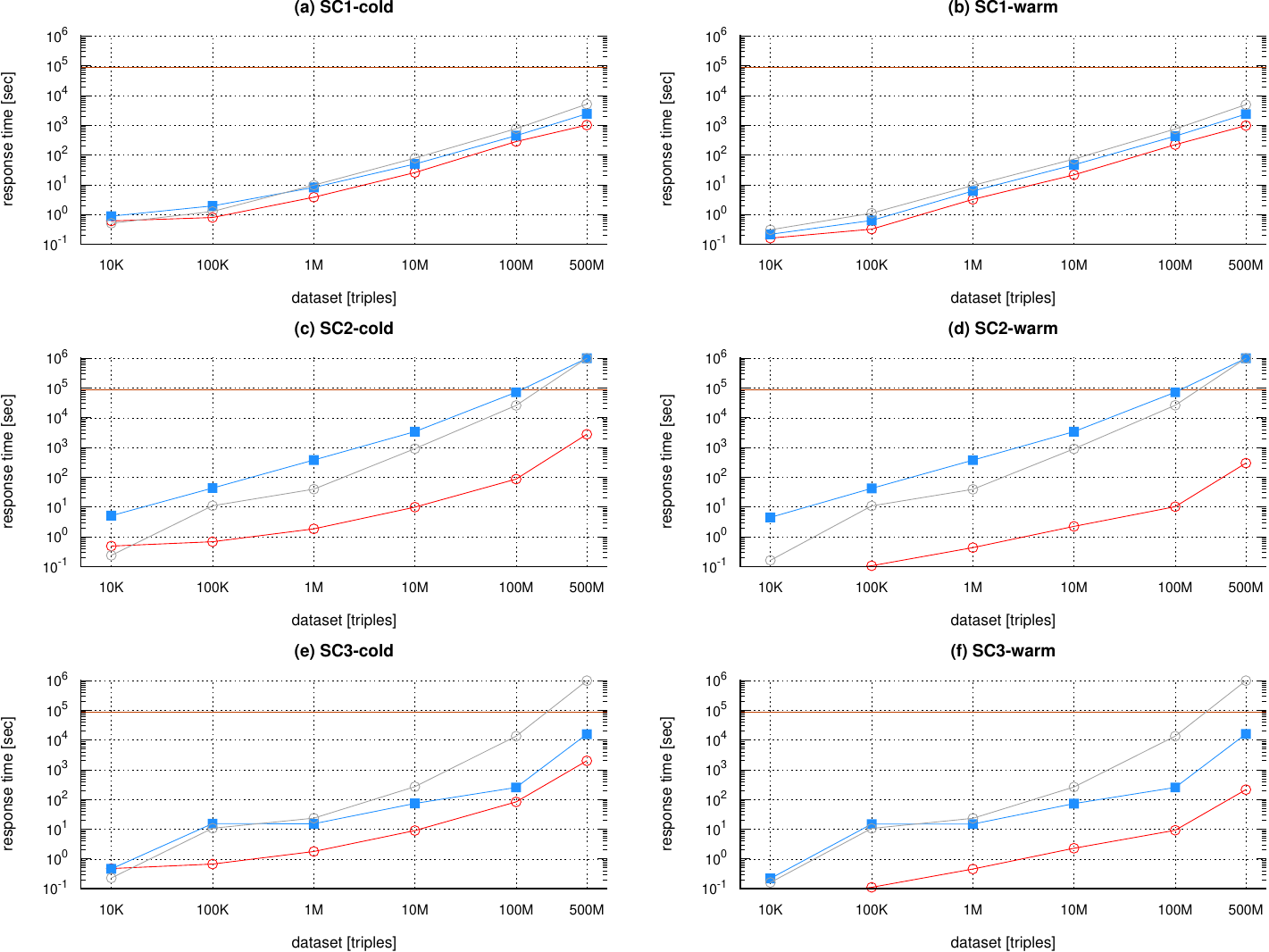}}
	\setcounter{subfigure}{1}
	\caption{Response times - Scalability workload}
	\label{fig:ScalabilityResponseTimes}
\end{figure*}
		
\iflong
	\section{Evaluating the Performance of RDF Stores with Limited Geospatial Capabilities}
	\label{sec:resultPoints}
	Apart from the RDF stores that we have already tested, there are also some RDF 
stores that provide geospatial capabilities only for points.
Indexing and evaluating queries only for simple geometry types (points) allows
to use different index and query evaluation methods than that for more complex
geometry types, like polygons and lines. For example, points can be indexed using 
two B-trees or a point quadtree, while polygons are usually indexed using an R-tree.
In order to find out if there is a performance trade-off between these two approaches
, this section evaluates  the performance of two
RDF stores that provide limited geospatial capabilities 
(support only points) and compare them with the 
geospatial RDF stores which are tested in the previous sections. 
For this purpose Virtuoso v7.1 and a proprietary RDF store 
(which we will call System Y) are used. Virtuoso and System Y support only point geometries. Thus, 
for the real world and synthetic workloads of Geographica 2, the parts of the datasets 
that contains only point geometries are kept  and only the queries which 
handle point geometries are rerun in the geospatial RDF stores Strabon, uSeekM,
Parliament, System X, GraphDB, RDF4J and the limited-functionality systems Virtuoso and System Y.

\subsection{Real World Workload}
The real world workload used in this case consists of only the corresponding 
datasets from DBpedia and GeoNames. Because Virtuoso and System Y 
do not provide any non-topological function only spatial selections
and spatial joins were tested. Also, the macro benchmark is not used in this section, since 
all scenarios use more complex geometry types than points.

\begin{figure*}[!t]
\begin{minipage}[b][][b]{\textwidth}
	\begin{minipage}[b][][b]{0.43\textwidth}
		\centering
		{	\tiny
	\scalebox{0.85}{
	\begin{tabular}{|c|c|c|c|c|c|c|c|c|}
		\hline
		\textbf{Workload} & \textbf{Strabon} & \textbf{uSeekM} & \textbf{Parliament} & \textbf{System X} & \textbf{GraphDB} & \textbf{RDF4J} & \textbf{Virtuoso} & \textbf{System Y}  \\
		\hline
		\textbf{Real world} & 65 & 45  & 36 & \color{darkred} 68 & 62 & 26 & 9 & \color{darkgreen}8   \\
		\textbf{Synthetic} & 496 & 771 & \color{darkred} 874 & 418 & 116 & 306 & 63 & \color{darkgreen} 31   \\
		\hline
	\end{tabular}
}
}
		\captionof{table}{Storing times (sec.). Real world: GeoNames, DBpedia. Synthetic: PointsOfInterest}
		\label{table:StoringTimePoints}      
	\end{minipage}
	\hfill
	\begin{minipage}[b][][b]{0.43\textwidth}
		\centering
		{	\tiny
	\scalebox{0.85}{
	\begin{tabular}{|c|c|c|c|c|c|c|c|c|}
		\hline
		\textbf{Workload} & \textbf{Strabon} & \textbf{uSeekM} & \textbf{Parliament} & \textbf{System X} & \textbf{GraphDB} & \textbf{RDF4J} & \textbf{Virtuoso} & \textbf{System Y}  \\
		\hline
		\textbf{Real world} & 338 & 124  & 238 & 534 & 220 & \color{darkgreen}122 & 213 & \color{darkred}657  \\
		\textbf{Synthetic} & 2201 & 1871 & 2864 & 2865 & 733 & 1031 & \color{darkgreen}703 &  \color{darkred}3084  \\
		\hline
	\end{tabular}
}
}	
		\captionof{table}{Repository sizes (MB). Real world: GeoNames, DBpedia. Synthetic: PointsOfInterest}
		\label{table:RepositorySizePoints}
	\end{minipage}
\end{minipage}
\end{figure*}

\subsubsection{Dataset Storage}
The storage times for the real world workload are presented in 
Table~\ref{table:StoringTimePoints} and Table~\ref{table:RepositorySizePoints}
presents the storage space required in each case. For this subset of Geographica 2, 
we stored only the corresponding datasets of DBpedia and GeoNames.
In this table we observe that Virtuoso and System Y
need considerably less time to store and index the real 
datasets than the other systems. They provide dedicated bulk 
loaders which achieve better storage times in comparison to storage times of the
full geospatial RDF stores which either use Sesame, RDF4J and Jena Java API for loading data
(e.g., uSeekM, GraphDB, Parliament) or they provide bulk loaders that perform complex
processing of the input data (e.g., Strabon, System X).

The space needs of the two RDF stores with point geometry
functionality differ significantly. System Y allocates a large amount of space, even more than the
full geospatial RDF stores. Virtuoso, on the other hand, needs little space which, however, is almost twice as much as RDF4J and uSeekM.

\begin{figure*}[!t]
	\centering  
		\subfigure
		{\includegraphics[width=0.3\textwidth]{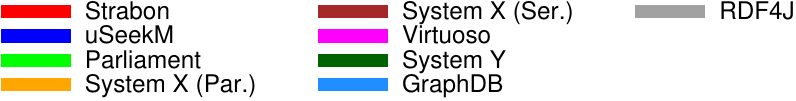}}
		\setcounter{subfigure}{0}
		\\
		\subfigure[][Cold caches]
		{\includegraphics[width=0.45\textwidth]{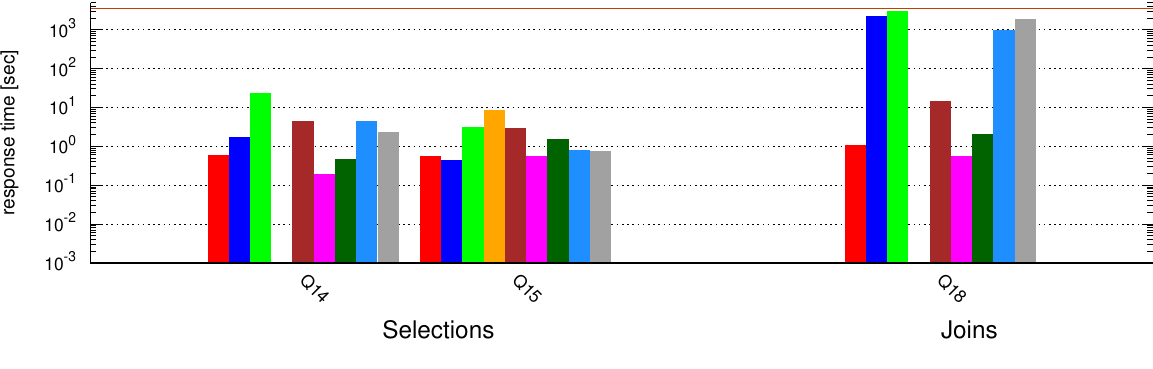}
		\label{fig:RealCold}}
		\hspace{2em}
		\subfigure[][Warm caches]
		{\includegraphics[width=0.45\textwidth]{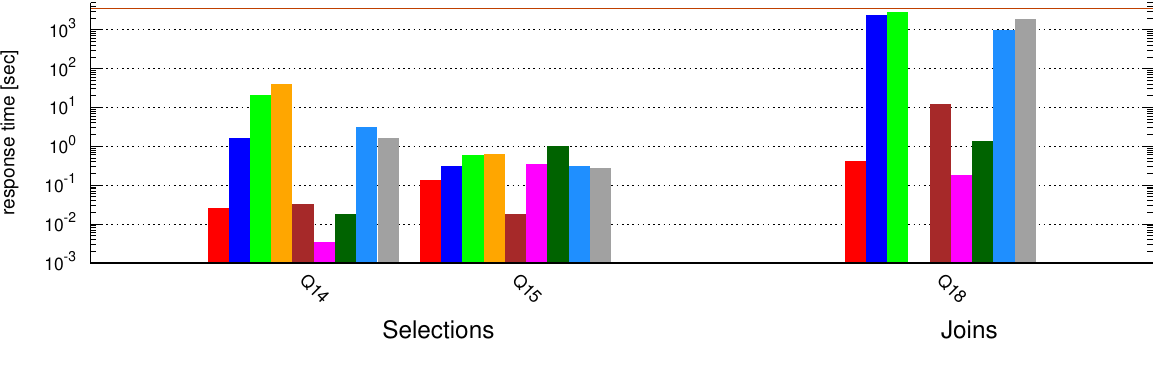}
		\label{fig:RealWarm}}
		\hspace{-1em}
		\\
		\caption{Response times, real world workload}
		\label{fig:RealResponseTimesPoints}
\end{figure*}

\subsubsection{Queries}
Given the spatial selections that are supported by Virtuoso and
System Y, queries Q14 and Q15 (spatial selections) have been tested. 
As described in Section~\ref{sec:resultsRealMicro}, these queries
ask for points that lie within a given distance from a given point, but each query uses  
a different function. Also Query Q18 (spatial join) was tested which 
asks for pairs of points of the datasets GeoNames and DBpedia that are equal.
Virtuoso does not offer functions to create a buffer. Instead it provides the 
functions \texttt{bif:st\_within}, 
\texttt{bif:st\_intersects}, and \texttt{bif:st\_contains} which receive a third 
argument that is a tolerance value for matching in units of linear distance.
In order to emulate Query Q15 in Virtuoso, instead of creating a buffer of the 
given point with radius $r$ and asking for points inside the buffer, we can ask 
for points inside the given point with tolerance $r$.

The response times of these queries are reported in 
Figure~\ref{fig:RealResponseTimesPoints}. Query Q14 has a complex filter clause 
(point inside a point buffer) but if the point buffer is computed the spatial 
index can be used to evaluate the query. Thus, the majority of RDF stores that 
utilize the spatial index to evaluate this query (Virtuoso, System Y, and 
Strabon) respond to it faster than uSeekM, Parliament, System X serial, GraphDB and RDF4J 
which do not use an index. System X parallel did not complete the evaluation of Q14.
Especially Virtuoso, which is not burdened with the cost of evaluating a topological relation over complex 
geometries (like a buffer of a point), achieves the best response time. Query Q15 
does not favor the use of spatial index but its filter clause (distance between 
points) is executed faster than the filter clause of query Q14 (point inside point 
buffer). Thus, uSeekM, Parliament, GraphDB and RDF4J need less time to respond to this query than 
Query Q15. uSeekM needs the least time of all systems. Virtuoso and 
System Y need slightly more time to respond to query Q15 than Q14, 
because they do not use their spatial indexes.
In the case of spatial join (query Q18) Virtuoso has the fastest response time while 
Strabon comes second. System X (in parallel mode) needs more time than the
one hour limit to evaluate this join, as well as in the full micro-benchmark.

\subsection{Synthetic Workload}
 
The synthetic workload used the generator, which is 
described in Section~\ref{sec:syntheticDataset}, to generate a dataset
and only the generated points of interest were stored.
Because only points were used we chose to set the
generator parameters $n=1024$ and $k=10$ and generate a 
bigger dataset that contains about seven million triples.

\subsubsection{Dataset Storage}
The corresponding storage times and allocated storage space are shown in Tables 
\ref{table:StoringTimePoints} and \ref{table:RepositorySizePoints}. 
As well as for the real world workload, Virtuoso and System Y need less time to store
the dataset of the synthetic workload than the full geospatial RDF stores.
For the synthetic dataset which is bigger than the real world dataset the low
requirements in storage space of Virtuoso, GraphDB and RDF4J are more emphasized. While Virtuoso is
slightly more compact than GraphDB, both stores need
about two times less space than uSeekM and three times less space than the other RDF stores. On the other hand, System Y has higher space requirements than
the full geospatial RDF stores.

\begin{figure}[t]
	\centering    
	\includegraphics[width=.6\columnwidth]{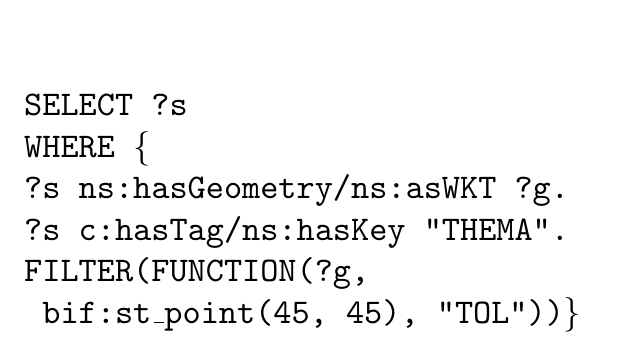}
	\captionof{table}{Query template for synthetic queries of Virtuoso}
	\label{tbl:syntheticQueryTemplateOnlyPoints}
\end{figure}

\subsubsection{Queries}
For this subset of Geographica 2 only spatial selections using
topological relation \texttt{geof:sfWithin} were executed. 
Since functions in Virtuoso cannot receive a rectangle as argument, the respective
queries that were run by Virtuoso are produced by instantiating the template in 
Table~\ref{tbl:syntheticQueryTemplateOnlyPoints}. The parameter \texttt{TOL} is 
the tolerance value that will be used by Virtuoso for evaluating the 
topological relation defined by parameter \texttt{FUNCTION}. So a
circle is considered by \texttt{bif:st\_within} and the radius of the circle
(defined by the parameter TOL)
is instantiated to achieve each time the proper spatial selectivity.

The response times for these queries are presented 
in Figure~\ref{fig:SyntheticResponseTimesPoints}.
For high thematic selectivity \texttt{tag=1} Virtuoso is the fastest system while Strabon comes second.
For low thematic selectivity \texttt{tag=1024} GraphDB and Strabon are at the top.
Both System Y and Virtuoso executed all queries by starting with the 
spatial part of the query and then continuing with the thematic part and this
is why their performance is affected more by the spatial  selectivity of
the query than by the thematic.
For example, when the value of the parameter THEMA is 1 
(Figures~\ref{fig:SyntheticPointsWithin1Cold}, \ref{fig:SyntheticPointsWithin1Warm})
Virtuoso needs the shortest time to evaluate the spatial selections. But when the value of 
the parameter THEMA is 1024 (Figures~\ref{fig:SyntheticPointsWithin1024Cold}, 
\ref{fig:SyntheticPointsWithin1024Warm}) Virtuoso does not exploit the fact that
few points satisfy the thematic part of the query and its response time increases
while the spatial selectivity is increased and more points satisfy the spatial predicate.
Thus, GraphDB and Strabon, which changes its execution path if the spatial selectivity is 
increased, respond to these queries faster.

\subsection{Summary}
This section compared the performance of RDF systems that implement GeoSPARQL 
with general purpose RDF systems which provide limited spatial functionality.
Regarding data storage, Virtuoso and System Y provide the best bulk loading capabilities 
while Virtuoso and GraphDB, have very low space requirements. 
Regarding query evaluation Virtuoso, Strabon and GraphDB have the best performance. Finally,
the query optimizers of Virtuoso and System Y do not take into account the selectivity of a spatial predicate. However, this does not lead to bad performance because of their fast 
search mechanisms, even if the thematic selectivity of a query is greater than the spatial.

\begin{figure*}[!t]
	\centering
		\subfigure{\includegraphics{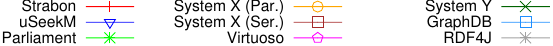}}
		\\
		\setcounter{subfigure}{0}
		\subfigure[][\shortstack{Within\\ \normalsize{ tag 1, cold caches}}]
		{\includegraphics[scale=1.2]{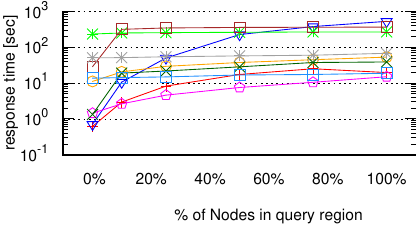}
		\label{fig:SyntheticPointsWithin1Cold}}
		\subfigure[][\shortstack{Within\\ \normalsize{tag 1024, cold caches}}]
		{\includegraphics[scale=1.2]{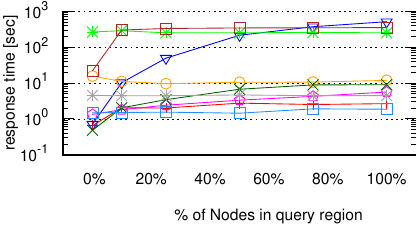}
		\label{fig:SyntheticPointsWithin1024Cold}}
		\hspace{-1em}
		\\
		\subfigure[][\shortstack{Within\\ \normalsize{ tag 1, warm caches}}]
		{\includegraphics[scale=1.2]{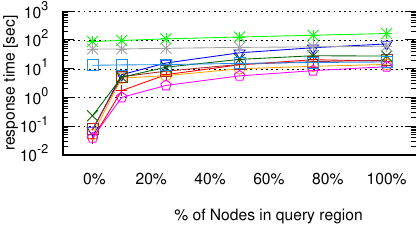}
		\label{fig:SyntheticPointsWithin1Warm}}
		\subfigure[][\shortstack{Within\\ \normalsize{ tag 1024, warm caches}}]
		{\includegraphics[scale=1.2]{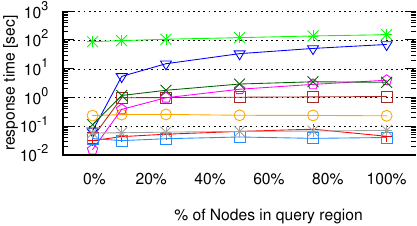}
		\label{fig:SyntheticPointsWithin1024Warm}}
		\hspace{-1em}
		\\
		\caption{Response times - synthetic workload (selections)}
		\label{fig:SyntheticResponseTimesPoints}
\end{figure*}

\fi

\section{Summary and Future Work}
\label{sec:conclusions}

In this section, we summarize the work presented in this article, and discuss 
the limitations and future extensions of our work.

\subsection{Summary}
\label{subsec:conclusions}

This article presents a benchmark for evaluating geospatial RDF stores. 
First, it presents a functional comparison of well known geospatial RDF stores. 
Then, it compares their performance according to three workloads: the real world,
the synthetic and the scalability workload.

The real world workload is based on real data and is separated into two
parts. The micro part tests geospatial operations in isolation and aims
at stressing the  spatial module of RDF stores. A conclusion that
can be drawn from this part is that integration of a spatial
module in most of the geospatial RDF stores is not mature. 
Most of the poor performance issues observed were either because the spatial index
was not properly utilized or due to inefficiencies of the spatial relation 
evaluation engine (e.g., not being optimized for complex geometries). 

The macro part of the real world workload evaluates
the performance of RDF stores in simulations of real application scenarios. Various application scenarios were specified that range from simple scenarios 
(e.g., "Geocoding", "Reverse Geocoding", "Map Search and Browsing") to 
more complicated scenarios that serve domain expert needs (e.g., "Rapid Mapping for
Fire Monitoring", "Computing Statistics of Geospatial Datasets"). 
uSeekM has the best performance for queries that consist of simple
spatial operations (e.g., spatial selections), like "Geocoding", "Reverse Geocoding", 
"Map Search and Browsing", and "Computing Statistics" of Geospatial Datasets. For more
complex applications that include both spatial joins or spatial aggregations, like
"Rapid Mapping for Fire Monitoring" Strabon is the only RDF store that performed well. 
System X and Parliament performed well only for some scenarios, e.g. "Geocoding" and "Reverse Geocoding"
respectively, but they had always worse performance than uSeekM or Strabon.
So, every scenario can be well served by at least one RDF store. 
This means that there are already
implementations capable of being used in real applications and bringing the merits
of linked data in the geospatial domain.

The synthetic workload uses synthetic data of arbitrary size and queries with various 
thematic and spatial selectivities and tests whether spatial query processing is 
deeply integrated in their query engines. The results of this workload
highlight the importance of spatial statistics and using
them to select appropriate query execution paths. RDF stores that do so manage to have
a good performance for all combinations of spatial and thematic selectivity. While,
other RDF stores that do not take into consideration the spatial selectivity of a 
query and stick to only one type of execution pat (e.g., always execute the
spatial part of a query and then the thematic part) do not always achieve good performance.

The scalability workload is based on a set of increasingly larger subsets of the union of two real world
datasets, OSM and CORINE 2012. Three systems participated in this test and they were
selected based on being actively maintained and being representative of each one of
the different architectures of the RDF stores identified, namely RDF frameworks (RDF4J), NoSQL RDF stores (GraphDB) and 
hybrid RDF stores with RDBMS as a backend (Strabon). One spatial selection query and two spatial join queries,
a demanding and a moderate one, were used to stress the systems against datasets up to 500M triples. The
infrastructure used was a small system by today's standards and helped each system to show its limits
early on. Strabon which belongs to the hybrid architecture proved to be the most efficient one in answering
all queries but faced problems with its bulk loader over 100M triples. GraphDB achieved exceptional performance in bulk loading
and storage size but, as RDF4J, was not able to answer the three queries fast enough. 
Programmatic operation for GraphDB's stores
with the GeoSPARQL plugin enabled was not possible because of runtime errors, so a question remains about how well GraphDB would have performed
in queries using spatial predicates. 
Although RDF4J  was not able to manage more than 100M-triples datasets with geospatial
data, it proved that it can be considered as the basis for building more complete horizontally scalable geospatial RDF stores which
will provide a better spatial indexing mechanism both performance and storage wise. 
 
Finally, a comparison between RDF stores that provide full geospatial 
functionality and RDF stores with limited geospatial capabilities was performed, trying to find out 
whether ad-hoc implementations supporting only points would perform better  than 
full geospatial implementations. RDF stores with limited geospatial 
capabilities performed very well, especially at bulk loading and at spatial 
selections. However, the performance difference with some 
full geospatial RDF stores is not as much as expected. Especially in cases
when the spatial part of a query selects a lot of spatial features, some full
geospatial RDF stores become faster than the RDF stores that support only points.

\subsection{Limitations and Future work}
\label{subsec:limitations}

The real world and synthetic workloads used in Geographica are relatively small and they covered only a limited geographic extent, such as Greece or New York. However, as the results of the experimental evaluation showed, they were enough to stress all systems that were evaluated. By adding the scalability workload we raised considerably the size of the datasets to 500M triples with 100GB of geospatial data that covered many European countries and contained highly complex geometries. By replacing OWLIM with its successor GraphDB and introducing the geospatially enabled RDF4J successor of Sesame RDF store we included the newest developments in systems of this area. 

In future work, we plan to include in the benchmark the newest Virtuoso version which offers some GeoSPARQL features;
we have not been able to do it in this version due to problems with the current implementation as we discussed in 
Section \ref{sec:geographica-functional:limited}.

Given that there are today institutions such as cartographic agencies (e.g., Kadaster in the Netherlands\footnote{\url{https://www.kadasterdata.nl/}}) that manage TBs of geospatial data
and make some of it available as linked data, it is important to develop RDF stores that can manage
\textit{big linked geospatial data}~\cite{DBLP:journals/internet/KoubarakisBPSS17}.
This is currently done in the European project Extreme Earth\footnote{\url{http://earthanalytics.eu/}} 
that our group coordinates. Extreme Earth studies big linked geospatial data coming from the
Earth observation programme Copernicus\footnote{\url{https://www.copernicus.eu/en}} of the
European Union.

\subsection{Acknowledgements}
\label{subsec:acknowledgements}
We would like to express our gratitude to the development and support groups of all the systems
that were included in this work. We are grateful to people from the RDF4J project, 
the Ontotext group for the GraphDB system, the OpenLink Software group for Virtuoso Open-Source server, and
the developers of System X who provided insight and directions for achieving the best possible results in 
our experiments.

\iflong
	\bibliographystyle{elsarticle-num}
	\bibliography{bibliography}
\fi

\end{document}